\documentclass[11pt]{report}
\usepackage{amsmath}
\usepackage{sw20mitt}
\usepackage{graphicx}
\usepackage{amsfonts}
\usepackage{amssymb}
\newtheorem{theorem}{Theorem}
\newtheorem{acknowledgement}[theorem]{Acknowledgement}

\begin{document}

\title{Investigations in quantum games using EPR-type set-ups}
\author{Azhar Iqbal}
\prevdegrees{BSc (Hons), University of Sheffield, UK, 1995}
\department{Centre for Mathematics}
\thisdegree{Doctor of Philosophy}
\university{University of Hull, Hull, UK}
\degreemonth{January}
\degreeyear{2006}
\date{January 2006}
\chairmanname{Dr. Timothy Scott}
\chairmantitle{Head of the Centre for Mathematics}
\super{Dr. Timothy Scott}
\supertitle{Senior Lecturer}
\maketitle
\tableofcontents
\begin{abstract}
Research in quantum games has flourished during recent years. However, it
seems that opinion remains divided about their true quantum character and
content. For example, one argument says that quantum games are nothing but
`disguised' classical games and that to quantize a game is equivalent to
replacing the original game by a different classical game. The present thesis
contributes towards the ongoing debate about quantum nature of quantum games
by developing two approaches addressing the related issues. Both approaches
take Einstein-Podolsky-Rosen (EPR)-type experiments as the underlying physical
set-ups to play two-player quantum games. In the first approach, the players'
strategies are unit vectors in their respective planes, with the knowledge of
coordinate axes being shared between them. Players perform measurements in an
EPR-type setting and their payoffs are defined as functions of the
correlations, i.e. without reference to classical or quantum mechanics.
Classical bimatrix games are reproduced if the input states are classical and
perfectly anti-correlated, as for a classical correlation game. However, for a
quantum correlation game, with an entangled singlet state as input,
qualitatively different solutions are obtained. Reproducing the properties of
quantum correlation games appears to be conceptually impossible within the
framework of classical games. The second approach in the present thesis is
based on the fact that all derivations of the Bell inequalities assume local
hidden variable (LHV) models which produce a set of positive-definite
probabilities for detecting a particle with a given spin orientation. In
recent years it has been shown that when the predictions of a LHV model are
made to violate the Bell inequalities the result is that some probability
measures assume negative values. With the requirement that classical games
result when the predictions of a LHV model do not violate the Bell
inequalities, our analysis looks at the impact which the emergence of negative
probabilities has on the solutions of two-player games which are physically
implemented using the EPR-type experiments.
\end{abstract}

\vspace*{3.35in}

\begin{center}
{\Large Dedication}
\end{center}

\textit{Dedicated to my wife Ayesha whose encouragement and support was
essential in the writing of this thesis.}

\begin{description}
\item \bigskip
\end{description}

\newpage

\vspace*{3.35in}

The research presented in this thesis is drawn from the following publications:

\begin{itemize}
\item  Azhar Iqbal and Stefan Weigert, Quantum correlation games. J. Phys. A:
Math. Gen. \textbf{37,} 5873-5885 (2004).

\item  Azhar Iqbal, Playing games with EPR-type experiments. J. Phys. A: Math.
Gen. \textbf{38,} 9551-9564 (2005).
\end{itemize}

\pagebreak 

\chapter{Introduction}

Information processing \cite{Shannon} has been traditionally considered a
purely mathematical task that is independent of the carriers of the
information. This view underwent dramatic revision and change in the 1990s
with the finding that highly efficient ways of processing exist for certain
problems if information is stored and processed quantum mechanically.
Identifying a marked object in a database \cite{Grover} and factorization of
large integer numbers \cite{Shor} were recognized to have more efficient
solutions than are possible classically. This finding gave birth to the new
field of \emph{quantum information and computation }\cite{Nielsen,Hoi-Kwong},
in which classical bits are generalized to \emph{qubits} which allow linear
combination of classically incompatible states. the use of qubits as carriers
makes possible much faster processing of information, at least for certain problems.

In contrast to the recently developed studies of quantum information and
computation, game theory \cite{Neumann,Rasmusen} is an established branch of
mathematics which helps players participating in games to take decisions
rationally. The finding that information theory can benefit from a
quantum-mechanical implementation has motivated the proposal
\cite{MeyerDavid,Eisert} that game theory is another promising candidate to
gain a similar benefit. The roots of this suggestion can perhaps be traced
back to the understanding that information is\emph{ }a central element of
\emph{any} game. The outcome of a game is often decided by what (and how much)
information players possess at different times during the course of playing
the game. With information given a central place in the playing of a game, the
possibility of efficient information processing has consequences for the
solutions of the game when it is played using qubits. That is, the
solutions/outcomes of a game are dependent on the nature of the physical
objects used in its playing.

Players playing classical games share coins or dice and select from a set of
available strategies by performing moves, or actions, on the coins or dice.
Rational players select strategies that maximize their payoffs. Using this
analogy, the players in a quantum game are endowed with the capacity to make
quantum mechanical moves, or actions, on the qubits which they share.

The following are some of the stated reasons \cite{Eisert} why it is
interesting to play games in quantum regime:

\begin{itemize}
\item  Game theory involves concepts and methods of probability theory
\cite{Chung} to analyze games, in the process of finding their solutions.
Playing quantum games provides an opportunity to generalize conventional
probability to quantum probability \cite{QProbability}.

\item  Playing games has an intimate connection with quantum communication
concepts \cite{QCommunication}. For example, the quantum-mechanical protocols
for eavesdropping \cite{Ekert,Gisin} and optimal cloning \cite{WernerRF} can
readily be formulated as games between players.

\item  It is possible to re-formulate certain quantum algorithms as games
between classical and quantum players \cite{MeyerQGsQAlgorithms}. Because only
a few quantum algorithms are known to date, it has been suggested that quantum
games may shed new light on the working of quantum algorithms, possibly
helping to find new ones.

\item  Quantum mechanics has been shown to assure fairness in remote gambling
\cite{Goldenberg}. Gambling and playing games are related, thus making quantum
mechanics relevant for the latter.

\item  Molecular level interactions are dictated by quantum mechanics. If
Dawkins' dictum of the `Selfish Gene' \cite{Dawkins} is a reality then the
games of survival are already being played in such interactions. Quantum
mechanics naturally becomes relevant for those games.

\item  It has been suggested \cite{Pietarinen} that games can provide a useful
set of tools in giving a semantics to quantum logic and can be brought to bear
on questions concerning the interpretation and the nature of the concept of
uncertainty in the foundations of quantum theory.
\end{itemize}

Research in quantum games has seen noticeable growth \cite{ArxivSearch} during
recent years. Though game-like descriptions of surprising quantum-mechanical
situations can be found in the literature dating back many
decades\footnote{Refer to the literature review of quantum games in the
Section ($4.5$).}, quantum games emerged as a new field of research with
Meyer's publication of a quantum penny-flip game \cite{MeyerDavid}. Meyer's
game demonstrated the advantage that quantum strategies can attain over
classical ones. Shortly afterwards, Eisert et al. \cite{Eisert} put forward a
quantum version of the well-known game of Prisoners' Dilemma. Other
suggestions and schemes include a proposal for multi-player quantum games
\cite{Benjamin Multiplayer}, the quantization of Battle of Sexes
\cite{Marinatto}, a study of evolutionary stability in quantum games
\cite{IqbalToor}, the experimental realization of quantum games on a quantum
computer \cite{Du}, the quantum Monty Hall problem \cite{FlitneyAbbott}, the
quantum Parrondo's game \cite{FlitneyAbbott1} and a discussion of quantum
advantage in the presence of a corrupt source \cite{Ozdemir et al}. The list
is by no means exhaustive and can easily be extended considerably by a more
detailed literature review. However, it demonstrates the rapidly growing
research interest in the new field.

There have been many developments in quantum games during recent years
\cite{ArxivSearch} but opinion remains divided \cite{Enk,Meyer's
Reply,Benjamin1,Eisert's Reply,EnkPike} about their `true' quantum character.
For example, one argument \cite{EnkPike}\ considers quantum games only as
`disguised' classical games. The argument says that to quantize a game is
equivalent to replacing the original game by a \emph{different} classical game.

The present thesis defends quantum games by presenting a reply to the
criticisms surrounding this emerging field. It is argued that the possibility
of constructing a classical game which is able to reproduce the overall effect
of a quantum game cannot be used to deny the quantum content \cite{Enk} of
quantum games. The question which quantum game theory asks is how the quantum
aspects of distributed physical systems which are used to physically implement
games can leave their mark on game-theoretical solutions. The possibility of
classical constructions which can describe the overall situation of quantum
games does not make the question disappear.

The Einstein-Podolsky-Rosen (EPR) paradox \cite{EPR}\ is widely believed to
provide an example of a phenomenon having a truly quantum character and
content. The paradox involves two spatially-separated observers making local
measurements on singlet states. The paradox motivated the EPR experiments
\cite{EPR,Bell,CHSH2,Aspect et al}\ whose set-up provides an almost natural
arrangement for playing two-player games. This thesis argues that questions
about the quantum content of quantum games can be addressed by using the
EPR-type experiments to play two-player games. The thesis puts forward two
proposals to play quantum games. In both the proposals the set-up of EPR-type
experiments is used in playing two-player games.

In the first proposal, a two-player classical game is firstly re-expressed in
terms of EPR-type experiments, when they are performed on classical, but
anti-correlated, pairs of particles \cite{AsherPeres}. In the next step the
experiments are performed using anti-correlated quantum mechanical pairs of
particles. The resulting effects on the players' payoff functions, and on the
solutions of the game, are then explored in relation to the quantum
correlations existing between the parts of the pairs used in the experiments.

The second proposal follows a different line and is based on the analysis of
probabilities involved in a simple two-player game, written as a bimatrix,
when it is physically implemented by tossing four biased coins. In each turn
both players receive two coins; selecting one coin to toss is a player's
strategy. Players' payoffs are obtained after repeating the tosses in large
number. The construction allows us to consider the impact on the solutions of
the game of the feature that certain probability measures can assume negative
values \cite{FenymanNProb,FeynmanNProb1}. Essentially, the motivation of the
argument arises from the recent results \cite{Rothman,Han et al,Cereceda}
reporting that the requirement that certain hidden variable models should
predict the outcomes of EPR-type experiments forces the conclusion that some
of the involved probability measures must have negative values. In this
approach, for obvious reasons, it is required that the existence of a hidden
variable model leading to positive-definite probabilities \cite{Rothman}
always result in the corresponding classical game.

The concluding chapter summarizes the results and suggests future directions
of research.

\chapter{Introduction to game theory}

\emph{Game theory} \cite{Neumann,Rasmusen} is an established discipline of
mathematics. It is a vast subject having a rich history and content. Only
recently have its concepts and methods been brought into the realm of quantum
mechanics, thus giving rise to the new field of quantum games. In the
following review some of the fundamental concepts are described. The review is
not meant to be an exhaustive introduction to the field of game theory; a
selection of topics from the theory is made in order to serve as a background
for later chapters of this thesis.

\section{History}

Games such as chess, warfare and politics have been played throughout history.
Whenever individuals meet who have conflicting desires and priorities then
games are likely to be played. The analysis and understanding of games has
existed for a long time but the emergence of game theory as a formal study of
games is a relatively recent event. Roughly speaking, game theory is the
analysis of rational behavior when participants' actions are strategically
interdependent and when a participant's strategy depends on what his opponents
do. This situation appears quite similar to the typical scenario in
\emph{decision theory} \cite{DecisionTheory}. There is a difference, however.
A decision maker in decision theory chooses among a set of alternatives in the
light of their possible consequences, whereas decisions in game theory are
made in an environment where various players interact strategically.

The roots of game theory, as a formal discipline, can be traced back to the
year 1713 when James Waldegrave and Pierre-R\'{e}mond de Montmort
\cite{GameTheoryHistory} provided the first known minimax mixed strategy
solution to a two-player game. During the 1920s \'{E}mile Borel made
investigations on strategic games and defined what is known as the
\emph{normal form} of a game. It is a matrix representation of a game in which
a player can work out the best strategy without considering the sequence of moves.

In 1928 John von Neumann \cite{NeumannHistoryGames} gave the first formal
proof of the minimax theorem for two-player games. Essentially, the theorem
states that for each player in a zero-sum game a unique mixed strategy exists
such that payoffs are equalized regardless of the other player's actions. In
1944 Neumann and Morgenstern \cite{Neumann} published their pioneering book
``The Theory of Games and Economic Behaviour''. As well as expounding
two-person zero-sum games this book is considered the seminal work in
cooperative games. The book presented the account of axiomatic utility that
led to its widespread adoption within economics. The book established game
theory as a field of investigation in economics and mathematics.

Minimax theory finds the best strategy for a player in a zero-sum game,
independently of the strategies played by other players. Most often a player's
best strategy does depend on what strategies the other players play. In 1950
John Nash \cite{JohnNash} extended the minimax theory to $N$-player
noncooperative games and introduced the concept of a Nash equilibrium
\cite{JohnNash1}. The Nash equilibrium theory states that a set of strategies
can be found such that no player is left with a motivation to deviate
unilaterally from it.

The 1970s saw game theory being successfully applied to problems of
evolutionary biology. The concept of utility from economics was given an
interpretation in terms of Darwinian fitness. Players were dissociated from
their capacity to act rationally and the concept of \emph{evolutionary
stability} was born. In 1973 John Maynard Smith and G. R. Price
\cite{SmithPrice} introduced the concept of \emph{evolutionarily stable
strategy} \cite{Smith} as a strategy that cannot be invaded if a population
adapts it.

During recent times, Reinhard Selten further refined the concept of a Nash
equilibrium with his concept of trembling hand perfect and subgame perfect
equilibria. Also, John Harsanyi developed the analysis of games with
incomplete information. In 1994, John Nash, Reinhard Selten and John Harsanyi
\cite{HarsanyiSelten} won the Nobel Prize \cite{NobelPrize}\ in Economics for
their work on game theory.

\section{Fundamental concepts}

To describe a situation in which decision-makers interact we need to specify
who the decision-makers are, what each decision-maker can do, as well as each
decision-maker's payoff from each possible outcome. A \emph{game} consists of
a set of \emph{players}, a set of actions (sometimes called \emph{strategies})
for each player, and a \emph{payoff function} that gives the player's payoff
to each list of the players' possible actions. An essential feature of this
definition is that each player's payoff depends on the list of all the other
players' actions. In particular, a player's payoff does not depend only on her
own action.

There are two distinct but related ways of describing a game mathematically.
The first one is known as the \emph{normal }or \emph{strategic form} which is
representation of a game consisting of

\begin{itemize}
\item  A finite set of players: $\left\{  A,B,\cdots,G\right\}  $

\item  Strategy spaces of players, denoted by $S_{A},S_{B}$ etc.

\item  Payoff functions of players. For example for the player it is the
function $P_{A}:S_{A}\longrightarrow\mathbf{R}$
\end{itemize}

A \emph{strategy} is a complete plan of action for every stage of the game,
regardless of whether that stage actually arises in play. Each player is given
a set of strategies. A \emph{strategy space} for a player is the set of all
strategies available to that player. A \emph{pure strategy} refers to a
situation when a player chooses to take one action with probability $1$. A
\emph{mixed strategy} describes a strategy involving a probability
distribution which corresponds to how frequently each move is chosen. For
example, a mixed strategy of player $A$ is a convex combination of pure strategies:%

\begin{equation}
s_{A}=\sum_{i}c_{i}S_{Ai}%
\end{equation}
where $S_{Ai}$ are $A$'s pure strategies and $c_{i}$ are real constants with
$\sum_{i}c_{i}=1$. A \emph{totally mixed strategy} is a mixed strategy in
which the player assigns strictly positive probability to every pure strategy.
The concept of strategy is sometimes (wrongly) confused with that of a
\emph{move}. A move is an action taken by a player during some moment in a
game (e.g., in chess, moving white's Bishop from a2 to b3). A strategy on the
other hand is a complete algorithm for playing the game, implicitly listing
all moves and counter-moves for every possible situation throughout the game.

An \emph{extensive form} of a game specifies the players of a game, every
opportunity each player has to move, what each player can do at each move,
what each player knows for every move, and the payoffs received by every
player for every possible combination of moves. In comparison to the normal
form, the extensive form of a game captures the order of play and reveals how
equilibria are determined. The extensive form is the most detailed way of
describing a game. A \emph{game tree} represents a game in such a way that
each \emph{node} (called a \emph{decision node}) represents every possible
stage of the game as it is played.

A payoff function for a player is a mapping from the cross-product of players'
strategy spaces to the player's set of payoffs, which is normally within the
set of real numbers.

A game is \emph{zero-sum} if the total payoff to all players in the game, for
every combination of strategies, always adds up to zero. That is, a player
gains only at the expense of others. Two-person zero-sum games are sometimes
called strictly competitive games. A game is \emph{non-zero-sum} when gain by
one player does not necessarily correspond to a loss by another.

A \emph{payoff matrix} for a two-player game is an $m\times\ n$ matrix of real numbers:%

\begin{equation}%
\begin{array}
[c]{c}%
\text{Player }A
\end{array}%
\begin{array}
[c]{c}%
S_{1}\\
S_{2}\\
\vdots\\
S_{m}%
\end{array}
\overset{\overset{%
\begin{array}
[c]{c}%
\text{Player }B
\end{array}
}{%
\begin{array}
[c]{cccc}%
S_{1}^{\prime} & S_{2}^{\prime} & \cdots &  S_{n}^{\prime}%
\end{array}
}}{\left(
\begin{array}
[c]{cccc}%
a_{11} & a_{12} & \ldots &  a_{1n}\\
a_{21} & a_{22} & \cdots &  a_{2n}\\
\vdots & \vdots & \vdots & \vdots\\
a_{m1} & a_{m2} & \cdots &  a_{mn}%
\end{array}
\right)  } \label{TwoPlayerPayoffMatrix}%
\end{equation}
The matrix shows what payoff each player will receive at the outcome of the
game. The payoff for each player depends, of course, on the combined actions
of both players. In the matrix (\ref{TwoPlayerPayoffMatrix}) player $A$'s
strategies are designated down the left hand column and player $B$'s
strategies are designated along the top row.

A \emph{cooperative game} is a game in which two or more players do not
compete but rather strive toward a unique objective and therefore win or lose
as a group. In a \emph{non-cooperative game} no outside authority assures that
players stick to the same predetermined rules, and so binding agreements are
not feasible. In these games players may cooperate but any cooperation must be
self-enforcing. In a \emph{game of complete information} the knowledge about
other players is available to all participants i.e. every player knows the
payoff functions and the strategies available to other players.

\emph{Best response} is any strategy that yields the highest possible payoff
in response to the strategy of other players. \emph{Dominant strategy
equilibrium} is a strategy profile in which each player plays best-response
that does not depend on the strategies of other players.

In a zero-sum game between players $A$ and $B$ player $A$ should attempt to
minimize player $B$'s maximum payoff while player $B$ attempts to maximize his
own minimum payoff. When they do so a surprising conclusion comes out i.e. the
minimum of the maximum (mini-max) payoffs equals the maximum of the minimum
(max-min) payoffs. Neither player can improve his position, and so these
strategies form an equilibrium of the game.

The \emph{minimax theorem} states that for every two-person, zero-sum game,
there always exists a mixed strategy for each player such that the expected
payoff for one player is the same as the expected cost for the other. In other
words, there is always a rational solution to a precisely defined conflict
between two people whose interests are completely opposite. It is a rational
solution in that both parties can convince themselves that they cannot expect
to do any better, given the nature of the conflict.

Now consider a game in which each player chooses the action that is best for
her, given her beliefs about the other players' actions. How do players form
beliefs about each other? We consider here the case in which every player is
experienced i.e. she has played the game so many times that she knows the
actions the other players will choose. Thus we assume that every player's
belief about the other players' actions is correct. The notion of equilibrium
that embodies these two principles is called a \emph{Nash equilibrium
}\cite{JohnNash,JohnNash1}\emph{ }(after John Nash, who suggested it in the
early 1950s).

For a Nash equilibrium we need the concept of a \emph{strategy profile}. It is
a set of strategies for each player which fully specifies all the actions in a
game. A strategy profile must include one and only one strategy for every
player. For example, $(s_{A}^{\ast},s_{B}^{\ast},s_{C}^{\ast})$ is a strategy
profile for a three-player game in which $s_{A}^{\ast},s_{B}^{\ast}$ and
$s_{C}^{\ast}$ are strategies for the players $A,B$ and $C$, respectively. The
payoff to player $A$, with this strategy profile, is denoted by $P_{A}%
(s_{A}^{\ast},s_{B}^{\ast},s_{C}^{\ast})$.

Consider a set of players $A,B\cdots G$ playing a game. A strategy profile
$(s_{A}^{\ast},s_{B}^{\ast}\cdots s_{G}^{\ast})$ is said to be a Nash
equilibrium if and only if for any player $A,B\cdots G$ we have%

\begin{equation}
\left.
\begin{array}
[c]{c}%
P_{A}(s_{A},s_{B}^{\ast}\cdots s_{G}^{\ast})\leq P_{A}(s_{A}^{\ast}%
,s_{B}^{\ast}\cdots s_{G}^{\ast})\\
P_{B}(s_{A}^{\ast},s_{B}\cdots s_{G}^{\ast})\leq P_{B}(s_{A}^{\ast}%
,s_{B}^{\ast}\cdots s_{G}^{\ast})\\
\cdots\\
P_{G}(s_{A}^{\ast},s_{B}^{\ast}\cdots s_{G})\leq P_{G}(s_{A}^{\ast}%
,s_{B}^{\ast}\cdots s_{G}^{\ast})
\end{array}
\right\}  \label{Defining inequalities of NE}%
\end{equation}
for all $s_{A}\in S_{A}$ and $s_{B}\in S_{B}$ etc. In other words, a Nash
equilibrium is a strategy profile such that no player has an incentive to
unilaterally change her action. Players are in equilibrium if a change in
strategy by any one of them would lead that player to earn less than if she
remained with her current strategy. Note that nothing in the definition
suggests that a strategic game necessarily has a Nash equilibrium or that, if
it does, it has only a single Nash equilibrium. A strategic game may have no
Nash equilibrium, may have a single Nash equilibrium, or may have many Nash
equilibria. A strategy profile is a \emph{strict Nash equilibrium} if for it
the inequalities (\ref{Defining inequalities of NE}) hold strictly. An outcome
of a game is \emph{Pareto optimal} \cite{Rasmusen} if there is no other
outcome that makes every player at least as well off and at least one player
strictly better off. That is, a Pareto optimal outcome cannot be improved upon
without hurting at least one player. Often, a Nash equilibrium is not Pareto
optimal, implying that the players' payoffs can all be increased.

\section{Examples of games}

In the following sections three games and their solutions are given. Two of
the games, Prisoners' Dilemma and Matching Pennies, are selected because the
earliest proposals \cite{MeyerDavid,Eisert} of quantization were made for
those games. In both of these games players' moves are performed
simultaneously and they belong to the class of non-cooperative games. The
third game, Model of Entry, has a somewhat different information structure,
with players' making sequential moves, which is reminiscent of what happens in
the game of chess.

\subsection{Prisoners' Dilemma}

Prisoners' Dilemma \cite{Rasmusen} is a widely known noncooperative game. Its
name comes from the following situation: two criminals are arrested after
having committed a crime together. Each suspect is placed in a separate cell
and may choose between two strategies, namely confessing $(D)$ and not
confessing $(C)$, where $C$ and $D$ stand for cooperation and defection. If
neither suspect confesses, i.e. $(C,C)$, they go free; this is represented by
$3$ units of payoff for each suspect. When one prisoner confesses ($D$) and
the other does not $(C)$, the prisoner who confesses gets $5$ units of payoff,
while the prisoner who did not confess gets $0$, represented by his ending up
in the prison. When both prisoners confess, i.e. $(D,D)$, both are given a
reduced term, but both are convicted, which we represent by giving each $1$
unit of payoff, better than getting $0$ if the other prisoner confesses, but
not so good as going free i.e. a payoff of $5$. The game has the normal-form representation:%

\begin{equation}%
\begin{array}
[c]{c}%
\text{Alice}%
\end{array}%
\begin{array}
[c]{c}%
C\\
D
\end{array}
\overset{\overset{%
\begin{array}
[c]{c}%
\text{Bob}%
\end{array}
}{%
\begin{array}
[c]{cc}%
C & D
\end{array}
}}{\left(
\begin{array}
[c]{cc}%
(3,3) & (0,5)\\
(5,0) & (1,1)
\end{array}
\right)  } \label{PDmatrix1}%
\end{equation}
where Alice and Bob are the prisoners and the first and the second entries in
parentheses correspond to Alice's and Bob's payoff, respectively. The origin
of the dilemma stems from the fact that for either choice of the opponent it
is advantageous to defect $(D)$. But when both defect, i.e. $(D,D)$, the
payoff remains less than in the case when both cooperate $(C,C)$. In its
generalized form the PD is represented as:%

\begin{equation}%
\begin{array}
[c]{c}%
\text{Alice}%
\end{array}%
\begin{array}
[c]{c}%
C\\
D
\end{array}
\overset{\overset{%
\begin{array}
[c]{c}%
\text{Bob}%
\end{array}
}{%
\begin{array}
[c]{cc}%
C & D
\end{array}
}}{\left(
\begin{array}
[c]{cc}%
(r,r) & (s,t)\\
(t,s) & (u,u)
\end{array}
\right)  } \label{PDmatrix}%
\end{equation}
where $s<u<r<t$. PD has $(D,D)$ as the pure-strategy equilibrium.

\subsection{Matching Pennies}

Player $A$ chooses ``heads'' $(H)$ or ``tails'' $(T)$. Without knowing player
$A$'s choice, player $B$ also chooses ``heads'' or ``tails''. If the two
choices are alike, then player $B$ wins a penny from player $A$; otherwise,
player $A$ wins a penny from player $B$. The normal form of this game is the matrix%

\begin{equation}%
\begin{array}
[c]{c}%
\text{Player }A
\end{array}%
\begin{array}
[c]{c}%
H\\
T
\end{array}
\overset{\overset{%
\begin{array}
[c]{c}%
\text{Player }B
\end{array}
}{%
\begin{array}
[c]{ccc}%
H &  & T
\end{array}
}}{\left(
\begin{array}
[c]{cc}%
(-1,1) & (1,-1)\\
(1,-1) & (-1,1)
\end{array}
\right)  }%
\end{equation}
where each row represents player $A$'s strategy, and each column a strategy of
player $B$. Fig. ($2$-$1$) shows the game tree of this game with terminal
vertices representing the players' payoffs.%

%TCIMACRO{\FRAME{dtbpFU}{2.5071in}{1.3733in}{0pt}{\Qcb{Figure $2$-$1$: Game
%tree of Matching Pennies. The terminal vertices represent the players'
%payoffs.}}{\Qlb{MatchingPenniesFig}}{pennyflip.eps}%
%maintain-aspect-ratio TRUE;  display "PICT";  valid_file "F";
%width 2.5071in;  height 1.3733in;  depth 0pt;  original-width 8.0678in;
%original-height 10.3259in;  cropleft "0.2335";  croptop "0.8591";
%cropright "0.8871";  cropbottom "0.5819";
%filename 'PennyFlip.eps';file-properties "XNPEU";}}}%
%BeginExpansion
\begin{center}
\includegraphics[
trim=1.883831in 6.008641in 0.910855in 1.454919in,
height=1.3733in,
width=2.5071in
]%
{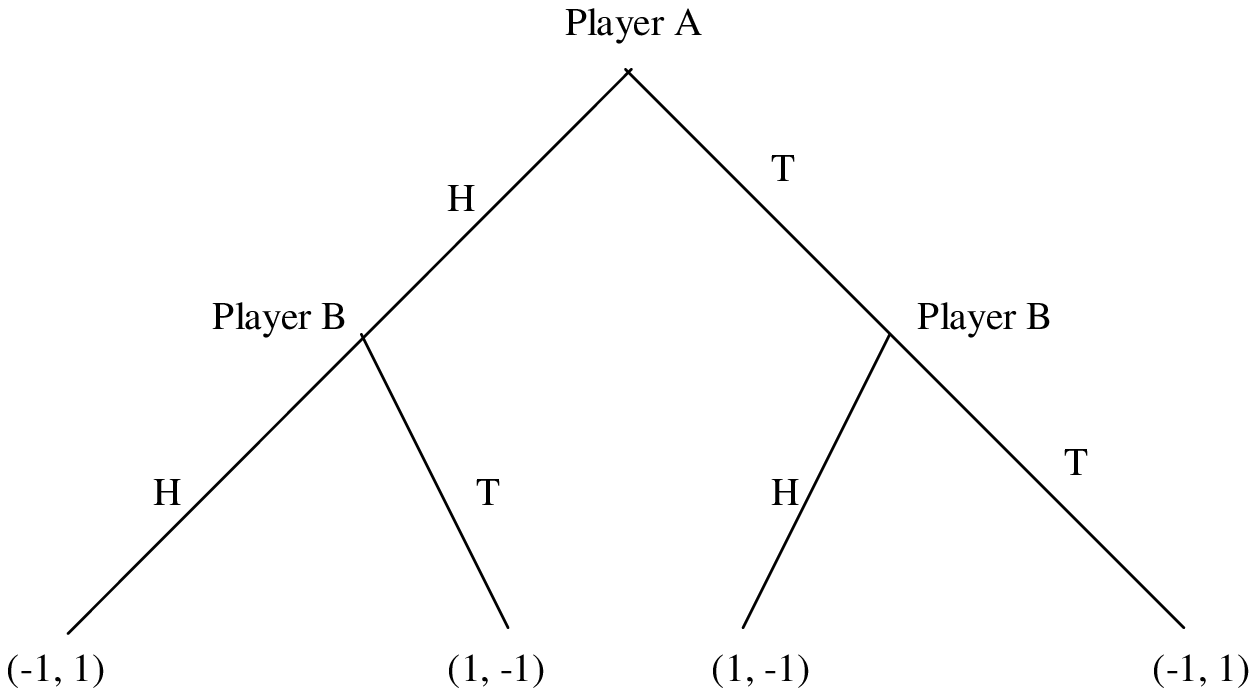}%
\\
Figure $2$-$1$: Game tree of Matching Pennies. The terminal vertices represent
the players' payoffs.
\label{MatchingPenniesFig}%
\end{center}
%EndExpansion

The game has no pure strategy equilibrium. To find the mixed strategy
equilibrium suppose $p$ and $q$ are the probabilities with which players $A$
and $B$ play the strategy $H$, respectively. Let $(p^{\ast},q^{\ast})$ be an equilibrium:%

\begin{align}
P_{A}(p^{\ast},q^{\ast})-P_{A}(p,q^{\ast})  &  =2(p^{\ast}-p)(1-2q^{\ast
})\nonumber\\
P_{B}(p^{\ast},q^{\ast})-P_{B}(p^{\ast},q)  &  =-2(q^{\ast}-q)(1-2p^{\ast})
\end{align}
giving $p^{\ast}=q^{\ast}=1/2$.

\subsection{Model of Entry}

Firm $A$ is an incumbent monopolist in an industry. Firm $B$ has the
opportunity to enter the industry. After firm $B$ makes the decision to enter,
firm $A$ will have the chance to choose a pricing strategy. It can choose
either to \textit{Fight} the entrant or to \textit{Accommodate} it with higher prices.%

%TCIMACRO{\FRAME{dtbpFU}{1.9588in}{1.7253in}{0pt}{\Qcb{Figure $2$-$2$: Game
%tree of Model of Entry.}}{\Qlb{ModelofEntry}}{modelofentry.eps}%
%maintain-aspect-ratio TRUE;  display "PICT";  valid_file "F";
%width 1.9588in;  height 1.7253in;  depth 0pt;  original-width 7.8672in;
%original-height 11.336in;  cropleft "0.3792";  croptop "0.8395";
%cropright "0.7628";  cropbottom "0.6055";
%filename 'ModelofEntry.eps';file-properties "XNPEU";}}}%
%BeginExpansion
\begin{center}
\includegraphics[
trim=2.983242in 6.863948in 1.866100in 1.819428in,
height=1.7253in,
width=1.9588in
]%
{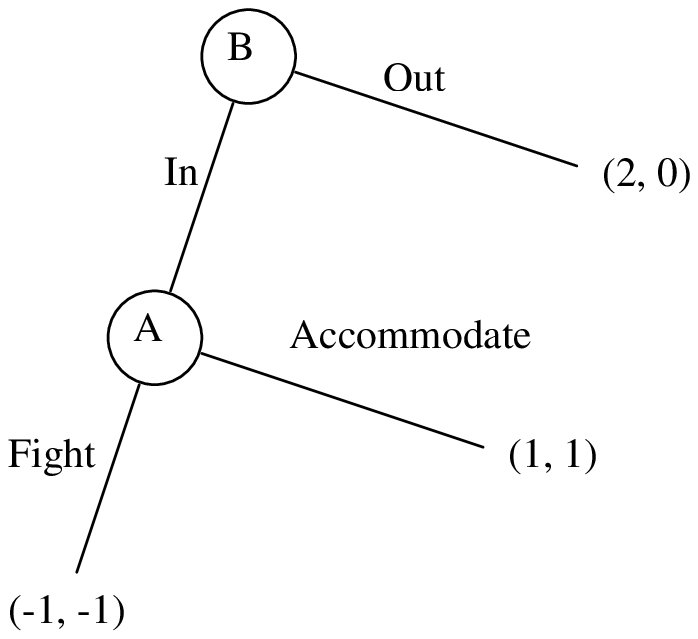}%
\\
Figure $2$-$2$: Game tree of Model of Entry.
\label{ModelofEntry}%
\end{center}
%EndExpansion

Fig. ($2$-$2$) is the game tree for Model of Entry and its normal form has the representation:%

\begin{equation}%
\begin{array}
[c]{c}%
\text{Player }A
\end{array}%
\begin{array}
[c]{c}%
\text{Fight}\\
\text{Accommodate}%
\end{array}
\overset{\overset{%
\begin{array}
[c]{c}%
\text{Player }B
\end{array}
}{%
\begin{array}
[c]{ccc}%
\text{Out} &  & \text{In}%
\end{array}
}}{\left(
\begin{array}
[c]{cc}%
(2,0) & (-1,-1)\\
(2,0) & (1,1)
\end{array}
\right)  } \label{Model of Entry Matrix}%
\end{equation}
The game has two pure-strategy equilibria i.e. $($Fight, Out$)$ and
$($Accommodate, In$)$.

\chapter{Quantum mechanics}

Being one of the pillars of modern physics, quantum mechanics
\cite{AsherPeres} has an impressive amount of supporting experiments, and many
technological applications are based on it. Quantum mechanics can be
approached from several directions and viewpoints. Below, some of the basic
concepts and definitions of quantum mechanics are described, to give a
necessary background for the topics in quantum games which are discussed in
later chapters of this thesis.

\section{Mathematical preliminaries}

The most popular approach uses the concept of a \emph{vector space}. In Dirac
notation, a complex vector space is a set $V$ consisting of elements of the
form $\left|  \alpha\right\rangle $ called \emph{kets.} It is a set in which

\begin{itemize}
\item  A vector $\left|  \alpha\right\rangle +\left|  \beta\right\rangle $ can
be associated with each pair of vectors $\left|  \alpha\right\rangle ,\left|
\beta\right\rangle \in V$. This association satisfies the following axioms:
\end{itemize}

\begin{enumerate}
\item [ a)]For all $\left|  \alpha\right\rangle ,\left|  \beta\right\rangle
,\left|  \gamma\right\rangle \in V$ we have $\left|  \alpha\right\rangle
+(\left|  \beta\right\rangle +\left|  \gamma\right\rangle )=(\left|
\alpha\right\rangle +\left|  \beta\right\rangle )+\left|  \gamma\right\rangle $

\item[ b)] There is a null vector $\left|  0\right\rangle \in V$ such that
$\left|  0\right\rangle +\left|  \alpha\right\rangle =\left|  \alpha
\right\rangle +\left|  0\right\rangle =\left|  \alpha\right\rangle $ for all
$\left|  \alpha\right\rangle \in V$

\item[ c)] For each $\left|  \alpha\right\rangle \in V$ there exists an
inverse vector denoted by $-\left|  \alpha\right\rangle $ such that $\left|
\alpha\right\rangle +(-\left|  \alpha\right\rangle )=\left|  0\right\rangle $

\item[ d)] For all $\left|  \alpha\right\rangle ,\left|  \beta\right\rangle
\in V$ we have $\left|  \alpha\right\rangle +\left|  \beta\right\rangle
=\left|  \beta\right\rangle +\left|  \alpha\right\rangle $
\end{enumerate}

\begin{itemize}
\item  A vector $\lambda\left|  \alpha\right\rangle $ can be associated with
each $\left|  \alpha\right\rangle \in V$ and $\lambda\in\mathbf{C}$, where
$\mathbf{C}$ is the field of complex numbers. For all $\lambda,\mu
\in\mathbf{C}$ and $\left|  \alpha\right\rangle ,\left|  \beta\right\rangle
\in V$ this association satisfies the following axioms:
\end{itemize}

\begin{enumerate}
\item [ a)]$\lambda(\left|  \alpha\right\rangle +\left|  \beta\right\rangle
)=\lambda\left|  \alpha\right\rangle +\lambda\left|  \beta\right\rangle $

\item[ b)] $(\lambda+\mu)\left|  \alpha\right\rangle =\lambda\left|
\alpha\right\rangle +\mu\left|  \alpha\right\rangle $

\item[ c)] $\lambda(\mu\left|  \alpha\right\rangle )=(\lambda\mu)\left|
\alpha\right\rangle $

\item[ d)] $1\left|  \alpha\right\rangle =\left|  \alpha\right\rangle $

\item[ e)] $0\left|  \alpha\right\rangle =\left|  0\right\rangle $
\end{enumerate}

The vector space provides the arena for doing mathematical actions on ket
vectors. Its structure guarantees that such actions do not result in yielding
a ket that does not reside in the vector space under consideration.

An \emph{operator} $\hat{O}$ maps each vector $\left|  \alpha\right\rangle \in
V$ into another vector $\left|  \beta\right\rangle \in V$:%

\begin{equation}
\left|  \beta\right\rangle =\hat{O}\left|  \alpha\right\rangle
\end{equation}
The operator $\hat{O}$ is \emph{linear} if, for any vectors $\left|
\alpha\right\rangle ,\left|  \beta\right\rangle \in V$ and for any
$\lambda,\mu\in\mathbf{C}$\textbf{,} the property%

\begin{equation}
\hat{O}(\lambda\left|  \alpha\right\rangle +\mu\left|  \beta\right\rangle
)=\lambda\hat{O}\left|  \alpha\right\rangle +\mu\hat{O}\left|  \beta
\right\rangle
\end{equation}
is true. A \emph{spanning set} for a vector space $V$ is a set of vectors
$\left|  \beta_{1}\right\rangle ,\left|  \beta_{2}\right\rangle ,...,\left|
\beta_{n}\right\rangle \in V$ such that any vector $\left|  \alpha
\right\rangle $ in $V$ can be written as a linear combination $\left|
\alpha\right\rangle =\sum_{i}\lambda_{i}\left|  \beta_{i}\right\rangle $ of
vectors in that set. A set of vectors $\left|  \alpha_{1}\right\rangle
,\left|  \alpha_{2}\right\rangle ,...\left|  \alpha_{n}\right\rangle \in V$ is
said to be \emph{linearly independent} if the relation%

\begin{equation}
c_{1}\left|  \alpha_{1}\right\rangle +c_{2}\left|  \alpha_{2}\right\rangle
,...+c_{n}\left|  \alpha_{n}\right\rangle =0
\end{equation}
holds if and only if $c_{1}=c_{2}=...=c_{n}=0$. It can be shown that any two
sets of linearly independent vectors which span $V$ contain the same number of
elements. Such a set is called a \emph{basis} for $V$. The number of elements
in a basis is defined to be the \emph{dimension} of $V$.

To any vector $\left|  \alpha\right\rangle \in V$ a \emph{dual vector} can be
associated; this is written as $\left\langle \alpha\right|  $ and is called a
\emph{bra}. The bra vector $\left\langle \alpha\right|  $ is a linear operator
from the vector space $V$ to the field $\mathbf{C}$, defined by $\left\langle
\alpha\right|  (\left|  \beta\right\rangle )=\left\langle \alpha\right|
\left.  \beta\right\rangle $. For any $\left|  \alpha\right\rangle ,\left|
\beta\right\rangle \in V$ the \emph{inner product} $\left\langle
\alpha\right|  \left.  \beta\right\rangle $ is defined to be a complex number
with the following properties:

\begin{enumerate}
\item $\left\langle \alpha\right|  \left.  \beta\right\rangle =\left\langle
\beta\right|  \left.  \alpha\right\rangle ^{\ast}$ where $\ast$ denotes
complex conjugate

\item $\left\langle \alpha\right|  \left.  \lambda\beta+\mu\gamma\right\rangle
=\lambda\left\langle \alpha\right|  \left.  \beta\right\rangle +\mu
\left\langle \alpha\right|  \left.  \gamma\right\rangle ,$ where $\left|
\alpha\right\rangle ,\left|  \beta\right\rangle ,\left|  \gamma\right\rangle
\in V$ and $\lambda,\mu\in\mathbf{C}$

\item $\left\langle \alpha\right|  \left.  \alpha\right\rangle \geq0$ for any
$\left|  \alpha\right\rangle \in V$, with equality if and only if $\left|
\alpha\right\rangle $ is a zero vector
\end{enumerate}

From these properties it follows that%

\begin{equation}
\left\langle \lambda\alpha\right|  \left.  \beta\right\rangle =\lambda^{\ast
}\left\langle \alpha\right|  \left.  \beta\right\rangle
\end{equation}

The \emph{norm} of a vector $\left|  \alpha\right\rangle $ is defined by%

\begin{equation}
\left\|  \left|  \alpha\right\rangle \right\|  =\sqrt{\left\langle
\alpha\right|  \left.  \alpha\right\rangle }%
\end{equation}

Any non-zero vector $\left|  \alpha\right\rangle $ can be \emph{normalized} by
dividing it by its norm. The normalized vector $\left|  \alpha\right\rangle
/\left\|  \left|  \alpha\right\rangle \right\|  $ has unit norm and is
therefore called a \emph{unit vector}. Vectors $\left|  \alpha\right\rangle $
and $\left|  \beta\right\rangle $ are called \emph{orthogonal} if their inner
product is zero. A vector space equipped with an inner product is called an
\emph{inner product space}.

In quantum mechanics states of physical systems are represented by unit
vectors in \emph{Hilbert space}. As is the case with quantum computation and
information \cite{Nielsen},\ the field of quantum games mostly deals with
finite dimensional complex vector spaces for which a Hilbert space becomes the
same as an inner product space.

Let $\left|  \alpha_{i}\right\rangle $ $(i=1,2,...,n)$ be a basis for an
$n$-dimensional Hilbert space $\mathcal{H}$. A vector $\left|  \alpha
\right\rangle $ in $\mathcal{H}$ can be written as%

\begin{equation}
\left|  \alpha\right\rangle =\sum_{i}a_{i}\left|  \alpha_{i}\right\rangle
\end{equation}
where $a_{i}=\left\langle \alpha_{i}\right|  \left.  \alpha\right\rangle $.
Thus $\left|  \alpha\right\rangle =\sum_{i}\left\langle \alpha_{i}\right|
\left.  \alpha\right\rangle \left|  \alpha_{i}\right\rangle =\sum_{i}\left|
\alpha_{i}\right\rangle \left\langle \alpha_{i}\right|  \left.  \alpha
\right\rangle =(\sum_{i}\left|  \alpha_{i}\right\rangle \left\langle
\alpha_{i}\right|  )\left|  \alpha\right\rangle $ and we have the completeness relation%

\begin{equation}
\sum_{i}\left|  \alpha_{i}\right\rangle \left\langle \alpha_{i}\right|
=\hat{I}%
\end{equation}
where $\hat{I}$ is the \emph{identity operator} defined by $\hat{I}\left|
\alpha\right\rangle =\left|  \alpha\right\rangle $.

Let $\left|  \alpha\right\rangle $ $\in V$ be a unit vector. An operator is
called a \emph{projector} if it projects a vector $\left|  \beta\right\rangle
\in V$ along the direction $\left|  \alpha\right\rangle $. For example
$P_{\alpha}=\left|  \alpha\right\rangle \left\langle \alpha\right|  $ is a
one-dimensional projector that acts on the vector $\left|  \beta\right\rangle
$ as follows:%

\begin{equation}
P_{\alpha}\left|  \beta\right\rangle =\left|  \alpha\right\rangle \left\langle
\alpha\right|  \left.  \beta\right\rangle =\left\langle \alpha\right|  \left.
\beta\right\rangle \left|  \alpha\right\rangle
\end{equation}
From the definition it follows that $P_{\alpha}^{2}=P_{\alpha}$.

Consider a linear operator $\hat{O}$ that acts on a non-zero vector $\left|
\alpha\right\rangle $ such that%

\begin{equation}
\hat{O}\left|  \alpha\right\rangle =\alpha\left|  \alpha\right\rangle
\label{eigenvalue eq}%
\end{equation}
where $\alpha\in\mathbf{C}$ and is called the \emph{eigenvalue} of $\hat{O}$
corresponding to the \emph{eigenvector} $\left|  \alpha\right\rangle $. Eq.
(\ref{eigenvalue eq}) is called an \emph{eigenvalue equation}. It is found
that an eigenvalue equation always has a solution.

For any linear operator $\hat{O}$ on a Hilbert space $\mathcal{H}$ a unique
linear operator $\hat{O}^{\dagger}$ on $\mathcal{H}$ can be found which is
called the \emph{adjoint} or \emph{Hermitian conjugate} of $\hat{O}$. It is
defined as follows. For any $\left|  \alpha\right\rangle ,\left|
\beta\right\rangle \in\mathcal{H}$%

\begin{equation}
\left\langle \alpha\right|  \left.  \hat{O}\beta\right\rangle =\left\langle
\hat{O}^{\dagger}\alpha\right|  \left.  \beta\right\rangle
\end{equation}
from which it follows that $\left\langle \hat{O}\alpha\right|  \left.
\beta\right\rangle =\left\langle \alpha\right|  \left.  \hat{O}^{\dagger}%
\beta\right\rangle $. The operator $\hat{O}$ is \emph{Hermitian} or
\emph{self-adjoint }if%

\begin{equation}
\hat{O}=\hat{O}^{\dagger}%
\end{equation}
Consider the scalar product $\left\langle \alpha\right|  \left.  \hat{O}%
\alpha\right\rangle $ when $\hat{O}$ is Hermitian. Because $\left\langle
\alpha\right|  \left.  \hat{O}\alpha\right\rangle ^{\ast}=\left\langle \hat
{O}^{\dagger}\alpha\right|  \left.  \alpha\right\rangle =\left\langle \hat
{O}\alpha\right|  \left.  \alpha\right\rangle =\left\langle \alpha\right|
\left.  \hat{O}\alpha\right\rangle $ so the eigenvalues of a Hermitian
operator are real.

An important class of operators in Hilbert space consists of \emph{unitary
operators}. An operator $\hat{U}$ is unitary if%

\begin{equation}
\hat{U}\hat{U}^{\dagger}=\hat{U}^{\dagger}\hat{U}=\hat{I}%
\end{equation}
From this definition it follows that $\hat{U}^{\dagger}=\hat{U}^{-1}$ i.e. the
adjoint of a unitary operator is the same as its inverse. The action of
unitary operators in Hilbert space is similar to that of rotations in
Euclidean space; it preserves both the length of a vector and the angle
between two vectors. It can be seen by considering two vectors $\left|
\alpha\right\rangle ,\left|  \beta\right\rangle $. Consider the inner product%

\begin{equation}
\left\langle \hat{U}\alpha\right|  \left.  \hat{U}\beta\right\rangle
=\left\langle \alpha\right|  \hat{U}^{\dagger}\hat{U}\left|  \beta
\right\rangle =\left\langle \alpha\right|  \left.  \beta\right\rangle
\end{equation}
which in the case $\left|  \alpha\right\rangle =\left|  \beta\right\rangle $
shows that a unitary operator does not change the norm of a vector.

In quantum games three unitary and Hermitian operators $\hat{\sigma}_{x}%
,\hat{\sigma}_{y}$ and $\hat{\sigma}_{z}$ are important. These are the
\emph{Pauli matrices} defined as follows:%

\begin{equation}
\hat{\sigma}_{x}=\left[
\begin{array}
[c]{cc}%
0 & 1\\
1 & 0
\end{array}
\right]  ,\text{ \ \ }\hat{\sigma}_{y}=\left[
\begin{array}
[c]{cc}%
0 & -i\\
i & 0
\end{array}
\right]  ,\text{ \ \ }\hat{\sigma}_{z}=\left[
\begin{array}
[c]{cc}%
1 & 0\\
0 & -1
\end{array}
\right]
\end{equation}
The Pauli matrices have two well-known properties. The first property is
$\hat{\sigma}_{x}^{2}=\hat{\sigma}_{y}^{2}=\hat{\sigma}_{z}^{2}=\hat{I}$,
where $\hat{I}$ is the identity matrix. The second property is%

\begin{equation}
\hat{\sigma}_{x}\hat{\sigma}_{y}=i\hat{\sigma}_{z},\text{ \ \ }\hat{\sigma
}_{y}\hat{\sigma}_{z}=i\hat{\sigma}_{x},\text{ \ \ }\hat{\sigma}_{z}%
\hat{\sigma}_{x}=i\hat{\sigma}_{y}%
\end{equation}

\section{Tensor products}

Quantum mechanics offers the method of \emph{Tensor products}
\cite{AsherPeres} to combine together two Hilbert spaces to form a bigger
Hilbert space. It is an important concept for the understanding of the
multiparticle quantum systems which are fundamental for playing quantum games.
Quantum games are most often played while players share multiparticle quantum systems.

Consider two Hilbert spaces $\mathcal{H}_{A}$ and $\mathcal{H}_{B}$, having
dimensions $m$ and $n$, respectively. The tensor product between these Hilbert
spaces is written as $\mathcal{H}=\mathcal{H}_{A}\otimes\mathcal{H}_{B}$. The
space $\mathcal{H}$ is defined as follows. Suppose $\left|  \alpha
\right\rangle \in\mathcal{H}_{A}$ and $\left|  \beta\right\rangle
\in\mathcal{H}_{B}$. These two vectors can be associated with a vector
$\left|  \alpha\right\rangle \otimes\left|  \beta\right\rangle \in\mathcal{H}$
which is the tensor product between the vectors $\left|  \alpha\right\rangle $
and $\left|  \beta\right\rangle $. The product $\left|  \alpha\right\rangle
\otimes\left|  \beta\right\rangle $ has the following properties:

a) For any $\left|  \alpha\right\rangle \in\mathcal{H}_{A}$, $\left|
\beta\right\rangle \in\mathcal{H}_{B}$ and $\lambda\in\mathbf{C},$%
\begin{equation}
\lambda(\left|  \alpha\right\rangle \otimes\left|  \beta\right\rangle
)=(\lambda\left|  \alpha\right\rangle )\otimes\left|  \beta\right\rangle
=\left|  \alpha\right\rangle \otimes(\lambda\left|  \beta\right\rangle );
\end{equation}

b) For any $\left|  \alpha_{1}\right\rangle ,\left|  \alpha_{2}\right\rangle
\in\mathcal{H}_{A}$ and $\left|  \beta\right\rangle \in\mathcal{H}_{B},$%

\begin{equation}
(\left|  \alpha_{1}\right\rangle +\left|  \alpha_{2}\right\rangle
)\otimes\left|  \beta\right\rangle =\left|  \alpha_{1}\right\rangle
\otimes\left|  \beta\right\rangle +\left|  \alpha_{2}\right\rangle
\otimes\left|  \beta\right\rangle
\end{equation}

c) For any $\left|  \alpha\right\rangle \in\mathcal{H}_{A}$ and $\left|
\beta_{1}\right\rangle ,\left|  \beta_{2}\right\rangle \in\mathcal{H}_{B},$%

\begin{equation}
\left|  \alpha\right\rangle \otimes(\left|  \beta_{1}\right\rangle +\left|
\beta_{2}\right\rangle )=\left|  \alpha\right\rangle \otimes\left|  \beta
_{1}\right\rangle +\left|  \alpha\right\rangle \otimes\left|  \beta
_{2}\right\rangle
\end{equation}
The tensor product $\left|  \alpha\right\rangle \otimes\left|  \beta
\right\rangle $ is often simply written as $\left|  \alpha\right\rangle
\left|  \beta\right\rangle $.

\section{The qubit}

A system with two states is called a bit in classical information theory
\cite{Shannon}. These states are usually denoted by $0$ and $1$. The name
\emph{qubit} \cite{AsherPeres} comes from quantum bit and its role in quantum
information theory \cite{Preskill,Nielsen} is the same as that of the bit in
classical information theory \cite{Shannon}. A qubit is the simplest
nontrivial quantum system whose state can be described by a vector in
two-dimensional complex Hilbert space. An example of such a system is a
spin-$1/2$ particle, for instance the electron. Measurement of the
$z$-component of its spin always gives either up or down ($\pm1/2$) as the
result. The state of the electron becomes an eigenstate of the observable
after the measurement. In binary notation the two eigenstates can be denoted
by $\left|  0\right\rangle $ and $\left|  1\right\rangle $, representing spins
parallel ($+1/2$) and anti-parallel ($-1/2$) to the $z$-axis. These are
orthogonal states and are taken as the basis vectors of the two-dimensional
spin Hilbert space. A general spin state vector can be a superposition of the
eigenstates i.e.%

\begin{equation}
\left|  \psi\right\rangle =\alpha\left|  0\right\rangle +\beta\left|
1\right\rangle \label{QubitState}%
\end{equation}
where $\alpha,\beta\in\mathbf{C}$ satisfy $\left|  \alpha\right|  ^{2}+\left|
\beta\right|  ^{2}=1$. For an electron in this state the outcomes $+1/2$ and
$-1/2$ appear with probabilities $\left|  \alpha\right|  ^{2}$ and $\left|
\beta\right|  ^{2}$, respectively, when the spin is measured along the $z$-axis.

The state (\ref{QubitState}) is often represented as%

\begin{equation}
\left|  \psi\right\rangle =\cos(\theta/2)\left|  0\right\rangle +e^{i\phi}%
\sin(\theta/2)\left|  1\right\rangle \label{QubitState1}%
\end{equation}
where only the phase difference between $\alpha$ and $\beta$ is considered and
$\alpha$ is chosen real. For the range $\theta\in\lbrack0,\pi]$ and $\phi
\in\lbrack0,2\pi]$ of the parameters the state (\ref{QubitState1}) can
describe any state of the qubit. Although a qubit can exist in a continuum of
superpositions between the states $\left|  0\right\rangle $ and $\left|
1\right\rangle $ only a single bit of information can be extracted from a
single qubit. Apart from the spin-$1/2$ particles, the polarization states of
a photon and the energy levels of a two-level system can also represent a qubit.

In quantum games the concept of qubit helps in finding a quantum version of a
classical game. It is because many, if not all, classical games can be played
when participating players share the carriers of classical information, i.e.
bits. Replacing bits with qubits, with players having the capacity to do
actions on their qubits, allows one to construct a quantum version of the
classical game.

Some classical games can be more easily played when players share dice instead
of coins. Likewise, for certain quantum games it is helpful if they are played
when players share higher-dimensional quantum systems. For example, a quantum
version of the two-player game of the rock, scissors and paper can be more
easily played \cite{RSP}\ when the players share two qutrits ($3$-dimensional
quantum system), instead of two qubits.

\section{Postulates of quantum mechanics}

The physical set-up for playing quantum games is a quantum mechanical system.
Like it is the case with usual quantum systems, the physical set-up is assumed
isolated from its surrounding environment with its behaviour controlled
externally. That is, the system is not disturbed by events which are unrelated
to the control procedures. In quantum mechanics such a system can be modelled
by the following postulates \cite{Preskill,AsherPeres}.

\subsubsection{Postulate I}

A Hilbert space $\mathcal{H}$ is associated with a quantum system. The system
is completely described by its \emph{state vector }$\left|  \psi\right\rangle
$, which is a unit vector in the system's state space.

\subsubsection{Postulate II}

The evolution of an isolated quantum system is described by \emph{unitary
transformations}. The states $\left|  \psi_{1}\right\rangle $ and $\left|
\psi_{2}\right\rangle $ of the system at times $t_{1}$ and $t_{2}$,
respectively, are related by a unitary transformation $\hat{U}$, which depends
only on $t_{1}$ and $t_{2}$, such that $\hat{U}(\left|  \psi_{1}\right\rangle
)=\left|  \psi_{2}\right\rangle $.

\subsubsection{Postulate III}

A \emph{measurement} of a quantum system consists of a set $\left\{
M_{m}:m=1,2,...,k\right\}  $ of linear operators on $\mathcal{H}$, such that%

\begin{equation}
\sum\limits_{m=1}^{k}M_{m}^{\dagger}M_{m}=\hat{I} \label{QM Postulate 3}%
\end{equation}
The measurement results in one of the indices $m$. If the system is in the
state $\left|  \psi\right\rangle $ then the probability that $m$ is observed is%

\begin{equation}
p(m)=\left\langle M_{m}(\left|  \psi\right\rangle )\right|  \left.
M_{m}(\left|  \psi\right\rangle )\right\rangle =\left\langle \psi\right|
M_{m}^{\dagger}M_{m}\left|  \psi\right\rangle
\end{equation}
If $m$ is observed then the state $\left|  \psi\right\rangle $ transforms to%

\begin{equation}
\frac{M_{m}\left|  \psi\right\rangle }{\sqrt{p(m)}}%
\end{equation}

It can be noticed from (\ref{QM Postulate 3}) that $p$ is a probability
measure, since%

\begin{equation}
\sum_{m=1}^{k}p(m)=\sum_{m=1}^{k}\left\langle \psi\right|  M_{m}^{\dagger
}M_{m}\left|  \psi\right\rangle =\left\langle \psi\right|  \hat{I}\left|
\psi\right\rangle
\end{equation}

Postulate III relates to a generalized measurement. Usually the special type
of measurement is considered, for which $M_{m}$ is self-adjoint and $M_{m}%
^{2}=M_{m}$ for all $m$ and $M_{m}M_{n}=0$ for $m\neq n$. Such measurement is
called the \emph{projective measurement}. In the case of a projective
measurement there are mutually orthogonal subspaces $P_{1},...,P_{k}$ of
$\mathcal{H}$ such that $\mathcal{H}=\sum_{m}P_{m}$ and $M_{m}=\sum_{i}\left|
i\right\rangle \left\langle i\right|  $ for some orthogonal basis $\left|
i\right\rangle $ of $P_{m}$. Then $M_{m}$ becomes a projection onto $P_{m}$.

For an ensemble of quantum states $\left|  \psi_{i}\right\rangle $ the
\emph{density operator} \cite{AsherPeres,Preskill,Nielsen} is defined as%

\begin{equation}
\rho=\sum_{i}p_{i}\left|  \psi_{i}\right\rangle \left\langle \psi_{i}\right|
\end{equation}
where $p_{i}$ is the probability of finding the quantum system in state
$\left|  \psi_{i}\right\rangle $. The ensemble evolves unitarily in time as%

\begin{equation}
\left|  \psi(t_{2})\right\rangle =\hat{U}(t_{2},t_{1})\left|  \psi
(t_{1})\right\rangle
\end{equation}
which is described by the following evolution of the density operator:%

\begin{align}
\rho(t_{2})  &  =\sum_{i}p_{i}\left|  \psi(t_{2})\right\rangle \left\langle
\psi(t_{2})\right| \nonumber\\
&  =\sum_{i}p_{i}\hat{U}(t_{2},t_{1})\left|  \psi(t_{1})\right\rangle
\left\langle \psi(t_{1})\right|  \hat{U}^{\dagger}(t_{2},t_{1})\nonumber\\
&  =\hat{U}(t_{2},t_{1})\rho(t_{1})\hat{U}^{\dagger}(t_{2},t_{1})
\end{align}
The \emph{expectation value} of an operator $\hat{A}$, denoted by
$\left\langle \hat{A}\right\rangle $, gives the average value of the physical
observable $A$ as obtained after an infinite number of measurements of $A$ on
a system in the same state $\left|  \psi\right\rangle $. If $\rho$ is the
density operator corresponding to the state $\left|  \psi\right\rangle $ then%

\begin{equation}
\left\langle \hat{A}\right\rangle =\text{tr}(\hat{A}\rho)
\end{equation}

\section{Einstein-Podolsky-Rosen paradox}

In a famous paper \cite{EPR} entitled ``Can The Quantum Mechanical Description
of Physical Reality Be Considered Complete?'' published in 1935, Einstein,
Podolsky and Rosen (EPR) pointed out a paradox that now carries their
initials. EPR defined their \emph{elements of physical reality} by the
criterion: ``If, without in any way disturbing a system, we can predict with
certainty ... the value of a physical quantity, then there exists an element
of physical reality corresponding to this physical quantity.'' EPR argued
\cite{EPR} that the quantum mechanical description of physical reality cannot
be considered as \emph{complete}, because their criterion gives rather strange
predictions when applied to a composite system consisting of two particles
that have interacted once but are now separate from one another and do not interact.

In 1951 Bohm \cite{Bohm} suggested a simpler version of the original paradox,
which can be described as follows. A particle decays, producing two spin-$1/2$
particles whose total spin angular momentum is zero. These particles move away
from each other in opposite directions. Two observers, call them $A$ and $B$,
measure the components of their spins along various directions. Because the
total spin is zero, the measurement results will be opposite for $A$ and $B$
along a particular direction. Such anti-correlations are not difficult to
imagine in the context of classical physics, given the fact that both
particles \emph{possess} their anti-correlated values of the angular momentum.

What quantum theory says about this situation is quite different. Consider the
quantum state corresponding to a total spin of zero for the two particles,%

\begin{equation}
\left|  \psi\right\rangle =(\left|  01\right\rangle -\left|  10\right\rangle
)/\sqrt{2} \label{EPR State}%
\end{equation}
For this state suppose that $A$ measures the spin of his particle along the
$z$-axis and finds it to be $+1/2$; he can immediately conclude that the
result will be $-1/2$ if $B$ measures the spin of the other particle. From a
\emph{realistic} view the obvious question is: how is $A$'s result immediately
communicated to the other particle so as to guarantee that $B$ will always
obtain an exactly opposite result to that of $A$? The paradox deepens on
noting that (\ref{EPR State}) is a singlet state; i.e. $A$ can chose any other
direction he may like and still $B$'s measurement will always yield a result
opposite to that of $A$.

Though EPR concluded that the quantum description of physical reality is not
complete, they left open the question of whether or not a complete description
exists. In later years the so-called \emph{hidden variables} \emph{theories
}(HVTs) \cite{Belinfante} were developed. HVTs suggested that deterministic
theories can describe nature, with the exact values of all observables of a
physical system being fixed by hidden variables which are not directly
accessible to measurement. The HVTs were meant to construct deeper levels of
description of quantum phenomena in which the properties of individual systems
do have preexisting values revealed by the act of measurement. In this
approach quantum mechanics becomes \emph{only} a statistical approximation to
a HVT.

\section{Quantum entanglement}

The phenomenon of quantum entanglement is widely considered to be central to
the field of quantum computation and information \cite{Nielsen}. It is the
phenomenon considered responsible for faster quantum algorithms, though with
some disagreements. For example \cite{ApporvaPatel}, Grover's database search
algorithm \cite{Grover}, although discovered in the context of quantum
computation, can be implemented using \emph{any} system that allows a
superposition of states. In similar vein, much of the recent research in
quantum games uses entanglement for game-theoretical ends, although its exact
role is not clear. Certain quantum games \cite{Du1} have been suggested which
involve no entanglement but still outperform the corresponding classical
games. Even in Meyer's quantum penny-flip game \cite{MeyerDavid}, widely
believed to have started the field of quantum games, entanglement is not
generated at any stage of the game.

Mathematically, the entanglement is described as follows. For a system that
can be divided into two subsystems quantum mechanics associates two Hilbert
spaces $\mathcal{H}_{A}$ and $\mathcal{H}_{B}$ to the subsystems. Assume that
$\left|  i\right\rangle _{A}$ and $\left|  j\right\rangle _{B}$ $($where
$i,j=1,2,...)$ are two complete orthonormal basis sets for the Hilbert spaces
$\mathcal{H}_{A}$ and $\mathcal{H}_{B}$, respectively. The tensor product
$\mathcal{H}_{A}\otimes\mathcal{H}_{B}$ is another Hilbert space that quantum
mechanics associates with the system consisting of the two subsystems. The
tensor product states $\left|  i\right\rangle _{A}\otimes\left|
j\right\rangle _{B}$ (often written as $\left|  i\right\rangle _{A}\left|
j\right\rangle _{B}$) span the space $\mathcal{H}_{A}\otimes\mathcal{H}_{B}$.
Any state $\left|  \psi\right\rangle _{AB}$ of the composite system made of
the two subsystems is a linear combination of the product basis states
$\left|  i\right\rangle _{A}\left|  j\right\rangle _{B}$ i.e.%

\begin{equation}
\left|  \psi\right\rangle _{AB}=\underset{i,j}{\sum c_{ij}}\left|
i\right\rangle _{A}\left|  j\right\rangle _{B}%
\end{equation}
where $c_{ij}\in\mathbf{C}$. The normalization condition of the state $\left|
\psi\right\rangle _{AB}$ is $\sum_{i,j}\left|  c_{ij}\right|  ^{2}=1$. The
state $\left|  \psi\right\rangle _{AB}$ is called \emph{direct product
}(\emph{or separable})\emph{ state} if it is possible to factor it into two
normalized states from the Hilbert spaces $\mathcal{H}_{A}$ and $\mathcal{H}%
_{B}$. Assume that $\left|  \psi^{(A)}\right\rangle _{A}=\underset{i}{\sum
}c_{i}^{(A)}\left|  i\right\rangle _{A}$ and $\left|  \psi^{(B)}\right\rangle
_{B}=\underset{j}{\sum}c_{j}^{(B)}\left|  j\right\rangle _{B}$ are the two
normalized states from $\mathcal{H}_{A}$ and $\mathcal{H}_{B}$, respectively.
The state $\left|  \psi\right\rangle _{AB}$ is a direct product state when%

\begin{equation}
\left|  \psi\right\rangle _{AB}=\left|  \psi^{(A)}\right\rangle _{A}\left|
\psi^{(B)}\right\rangle _{B}=\left(  \sum_{i}c_{i}^{(A)}\left|  i\right\rangle
_{A}\right)  \left(  \sum_{j}c_{j}^{(B)}\left|  j\right\rangle _{B}\right)
\end{equation}
Now a state in $\mathcal{H}_{A}\otimes\mathcal{H}_{B}$ is called
\emph{entangled} if it is not a direct product state. In words, entanglement
describes the situation when the state of `whole' cannot be written in terms
of the states of its constituent `parts'.

\subsection{Entanglement in $2\otimes2$ systems}

The $2\otimes2$-dimensional quantum systems are of particular interest and
importance to the study of quantum games. In particular, such systems are
considered as the natural requirement for playing two-player quantum games.
Eisert et. al \cite{Eisert} used a $2\otimes2$ system to investigate the
impact of entanglement on a Nash equilibrium in Prisoners' Dilemma.

Let $\mathcal{H}_{A}$ and $\mathcal{H}_{B}$ be two-dimensional Hilbert spaces
with bases $\left\{  \left|  0\right\rangle _{A},\left|  1\right\rangle
_{A}\right\}  $ and$\ \left\{  \left|  0\right\rangle _{B},\left|
1\right\rangle _{B}\right\}  $, respectively. Then a basis for the Hilbert
space $\mathcal{H}_{A}\otimes\mathcal{H}_{B}$ is given by%
\begin{equation}
\left\{  \left|  0\right\rangle _{A}\otimes\left|  0\right\rangle _{B},\text{
\ }\left|  0\right\rangle _{A}\otimes\left|  1\right\rangle _{B},\text{
\ }\left|  1\right\rangle _{A}\otimes\left|  0\right\rangle _{B},\text{
\ }\left|  1\right\rangle _{A}\otimes\left|  1\right\rangle _{B}\right\}  .
\end{equation}
The most general state in the Hilbert space $\mathcal{H}_{A}\otimes
\mathcal{H}_{B}$ can be written as%

\begin{equation}
\left|  \psi\right\rangle _{AB}=\sum_{i,j=0}^{1}c_{ij}\left|  i\right\rangle
_{A}\otimes\left|  j\right\rangle _{B}
\label{General State in 2-dimensional Hilbert Space}%
\end{equation}
which is usually written as $\left|  \psi\right\rangle =\sum_{i,j}%
c_{ij}\left|  ij\right\rangle $. Here the indices $i$ and $j$ refer to states
in the Hilbert spaces $\mathcal{H}_{A}$ and $\mathcal{H}_{B}$, respectively.
The general normalized states in $\mathcal{H}_{A}$ and $\mathcal{H}_{B}$ are
$\left|  \psi^{(A)}\right\rangle _{A}=c_{0}^{(A)}\left|  0\right\rangle
+c_{1}^{(A)}\left|  1\right\rangle $ and $\left|  \psi^{(B)}\right\rangle
_{B}=c_{0}^{(B)}\left|  0\right\rangle +c_{1}^{(B)}\left|  1\right\rangle $,
respectively. The state $\left|  \psi\right\rangle _{AB}$ is a product state when%

\begin{equation}
\left|  \psi\right\rangle _{AB}=(c_{0}^{(A)}\left|  0\right\rangle
+c_{1}^{(A)}\left|  1\right\rangle )\otimes(c_{0}^{(B)}\left|  0\right\rangle
+c_{1}^{(B)}\left|  1\right\rangle )
\label{Separability Criterion for 2-qubit state}%
\end{equation}
where $\left|  c_{0}^{(A)}\right|  ^{2}+\left|  c_{1}^{(A)}\right|  ^{2}=1$
and $\left|  c_{0}^{(B)}\right|  ^{2}+\left|  c_{1}^{(B)}\right|  ^{2}=1$.

For example, consider the state%

\begin{equation}
\left|  \psi\right\rangle _{AB}=(\left|  00\right\rangle +\left|
11\right\rangle )/\sqrt{2} \label{2-qubit entangled state}%
\end{equation}
For this state the criterion (\ref{Separability Criterion for 2-qubit state})
implies the results%

\begin{equation}
c_{0}^{(A)}c_{0}^{(B)}=1/\sqrt{2},\text{ \ \ }c_{1}^{(A)}c_{1}^{(B)}%
=1/\sqrt{2},\text{ \ \ }c_{0}^{(A)}c_{1}^{(B)}=0,\text{ \ \ }c_{1}^{(A)}%
c_{0}^{(B)}=0
\end{equation}
These equations cannot be true simultaneously. The state (\ref{2-qubit
entangled state}) is, therefore, entangled.

\section{Bell's inequality}

Starting from the assumptions of realism and locality, in 1964 Bell
\cite{Bell} derived an inequality which was shown \cite{Aspect et al} later to
be violated by the quantum mechanical predictions for entangled states of a
composite system. Bell's theorem \cite{AsherPeres} is the collective name for
a family of results, all showing the impossibility of local realistic
interpretation of quantum mechanics. Later work \cite{PeresBell} has produced
many different types of Bell-type inequalities.

Entangled states are closely related to Bell's inequalities. The relationship
is described by Gisin's theorem \cite{Gisintheorem} which says that pure
entangled states of $2\otimes2$-dimensional quantum systems (two qubits)
always violate a Bell-type inequality. Recently, Gisin's theorem has been
extended \cite{Gisin3Qubit} to $2\otimes3$-dimensional quantum systems (three
qubits), making stronger the relationship between entanglement and Bell's inequalities.

Let $A(a)$ and $A(a^{\prime})$ be the two observables for observer $A$ in the
an EPR experiment. Similarly, let $B(b)$ and $B(b^{\prime})$ be the two
observables for the observer $B$. In general, the observables $A(a)$ and
$A(a^{\prime})$ are incompatible and cannot be measured at the same time, and
the same holds for $B(b)$ and $B(b^{\prime})$.

It is assumed that the two particles that reach observers $A$ and $B$ in EPR
experiments\ possess hidden variables which fix the outcome of all possible
measurements. These hidden variables are collectively represented by $\lambda
$, assumed to belong to a set $\Lambda$ with a probability density
$\rho(\lambda)$. The normalization implies%

\begin{equation}
\int_{\Lambda}\rho(\lambda)d\lambda=1
\end{equation}
Because a given $\lambda$ makes the four dichotomic observables assume
definite values, we can write%

\begin{equation}
A(a,\lambda)=\pm1;\text{ \ \ }A(a^{\prime},\lambda)=\pm1;\text{ \ \ }%
B(b,\lambda)=\pm1;\text{ \ \ }B(b^{\prime},\lambda)=\pm1
\end{equation}
That is, the physical reality is marked by the variable $\lambda$. Now
introduce a \emph{correlation function} $C(a,b)$ between two dichotomic
observables $a$ and $b$, defined by%

\begin{equation}
C(a,b)=\int_{\Lambda}A(a,\lambda)B(b,\lambda)\rho(\lambda)d\lambda
\end{equation}
For a linear combination of four correlation functions, define \emph{Bell's
measurable quantity} $\Delta$ as%

\begin{equation}
\Delta=C(a,b)+C(a^{\prime},b^{\prime})+C(a^{\prime},b)-C(a,b^{\prime})
\end{equation}
Only four correlation functions, out of a total of sixteen, enter into the
definition of $\Delta$. We can write%

\begin{align}
&  \left|  C(a,b)+C(a^{\prime},b^{\prime})+C(a^{\prime},b)-C(a,b^{\prime
})\right| \nonumber\\
&  \leq\int_{\Lambda}\left\{  \left|  A(a,\lambda)\right|  \left|
B(b,\lambda)-B(b^{\prime},\lambda)\right|  +\left|  A(a^{\prime}%
,\lambda)\right|  \left|  B(b,\lambda)+B(b^{\prime},\lambda)\right|  \right\}
\rho(\lambda)d\lambda
\end{align}
Since%

\begin{equation}
\left|  A(a,\lambda)\right|  =\left|  A(a^{\prime},\lambda)\right|  =1
\end{equation}
we have%

\begin{align}
&  \left|  C(a,b)+C(a^{\prime},b^{\prime})+C(a^{\prime},b)-C(a,b^{\prime
})\right| \nonumber\\
&  \leq\int_{\Lambda}\left\{  \left|  B(b,\lambda)-B(b^{\prime},\lambda
)\right|  +\left|  B(b,\lambda)+B(b^{\prime},\lambda)\right|  \right\}
\rho(\lambda)d\lambda\label{bellproof}%
\end{align}
Also $\left|  B(b,\lambda)\right|  =\left|  B(b^{\prime},\lambda)\right|  =1$,
so that%

\begin{equation}
\left|  B(b,\lambda)-B(b^{\prime},\lambda)\right|  +\left|  B(b,\lambda
)+B(b^{\prime},\lambda)\right|  =2
\end{equation}
and the inequality (\ref{bellproof}) reduces to%

\begin{equation}
\left|  C(a,b)+C(a^{\prime},b^{\prime})+C(a^{\prime},b)-C(a,b^{\prime
})\right|  \leq2 \label{bell inequality}%
\end{equation}
which is called CHSH form \cite{CHSH,CHSH1}\ of Bell's inequality.

\subsection{Violation of Bell's inequality}

It can be shown that Bell's inequality is violated if the observers $A$ and
$B$ have appropriate observables. In Ref. \cite{Benenti} an illustrative
example is discussed showing the violation of the inequality for certain
observables. For completeness of this section the example is reproduced below.

Let $A(\mathbf{a})$ and $B(\mathbf{b})$ be the spin observables for an
entangled state%

\begin{equation}
\left|  \psi\right\rangle =c_{00}\left|  00\right\rangle -c_{11}\left|
11\right\rangle \label{two-qubit entangled state}%
\end{equation}
where $c_{00}$ and $c_{11}$ are real. Assume that the observers $A$ and $B$
measure along the directions $\mathbf{a}$ and $\mathbf{b}$, respectively.
Consider the quantum-mechanical mean value of the correlation:%

\begin{equation}
C(\mathbf{a},\mathbf{b})=\left\langle \psi\right|  (\mathbf{\hat{\sigma}%
}^{(A)}.\mathbf{a})(\mathbf{\hat{\sigma}}^{(B)}.\mathbf{b})\left|
\psi\right\rangle \label{correlation along a and b}%
\end{equation}
where $\mathbf{\hat{\sigma}=}\left(  \hat{\sigma}_{x},\hat{\sigma}_{y}%
,\hat{\sigma}_{z}\right)  $. To evaluate (\ref{correlation along a and b}) for
the entangled state (\ref{two-qubit entangled state}) the following matrix
elements can be found \cite{Benenti} for $\mathbf{r=}\left(  x,y,z\right)  $%

\begin{align}
\left\langle 0\right|  \mathbf{\hat{\sigma}.r}\left|  0\right\rangle  &
=z,\text{ \ \ }\left\langle 1\right|  \mathbf{\hat{\sigma}.r}\left|
1\right\rangle =-z,\nonumber\\
\left\langle 0\right|  \mathbf{\hat{\sigma}.r}\left|  1\right\rangle  &
=x-iy,\text{ \ \ }\left\langle 1\right|  \mathbf{\hat{\sigma}.r}\left|
0\right\rangle =x+iy. \label{EPR Matrix Elements}%
\end{align}
Take $\mathbf{a}=(x_{a},y_{a},z_{a})$ and $\mathbf{b}=(x_{b},y_{b},z_{b})$.
Using (\ref{EPR Matrix Elements}) the correlation $C(\mathbf{a},\mathbf{b})$
is given as \cite{Benenti}%

\begin{align}
C(\mathbf{a},\mathbf{b})  &  =\left\langle \psi\right|  (\mathbf{\hat{\sigma}%
}^{(A)}.\mathbf{a})(\mathbf{\hat{\sigma}}^{(B)}.\mathbf{b})\left|
\psi\right\rangle \nonumber\\
&  =(c_{00}\left\langle 00\right|  +c_{11}\left\langle 11\right|
)(\mathbf{\hat{\sigma}}^{(A)}.\mathbf{a})(\mathbf{\hat{\sigma}}^{(B)}%
.\mathbf{b})(c_{00}\left|  00\right\rangle +c_{11}\left|  11\right\rangle
)\nonumber\\
&  =c_{00}^{2}\left\langle 0\right|  \mathbf{\hat{\sigma}}^{(A)}%
.\mathbf{a}\left|  0\right\rangle \left\langle 0\right|  \mathbf{\hat{\sigma}%
}^{(B)}.\mathbf{b}\left|  0\right\rangle +c_{11}^{2}\left\langle 1\right|
\mathbf{\hat{\sigma}}^{(A)}.\mathbf{a}\left|  1\right\rangle \left\langle
1\right|  \mathbf{\hat{\sigma}}^{(B)}.\mathbf{b}\left|  1\right\rangle
\nonumber\\
&  +2c_{00}c_{11}\text{Re}\left(  \left\langle 0\right|  \mathbf{\hat{\sigma}%
}^{(A)}.\mathbf{a}\left|  1\right\rangle \left\langle 0\right|  \mathbf{\hat
{\sigma}}^{(B)}.\mathbf{b}\left|  1\right\rangle \right) \nonumber\\
&  =z_{a}z_{b}+2c_{00}c_{11}(x_{a}x_{b}-y_{a}y_{b})
\end{align}
Consider now a set of directions $\mathbf{a}=(1,0,0),$ $\mathbf{a}^{\prime
}=(0,0,1),$ $\mathbf{b}=(x_{b},0,z_{b})$ and $\mathbf{b}^{\prime}%
=(-x_{b},0,z_{b})$. For these directions we obtain%

\begin{equation}
\left|  C(\mathbf{a,b})+C(\mathbf{a}^{\prime}\mathbf{,b}^{\prime
})+C(\mathbf{a}^{\prime}\mathbf{,b})-C(\mathbf{a,b}^{\prime})\right|
=2\left|  2c_{00}c_{11}x_{b}+z_{b}\right|
\end{equation}
Bell's inequality is violated if $(2c_{00}c_{11}x_{b}+z_{b})>1$. Assume that
$\theta_{b}$ is a small angle between $\mathbf{b}$ and the $z$-axis; then
$z_{b}=\cos\theta_{b}\approx1$ and $x_{b}=\sin\theta_{b}\approx\theta_{b}$
which gives the result%

\begin{equation}
(2c_{00}c_{11}x_{b}+z_{b})\approx1+2c_{00}c_{11}\theta_{b}%
\end{equation}
The inequality, therefore, is violated with small $\theta_{b}$, when both
$c_{00}$ and $c_{11}$ are real and $c_{00}c_{11}>1$.

\subsection{Local realism and the violation of Bell's inequality}

Several experiments \cite{Aspect et al,ShihAlley} have shown that Bell's
inequality can be violated by quantum observables. The violation is often
interpreted \cite{AsherPeres} as the decisive argument against hypothesis of
the existence of local objective reality in quantum physics. Though that
remains the majority view, it is not a conclusion supported by general
agreement; several authors have disagreed that violation of Bell inequalities
necessarily leads to conclusions about local realism. In the following we
briefly describe the arguments of Fine \cite{Fine,Fine1}, Pitowsky
\cite{Pitowsky}, DeBaere \cite{DeBaere}, Malley \cite{Malley,Malley1} and
Fivel \cite{Fivel,Fivel 1}.

A description of the local realism and violation of Bell's inequality is
included here, firstly because it touches on the questions about the true
quantum content/character of a quantum game and secondly because such a
description is intimately connected to recent suggestions for non-local games
\cite{Pseudo Telepathy,Pseudo Telepathy1}.

\subsubsection{Works of Fine, Pitowsky, DeBaere, Malley, Fivel, Gustafson and Atkinson}

Two mathematical theorems due to Arthur Fine \cite{Fine,Fine1} and Itamar
Pitowsky \cite{Pitowsky} are considered important in the continuing debate
about what are the real lessons to be learned from the violation of the Bell
inequalities. The theorems link the violation of Bell's inequalities to
certain facts about probability theory. One of the influential arguments is
due to Fine \cite{Fine,Fine1} who showed that in a correlation experiment
\cite{AsherPeres} which is used to test Bell's inequalities the following
statements are mutually equivalent.

\begin{enumerate}
\item [ a)]There is a deterministic hidden variable model for the experiment.

\item[ b)] There is a factorizable, stochastic model.

\item[ c)] There is one joint distribution for all observables of the
experiment, such that it yields the experimental probabilities.

\item[ d)] There are well-defined, compatible joint distributions for all
pairs and triples of commuting and non-commuting observables.

\item[ e)] Bell's inequalities hold.
\end{enumerate}

The equivalence of these statements means that violation of Bell's
inequalities has a lot to do with the absence of joint probability
distributions for incompatible observables in correlation experiments. In
similar vein, Pitowsky \cite{Pitowsky} showed that one can view an eight-tuple
of real numbers from the interval $\left[  0,1\right]  $, associated with the
experiments used to test Bell's inequalities, as a set of four single and four
joint probabilities defined on a single classical probability space if and
only if the eight-tuple satisfies Bell's inequalities. In Pitowsky's analysis
the probabilities in Bell experiments are not defined on a single probability
space because non-commuting observables are involved. These probabilities are,
in fact, defined on four different probability spaces. Because the
experimental results cannot be embedded in a single probability space it
follows that violation of Bell's inequalities is not unusual.
Non-commutativity of quantum observables pertaining to a single system appears
to play a crucial role \cite{DeBaere} in the violation of Bell's inequalities.

On analyzing the geometry underlying no-hidden-variable theorems, Fivel
\cite{Fivel} also came to similar conclusions. He showed that a hidden
variable measure determines a metric on the homogeneous space consisting of
the set of orientations of a measuring device (e.g., a Stern-Gerlach magnet)
when these are regarded as being produced by the action of a Lie group. For
this metric the corresponding triangle inequality becomes one of Bell's
inequalities. He then makes an important observation that when the homogeneous
space is identified with Hilbert space projectors this identification induces
another metric on the space which is locally convex. Now EPR's definition of
element of physical reality \cite{EPR} forces the square of the locally convex
metric equal to the metric determined by the hidden variable measure. But, in
fact, it is impossible because the square of a locally convex metric cannot be
a metric.

Recently, Fine and Malley \cite{Malley,Malley1} have argued that violations of
local realism can be found in all those situations where non-commutative
observables are involved, without the necessity of sophisticated correlation experiments.

The above arguments imply that one cannot argue \cite{Gustafson,Atkinson}
either locality or non-locality on the basis of satisfaction or violation of
Bell's inequality. Bell's claim that his formulation of a ``locality
condition'' is an \emph{essential} assumption for the validity of his
inequality has, therefore, been put under scrutiny and has sometimes even been
rejected \cite{Adenier}. Apart from the implications for local realism, Bell's
inequalities have been shown to be closely related to entanglement. For
example, Gisin \cite{Gisintheorem} showed that any pure entangled state of two
particles violates a Bell inequality for two-particle correlations.

\chapter{Quantum games}

\section{Introduction}

Speaking roughly, a quantum game can be thought of as strategic manoeuvring of
a quantum system by participating parties who are identified as players. The
players have the necessary means to perform actions on the quantum system and
knowledge is shared among them about what constitutes a strategy. Often the
strategy space is the set of possible actions that players can take on the
quantum system. The players' payoff functions, or utilities, are associated
with their strategies. Payoffs are obtained from the results of measurements
performed on the quantum system.

A two-player quantum game, for example, is a set \cite{Eisert1}:%

\begin{equation}
\Gamma=(\mathcal{H},\rho,S_{A},S_{B},P_{A},P_{B})
\end{equation}
consisting of an underlying Hilbert space $\mathcal{H}$ of the physical
system, the initial state $\rho$, the sets $S_{A}$ and $S_{B}$ of allowed
quantum operations for two players, and the payoff functions $P_{A}$ and
$P_{B}$. In most of the existing set-ups to play quantum games the initial
state $\rho$ is the state of one or more qubits, or qutrits \cite{RSP}, or
possibly qudits.

\section{\label{Earlier Works}Earlier works}

Several situations in quantum theory can be found which have connections to
game theory. To find the roots of quantum games some of the earlier works are
enlisted below which appear as the underpinning to many illustrative examples
designed to show extraordinary features of quantum theory. The list is not
exhaustive and many others works can be found.

\subsubsection{Blaquiere\textbf{: ``}Wave mechanics as a two-player
game\textbf{''}}

Perhaps Blaquiere's article \cite{Blaquiere} entitled ``Wave mechanics as a
two-player game'' is one of the earliest ones where game-theoretical ideas are
discussed in the context of quantum physics. Blaquiere addresses a question
concerned with a connection between dynamic programming and the theory of
differential games, on the one hand, and wave mechanics on the other. The
author argues that wave mechanics is born of a dynamic programming equation
which Louis de Broglie obtained in 1923. He then expresses the stationarity
principle in the form of a min-max principle, which he writes in the form of
sufficiency conditions for the optimality of strategies in a two-player
zero-sum differential game. Blaquiere finds that the saddle-point condition,
on which optimality of strategies is based, is an extension of Hamilton's
principle of least action.

\subsubsection{Wiesner\textbf{: ``}Quantum money''}

Wiesner's work on quantum money \cite{Wiesner} is widely believed
\cite{Wiedemann} to have started the field of quantum cryptography. Because
cryptographic protocols can be written in the language of game theory; it then
seems reasonable to argue that, apart from originating quantum cryptography,
Wiesner's work provided a motivation for quantum games. Wiesner suggested
using the uncertainty principle for creating:

\begin{enumerate}
\item [ a)]a means of transmitting two messages, either but not both of which
may be received.

\item[ b)] money that it is physically impossible to counterfeit.
\end{enumerate}

Wiesner's proposal, though made much earlier, remained unpublished until 1983.

\subsubsection{Mermin\textbf{: ``}$n$-player quantum game\textbf{''}}

In 1990 Mermin \cite{Mermin,Mermin1} presented an $n$-player quantum game that
can be won with certainty when it involves $n$ spin half particles in a
Greenberger-Horne-Zeilinger (GHZ) state \cite{GHZ}; no classical strategy can
win the game with a probability greater than $\frac{1}{2}+\frac{1}{2^{(n/2)}}$.

\section{\label{Quantum penny-flip game}Quantum penny-flip game}

As described above, many earlier works in quantum physics can be found which
have links to game theory in one way or the other. Nevertheless, the credit
for the emergence of quantum games as an independent domain of research is
usually (and justifiably) reserved for D. Meyer. Meyer \cite{MeyerDavid}
suggested a quantum version of a penny-flip game played between Picard and Q,
the two well-known characters from the famous American science fiction serial
Star Trek. Meyer suggested \cite{MeyerQGsQAlgorithms} the game in the hope
that game theory might be helpful in understanding the working of quantum
algorithms and perhaps even in finding new ones, a task generally considered
hard because only a few quantum algorithms are known to date.

Meyer \cite{MeyerDavid} describes his game as follows. The starship Enterprise
is facing some imminent catastrophe when Q appears on the bridge and offers to
rescue the ship if Captain Picard\footnote{Meyer considered the initials and
abilities of Picard and Q ideal for his illustration.} can beat him at a
simple game: Q produces a penny and asks Picard to place it in a small box,
head up. Then Q, followed by Picard, followed by Q, reaches into the box,
without looking at the penny, and either flips it over or leaves it as it is.
After Q's second turn they open the box and Q wins if the penny is head up. Q
wins every time they play, using the following strategy:%

\begin{align*}
&  \left|  0\right\rangle \overset{\text{Q}}{\underset{\hat{H}}{\longmapsto}%
}\frac{1}{\sqrt{2}}(\left|  0\right\rangle +\left|  1\right\rangle )\\
&  \overset{\text{Picard}}{\underset{\hat{\sigma}_{x}\text{ or }\hat{I}_{2}%
}{\longmapsto}}\frac{1}{\sqrt{2}}(\left|  0\right\rangle +\left|
1\right\rangle )\\
&  \overset{\text{Q}}{\underset{\hat{H}}{\longmapsto}}\left|  0\right\rangle
\end{align*}
Here $0$ denotes `head' and $1$ denotes `tail', $\hat{H}=\frac{1}{\sqrt{2}%
}\left(
\begin{array}
[c]{cc}%
1 & 1\\
1 & -1
\end{array}
\right)  $ is the Hadamard transformation \cite{Nielsen}, $\hat{I}_{2}$ means
leaving the penny alone and the action with $\hat{\sigma}_{x}=\left(
\begin{array}
[c]{cc}%
0 & 1\\
1 & 0
\end{array}
\right)  $ flips the penny over. Q's quantum strategy of putting the penny
into the equal superposition of `head' and `tail', on his first turn, means
that whether Picard flips the penny over or not, it remains in an equal
superposition which Q can rotate back to `head' by applying $\hat{H}$ again
since $\hat{H}=\hat{H}^{-1}$. So Q always wins when they open the box.

Q's classical strategy consists of implementing $\hat{\sigma}_{x}$ or $\hat
{I}_{2}$ on his turn. When Q is restricted to play only classically, flipping
the penny over or not on each turn with equal probability becomes an optimal
strategy for both the players. By adapting this classical strategy Q wins only
with probability $\frac{1}{2}$. By using a quantum strategy Q can, therefore,
win with probability $1$.

\section{Quantum Prisoners' Dilemma}

Meyer's paper attracted immediate attention and soon afterwards Eisert,
Wilkens, and Lewenstein \cite{Eisert} formulated a quantization scheme for the
famous game of Prisoners' Dilemma. Eisert et al.'s scheme suggests a quantum
version of a two-player game by assigning two basis vectors $\left|
S_{1}\right\rangle $ and $\left|  S_{2}\right\rangle $ in the Hilbert space of
a qubit. States of the two qubits belong to the two-dimensional Hilbert spaces
$\mathcal{H}_{A}$ and $\mathcal{H}_{B}$ respectively. A state of the game is
defined by a vector residing in the tensor-product space $\mathcal{H}%
_{A}\otimes\mathcal{H}_{B}$, spanned by the basis $\left|  S_{1}%
S_{1}\right\rangle ,\left|  S_{1}S_{2}\right\rangle ,\left|  S_{2}%
S_{1}\right\rangle $ and $\left|  S_{2}S_{2}\right\rangle $. The game's
initial state is $\left|  \psi_{ini}\right\rangle =\hat{J}\left|  S_{1}%
S_{1}\right\rangle $ where $\hat{J}$ is a unitary operator known to both the
players. Let Alice and Bob be the two prisoners. Alice's and Bob's strategies
are unitary operations $\hat{U}_{A}$ and $\hat{U}_{B}$, respectively, chosen
from a strategic space \c{S}. The state of the game changes to $\hat{U}%
_{A}\otimes\hat{U}_{B}\hat{J}\left|  S_{1}S_{1}\right\rangle $ after the
players' actions. Finally, the measurement consists of applying the reverse
unitary operator $\hat{J}^{\dagger}$, followed by a pair of Stern-Gerlach type
detectors. Before detection the final state of the game is $\left|  \psi
_{fin}\right\rangle =\hat{J}^{\dagger}\hat{U}_{A}\otimes\hat{U}_{B}\hat
{J}\left|  S_{1}S_{1}\right\rangle $. The players' expected payoffs can then
be written as the projections of the state $\left|  \psi_{fin}\right\rangle $
onto the basis vectors of the tensor-product space $\mathcal{H}_{A}%
\otimes\mathcal{H}_{B}$, weighted by the constants appearing in the matrix
representation of the two-player game. For example, Alice's payoff function reads%

\begin{equation}
P_{A}=r\left|  \left\langle S_{1}S_{1}\mid\psi_{fin}\right\rangle \right|
^{2}+s\left|  \left\langle S_{1}S_{2}\mid\psi_{fin}\right\rangle \right|
^{2}+t\left|  \left\langle S_{2}S_{1}\mid\psi_{fin}\right\rangle \right|
^{2}+u\left|  \left\langle S_{2}S_{2}\mid\psi_{fin}\right\rangle \right|  ^{2}
\label{Eisert's Alice's payoff}%
\end{equation}
Bob's payoff function is then obtained by the transformation
$s\rightleftarrows t$ in Eq. (\ref{Eisert's Alice's payoff}). Eisert et al.
\cite{Eisert,Eisert1} allowed the players' actions to be chosen from the space
\c{S} of unitary operators of the form%

\begin{equation}
U(\theta,\phi)=\left(
\begin{tabular}
[c]{ll}%
$e^{i\phi}\cos(\theta/2)$ & $\sin(\theta/2)$\\
$\text{-}\sin(\theta/2)$ & $e^{-i\phi}\cos(\theta/2)$%
\end{tabular}
\right)  \label{Eisert's unitary operators}%
\end{equation}
where%

\begin{equation}
\theta\in\lbrack0,\pi],\text{ \ \ }\phi\in\lbrack0,\pi/2] \label{Ranges}%
\end{equation}
They defined their unitary operator $\hat{J}=\exp\left\{  i\gamma S_{2}\otimes
S_{2}/2\right\}  $ with $\gamma\in\lbrack0,\pi/2]$ representing a measure of
the game's entanglement. At $\gamma=0$ the game reduces to its classical form.
Eisert et al. found that for maximally entangled game (i.e. $\gamma=\pi/2$)
the game has a unique Pareto optimal\emph{ }equilibrium $\hat{Q}\otimes\hat
{Q}$ where $\hat{Q}\sim\hat{U}(0,\pi/2)$.

\section{\label{LiteratureReview}Review of recent literature}

Meyer's \cite{MeyerDavid} and Eisert et al.'s work \cite{Eisert,Eisert1} is
often cited as having started the field of quantum games, though many
game-like examples can be found in the quantum physics literature. In many
cases these examples are `tailored'\emph{ }to illustrate extraordinary
correlations that may exist among spatially-separated quantum objects. The
contribution of Meyer's work consists in providing the basis for a systematic
discussion of an explicitly game-theoretical problem when it is implemented
quantum mechanically. Eisert et al. carried this theme further by proposing an
experimental set-up using quantum-correlated pairs of objects to play a
quantum version of Prisoners' Dilemma. In contrast to several earlier
proposals of game-like examples in quantum physics, Meyer's and Eisert. et
al.'s work brought game theory right into the domain of quantum information
and computation. The era of proposing tailored games to demonstrate the
advantages of quantum strategies over classical ones was left behind after
suggestions of general procedures to quantize known and understood games began
to arise. The focus of discussion shifted from specially-designed games to
general procedures and schemes for implementing quantum versions of games
which were already well-known from courses in game theory.

Below we present a literature review of the recent work on quantum games,
divided into three categories: specially-designed quantum games; quantum games
based on Meyer's and Eisert et al.'s schemes; and other work related to
quantum games.

\subsection{Specially-designed games}

As it is described above, certain quantum games are designed in such a way so
as to give advantage to quantum players against the classical players.
Sometimes, these appear as the games motivated by quantum mechanical
situations with their winning conditions tailored such that only quantum
players can be the winners. Following are some of the examples.

\subsubsection{Meyer: penny-flip quantum game}

See the section (\ref{Quantum penny-flip game}).

\subsubsection{Brassard, Broadbent and Tapp\textbf{: ``}Quantum
pseudo-telepathy\textbf{''}}

In classical computer science, communication complexity theory is an area that
aims at quantifying the amount of communication necessary to solve distributed
computational problems. Quantum communication complexity theory uses quantum
mechanics to reduce the amount of communication below that which would be
classically required. Brassard, Broadbent and Tapp \cite{Pseudo
Telepathy,Pseudo Telepathy1} called `pseudo-telepathy' the situation in which
two or more quantum players can accomplish a distributed task with no need for
communication whatsoever. The situation would be an impossible feat for
classical players, but entanglement allows its possibility. Using Mermin's
$n$-player game (see section (\ref{Earlier Works})), that exploits multi-party
entanglement, Brassard, Broadbent and Tapp recast the game in terms of
pseudo-telepathy. They derived an upper bound on the best possible classical
strategy for attempting to play the game. It allowed them to find how well an
imperfect quantum-mechanical apparatus must perform in order to exhibit a
behaviour that would be classically impossible to explain.

\subsubsection{Mermin\textbf{: ``}$n$-player quantum game\textbf{''}}

See the section (\ref{Earlier Works}).

\subsubsection{Vaidman: ``Three-player quantum game''}

In 1999 Vaidman \cite{Vaidman,Vaidman1} illustrated the GHZ paradox \cite{GHZ}
by using a game among three players. A team of three players is allowed to
make any preparations before they are separately taken to three different
remote locations. Then, at a certain time each player is asked one of two
possible questions: ``What is $X$?'' or ``What is $Y$?'' to which they must
quickly give one of the answers: ``$1$'' or ``$-1$''. According to the rules
of the game, either all players are asked the $X$ question, or only one player
is asked the $X$ question and the other two are asked the $Y$ question. The
team wins if the product of their three answers is $-1$ in the case of three
$X$ questions and is $1$ in the case of one $X$ and two $Y$ questions. It can
easily be shown that if the answer of each player is determined by some local
hidden variable (LHV) theory then the best strategy of the team will lead to a
$75\%$ probability to win. However, a quantum team equipped with ideal devices
can win with certainty. Each player performs a spin measurement of a
spin-$1/2$ particle (a $\hat{\sigma}_{x}$ measurement for the $X$ question and
a $\hat{\sigma}_{y}$ measurement for the $Y$ question) and gives the answer
$1$ for spin ``up'' and $-1$ for spin ``down''. Quantum theory ensures that if
the players have particles prepared in the GHZ state \cite{GHZ}, the team
always wins. In Vaidman's opinion \cite{Vaidman1}, constructing such devices
and seeing that quantum teams indeed win the game with a probability
significantly larger than $75\%$ will be a very convincing proof of Bell-type inequalities.

It is to be pointed out that the algebraic contradiction, which is obtained by
comparing the four equations describing the team's winning condition, assumes
that all the equations hold simultaneously. In fact, the four equations
represent four incompatible situations. In other words, with reference to
Fine's theorem \cite{Fine,Fine1}, the description of each player's answer by
an LHV theory is strictly equivalent to the assumption of a joint probability
distribution for incompatible observables whose rejection, in Fine's words
\cite{Fine}, is the ``very essence of quantum mechanics''.

\subsubsection{Grib and Parfionov: ``Macroscopic quantum game''}

Using a new approach to quantum games, Grib and Parfionov \cite{Grib,Grib1}
considered a game in which the acts of the participants do not have an
adequate description in terms of Boolean logic and the classical theory of
probabilities. They constructed a model of the game interaction using a
non-distributive ortho-complemented lattice. They proposed a quantization
scheme for the game and developed an algorithm to search for a Nash
equilibrium. They showed that, in contrast with the classical case, a discrete
set of equilibria is possible in the quantum situation.

In \cite{Grib2} Grib and Parfionov presented examples of two-player
macroscopic quantum games in which special rules become responsible for
breaking the distributive property of a lattice of yes-no questions. They
discussed examples of games using spin-$1/2$ particles and found new Nash equilibria.

\subsubsection{Other games}

In \cite{Vaidman1} Vaidman lists several proposals where bizarre features of
quantum mechanics have been explained through various games. It includes a
variation of the GHZ game given by Steane and van Dam \cite{Steane van Dam
Game}, a game based on the original Bell proof by Tsirelson \cite{Tsirelson
Game}, the ``quantum cakes'' game using non-maximally entangled state by Kwiat
and Hardy \cite{Kwiat Hardy Game} and Cabello's proposal \cite{Cabello Game}
for a two-party Bell-inequality proof which can be put into the form of a
game. In \cite{Vaidman1} Vaidman also presented a game which he called
``impossible necklace''.

\subsection{Quantum games based on Eisert et al.'s formalism}

Eisert et al.'s quantum Prisoners' Dilemma gives a general scheme to quantize
the two-player noncooperative games. Following works on quantum games are
based on their scheme.

\subsubsection{Benjamin and Hayden\textbf{: ``}Multi-player quantum
games\textbf{''}}

Generalizing the scheme of Eisert et al., Benjamin and Hayden \cite{Benjamin
Multiplayer} presented their first study of quantum games with more than two
players. They found that such games can possess a new form of equilibrium
strategy, one which has no analogue either in traditional games or even in
two-player quantum games. They showed that in such games, because of a sharing
of entanglement among many players, there may exist `pure' coherent equilibria
enabling new kinds of cooperative behaviour. It is a cooperative behaviour
that prevents players from successfully betraying one-another.

Benjamin and Hayden's work provides a means to analyze multi-party
entanglement by using game theory.

\subsubsection{Marinatto and Weber\textbf{: ``}A quantum approach to static
games of complete information\textbf{''}}

Marinatto and Weber \cite{Marinatto} extended the concept of a classical
two-person static game to the quantum domain by giving a Hilbert space
structure to the space of classical strategies. They studied the game of
Battle of the Sexes, showing that entangled strategies did lead to a unique
solution of the game. Building up from Eisert et al.'s quantization of
Prisoners' Dilemma \cite{Eisert,Eisert1} they proposed a scheme in which a
strategy is a state in a $2\otimes2$ -- dimensional Hilbert space. At the
start of the game the players are supplied with this strategy. In the next
phase players manipulate the strategy by their actions, identified as
`tactics'. Finally, the quantum state (strategy) is measured and the payoffs
which are awarded depend on the results of the measurement.

The players' actions are within a two-dimensional subspace in this scheme and
`tactics' are local actions on players' qubits. A final measurement is made
independently on each qubit, so that it takes into consideration the local
nature of the players' manipulations. It is achieved by selecting a
measurement basis that respects the division of the Hilbert space into two
equal parts. In a comment, Benjamin \cite{Benjamin2} observed that, overall,
Marinatto and Weber's quantization scheme is fundamentally very similar to
Eisert et al.'s previously proposed scheme \cite{Eisert}. Benjamin argued that
in the quantum Battle of Sexes the players are still faced with the dilemma,
just as they are in the classical game. He noted that their quantum Battle of
Sexes does not have a unique solution, though it may be easier to resolve this
dilemma in the quantum version of the game.

\subsubsection{Du et al.\textbf{: ``}Experimental realization of quantum games
on a quantum computer\textbf{''}}

Du et al. \cite{Du} generalized the quantum Prisoner's Dilemma to the case in
which players share non maximally entangled states. They showed that the game
exhibits an intriguing structure as a function of the amount of entanglement.
They identified two thresholds, separating a classical region, an intermediate
region and a fully quantum region. Moreover, they realized their quantum game
experimentally on a nuclear magnetic resonance quantum computer.

\subsubsection{Flitney and Abbot\textbf{: ``}Quantum Monty Hall problem'',
``Quantum Parrondo Paradox\textbf{''}}

Flitney and Abbott \cite{FlitneyAbbott,Flitney} presented a version of the
Monty Hall problem where the players are permitted to select quantum
strategies. If the initial state involves no entanglement the Nash equilibrium
in the quantum game offers the players nothing more than can be obtained with
a classical mixed strategy. However, if the initial state involves
entanglement of the qubits belonging to the two players, it is advantageous
for one player to have access to a quantum strategy while the other does not.
When both players have access to quantum strategies there is no Nash
equilibrium in pure strategies but there is a Nash equilibrium in quantum
mixed strategies that gives the same average payoff as the classical game.

Parrondo's Paradox is an interesting situation arising when two losing games
are combined to produce a winning one. Flitney and Abbott
\cite{FlitneyAbbott1,Flitney} studied a history-dependent quantum Parrondo
game where the rotation operators, representing the toss of a classical biased
coin, are replaced by general SU($2$) operators to transform the game into the
quantum domain. They showed that, in the initial state of the qubits,
superposition can be used to couple the games to produce interference effects
leading to quite different payoffs from the ones in the classical case.

\subsubsection{Iqbal and Toor\textbf{: ``}Evolutionarily stable strategies in
quantum games\textbf{''}}

Using Marinatto and Weber's quantization scheme\cite{Marinatto}, Iqbal and
Toor \cite{IqbalToor}\ investigated a refinement of the Nash equilibrium
concept within the context of quantum games. The refinement, called an
evolutionarily stable strategy (ESS), was originally introduced in 1970s by
mathematical biologists to model an evolving population using game-theoretical
techniques. They considered situations where quantization changes ESSs without
affecting the corresponding Nash equilibria.

\subsubsection{Kawakami\textbf{: ``}Communication and computation by quantum
games\textbf{''}}

Kawakami \cite{Kawakami} argued that in the classical Prisoners' Dilemma a
certain degree of communication between the two prisoners is performed at the
moment the payoff is given to them by the jailer. Motivated by this view,
Kawakami studied communication and information carriers in quantum games. He
showed, quite surprisingly, that communications in special quantum games can
be used to solve problems that cannot be solved by using communications in
classical games.

\subsubsection{Lee and Johnson\textbf{: ``}General theory of quantum
games\textbf{''}}

Presenting a general theory of quantum games, Lee and Johnson
\cite{LeeJohnson} showed that quantum games are more efficient than classical
games and can provide a saturated upper bound for efficiency. They
demonstrated that the set of finite classical games is a strict subset of the
set of finite quantum games. They deduced a quantum version of the minimax
theorem and the Nash equilibrium theorem. In \cite{Neil Johnson} Johnson
showed that the quantum advantage arising in a simplified multi-player quantum
game becomes a disadvantage when the game's qubit-source is corrupted by a
noisy ``demon''. Above a critical value of the corruption-rate, or
noise-level, the coherent quantum effects impede the players to such an extent
that the optimal choice of game changes from quantum to classical.

\subsubsection{Cheon and Tsutsui\textbf{: ``}Classical and quantum contents of
solvable game theory on Hilbert space\textbf{''}}

Cheon and Tsutsui \cite{Cheon} presented a general formulation of the quantum
game theory that accommodates, for the first time, all possible strategies in
the Hilbert space. Their theory is solvable for two-strategy quantum games.
They showed that their quantum games are equivalent to a family of classical
games supplemented by quantum interference. Their formulation extends Eisert
et al.'s formalism, giving a perspective on why quantum strategies outmaneuvre
classical strategies.

\subsubsection{Shimamura et al.\textbf{: ``}Quantum and classical correlations
between players in game theory''}

Shimamura, \"{O}zdemir, Morikoshi and Imoto \cite{Shimamura} studied the
effects of quantum and classical correlations on game theory. They compared
the quantum correlation present in a maximally entangled state with the
classical correlation generated through phase damping processes on the
maximally entangled state. Their study sheds light on the behaviour of games
under the influence of noisy sources. They observed that the quantum
correlation can always resolve the dilemmas in non-zero sum games and attain
the maximum sum of both players' payoffs, while the classical correlation
cannot necessarily resolve the dilemmas.

\subsection{Related work}

In the following, work is described which is closely related to the field of
quantum games. It is the work that has motivated quantum games and, in certain
cases, has been motivated by the developments in quantum games.

\subsubsection{Deutsch\textbf{: ``}Quantum theory of probability and
decisions\textbf{''}}

The probabilistic predictions of quantum theory are conventionally obtained
from a special probabilistic axiom. Deutsch \cite{Deutsch Q Probability}
argued that all the practical consequences of the probabilistic predictions of
quantum theory can also be made to follow from the non-probabilistic axioms of
quantum theory together with the non-probabilistic part of classical decision theory.

\subsubsection{Blaquiere\textbf{: ``}Wave mechanics as a two-player game''}

See section (\ref{Earlier Works}).

\subsubsection{Wiesner\textbf{: ``}Quantum money''}

See section (\ref{Earlier Works}).

\subsubsection{Piotrowski and Sladkowski\textbf{: ``}Quantum-like description
of markets and economics\textbf{''}}

Piotrowski and Sladkowski \cite{Piotrowski} proposed quantum-like description
of markets and economics. Their approach has roots in the developments in
quantum game theory. In \cite{Piotrowski1} they investigated quantum
bargaining games that are a special class of quantum market games without
institutionalized clearing houses. In \cite{Piotrowski2} they showed the
possibility of defining a risk inclination operator, acting in Hilbert space,
that has similarities with the quantum description of the harmonic oscillator.
They also formulated a \emph{quantum anthropic principle}.

\subsubsection{Pietarinen\textbf{: ``}Game-theoretic perspective of quantum
logic and quantum theory\textbf{''}}

Doubts have been expressed about any `substantial' insight that quantum logic
can provide into the nature of composite quantum systems. In order to
neutralize such pessimistic views, Pietarinen \cite{Pietarinen} suggested that
games (especially extensive games of imperfect information) can provide a
useful set of tools for giving a semantics to quantum logic. He suggested that
game theory can be brought to bear on questions concerning the interpretation
and nature of uncertainty in the foundations of quantum theory. He also
suggested \cite{Pietarinen} that the kinship between game theory and quantum
logic implies that the propositional logic of informational independence is
useful in understanding non-locality and EPR-type paradoxes.

\subsubsection{Lassig\textbf{: ``}Game theory from statistical
mechanics\textbf{''}}

Looking at game theory from the viewpoint of statistical mechanics, Lassig
\cite{Lassig} presented a systematic theory of stochastic effects in game
theory, including the effects of fluctuations. In a biological context, such
effects are relevant for the evolution of finite populations with
frequency-dependent selection. The states of the populations are also
time-dependent and are defined by a probability distribution over mixed
strategies. Lassig found that stochastic effects make the equation governing
the evolutionary dynamics take the form of a Schr\"{o}dinger equation in
imaginary time, thus justifying the use of the name `quantum game theory'.

Although ostensibly dealing with quantum games, Lassig's approach to quantum
games appears significantly different from the one presented within the
framework of quantum computation and information theory.

\subsubsection{Boukas\textbf{: ``}Quantum formulation of classical two-person
zero-sum games\textbf{''}}

In a relatively less known paper, Boukas \cite{Boukas} extended the concept of
a classical player, corresponding to a simple random variable on a probability
space of finite cardinality, to that of a quantum player, corresponding to a
self-adjoint operator on a quantum probability Hilbert space. Boukas proved
quantum versions of von Neumann's minimax theorem.

\section{\label{Criticism of quantum games}Criticism of quantum games}

Quantum games have been put under close scrutiny since their earliest
suggestions were put forward. The debate is continuing to date
\cite{IqbalJPALetter,LaMura,DuChengLi,Cheon}. In the following are described
some of the well-known critical comments and the replies made to those.

\subsection{Enk's comment on quantum penny-flip game}

Meyer's quantum penny-flip game can be played using one qubit. Enk \cite{Enk}
commented that the game involves no entanglement and can therefore be
simulated classically, though Meyer had reached a correct conclusion. Enk
argued that neither Bell's inequalities nor the Kochen-Specker theorem
\cite{KochenSpecker} exist for the game and thus the quantum game shows only a
rather unsurprising result i.e. the superiority of an extended set of
strategies over a restricted one.

Meyer replied \cite{Meyer's Reply} that the issue is not whether there exists
a classical simulation or not but rather how the complexity of that classical
simulation would scale if the size of the game increases. Disagreeing with
Enk's claim that Q's strategy is not quantum (because a classical model exists
for it) he indicated that Enk's reasoning implies\emph{ }that P's strategy is
\emph{also} not classical, because quantum models exist for flipping a
two-state system. Meyer referred to Lloyd's \cite{Lloyd} result that Grover's
algorithm \cite{Grover}, well-known in quantum computation and information
theory, can be implemented without entanglement by using a system allowing the
superposition of states \cite{ApporvaPatel}, despite claims \cite{Jozsa} that
the power of quantum computing derives from entanglement.

In 1964 Bell \cite{Bell} constructed a hidden variable model of spin
measurements on a single particle, reproducing the quantum mechanical
predictions for expectation value of the spin in \emph{any} direction. In
spite of Bell's construction the question about how far a two-dimensional
quantum system of a qubit can be claimed to have a `true' quantum character
has remained a matter of active debate. For example, Fivel \cite{Fivel 1}
reported an alleged ambiguity in Bell's claim that the distinction between
quantum mechanics and hidden variable theories cannot be found in the
behaviour of single particle beams; Khrennikov \cite{Khrennikov} indicated
that a realistic pre-quantum model does not exist even for the two-dimensional
Hilbert space.

\subsection{Criticism of Eisert et al.'s quantum Prisoners' Dilemma}

Eisert et al.'s quantum Prisoners' Dilemma (PD) has been criticized twice;
from Benjamin and Hayden, and from Enk and Pike.

\subsubsection{Benjamin and Hayden}

Eisert et al. obtained $\hat{Q}\sim\hat{U}(0,\pi/2)$ as the new quantum
equilibrium in Prisoners' Dilemma, when both players have access to a
two-parameter set (\ref{Eisert's unitary operators}) of unitary $2\otimes2$
operators. Benjamin and Hayden \cite{Benjamin1} observed that when their
two-parameter set is extended to include all local unitary operations (i.e.
all of $SU(2)$) the strategy $\hat{Q}$ does not remain an equilibrium. They
showed that in the full space of deterministic quantum strategies there exists
no equilibrium for the quantum Prisoners' Dilemma. Also, they observed that
the set (\ref{Eisert's unitary operators}) of two-parameter quantum strategies
is not closed under composition, although this closure appears to be a
necessary requirement for \emph{any} set of quantum strategies. It can be
explained as follows. Eisert et al. permitted both players the same strategy
set but introduced an arbitrary constraint into that set. This amounts to
permitting a certain strategy while forbidding the logical counter strategy
which one would intuitively expect to be \emph{equally} allowed. Benjamin and
Hayden argued that $\hat{Q}$ emerges as the ideal strategy only because of
restricting the strategy set arbitrarily.

Agreeing with the factual content of Benjamin and Hayden's comment
\cite{Benjamin1}, Eisert et al. nevertheless replied \cite{Eisert's Reply}
that in order to prove the existence of a quantum extension for which
$(\hat{Q},\hat{Q})$ is a Pareto-optimal Nash equilibrium, it is indeed
sufficient to explicate just one set of strategies for which this is the case;
they asserted that the strategy set introduced by them does just this. They
indicated that nowhere in their proposal had they claimed that $(\hat{Q}%
,\hat{Q})$ was a Pareto-optimal Nash equilibrium under all circumstances; if
analyzed in a different strategic space the game's solution acquires a
different character.

\subsubsection{Enk and Pike}

Enk and Pike \cite{EnkPike} commented that the quantum solutions of PD, found
by Eisert et al. \cite{Eisert}, are neither quantum mechanical nor do they
solve the classical game. They argued that it is possible to capture the
`essence' of quantized PD by simply extending the payoff matrix of the
classical game. That is done by including an additional purely classical move
corresponding to $\hat{Q}$, which in Eisert et al.'s scheme appears as a new
quantum-mechanical `solution-move' removing the dilemma inherent in the game.
Enk and Pike maintained that, since Eisert's quantum solution to PD can be
reconstructed in such a classical way, the only defence that remains for the
quantum solution is its efficiency, which unfortunately does not have a role
in PD because of its being a one-shot game.

In the same paper Enk and Pike \cite{EnkPike} also suggested that a quantum
game exploiting non-classical correlations in entangled states, similar to
those that violate Bell's inequality, should be investigated. They indicated
that the non-classical correlations have no role in Eisert et al.'s set-up,
and in other quantization procedures derived from it, although entangled
states may be present. This is because various qubits, after their local
unitary manipulations, are brought together, during the final stage of the
game to make the measurement which produces the payoffs.

\chapter{\label{Quantum Correlation Games}Quantum correlation games}

\section{Introduction}

Quantum games have attracted significant research interest during the last few
years. However, opinions about the true \emph{quantum} character and content
of quantum games still remains divided, as it is discussed in the section
(\ref{Criticism of quantum games}).

The statistical predictions of quantum mechanics are different from the
predictions of a local realistic theory and the Bell inequalities express the
constraints on dichotomic variables imposed by the principle of local
causality. The most explicit way of showing that a game can give different
solutions depending on whether it is played quantum mechanically or
classically is to design an experimental set-up, statistical in nature, in
which the constraints expressed by Bell inequalities, when satisfied, always
result in the classical game. EPR-type experiments \cite{Aspect et al} provide
such a realization. This gives rise to the immediate problem of how to
transform the EPR-type experiments into an arrangement for playing a
two-player game. In this arrangement the spatially separated measurements in
the EPR-type experiments are associated with two players when they retain
their freedom to choose their strategies.

In the present chapter, motivated by this suggestion of using the EPR-type
set-up to play a two-player game, we associate a quantum game to a classical
game in a way that addresses the above mentioned criticism on quantum games.
We suggest that the following two constraints should be imposed on this association.

\begin{enumerate}
\item [(c1)]The players choose their moves (or actions) from the \emph{same}
set in both the classical and the quantized game.

\item[(c2)] The players agree on explicit expressions for their payoffs which
must \emph{not} be modified when switching between the classical and the
quantized version of the game.
\end{enumerate}

Games with these properties are expected to be immune to the criticism raised
above. In the new setting the only `parameter' is the \emph{input state} on
which the players act; its nature will determine the classical or quantum
character of the game. Our approach to quantum games, tailored to satisfy both
(c1) and (c2), is inspired by Bell's work \cite{Bell}: \emph{correlations} of
measurement outcomes are essential. Effectively, we will define payoff
relations in terms of correlations -- these payoffs will become sensitive to
the classical or quantum nature of the input, thus allowing for the existence
of modified Nash equilibria.

\section{Matrix games and payoffs}

Consider a matrix game for two players, called Alice and Bob. A large set of
identical objects are prepared in definite states, which are not necessarily
known to the players. Each object splits into two equivalent `halves' which
are handed over to Alice and Bob simultaneously. Let the players agree
beforehand on the following rules:

\begin{enumerate}
\item  Alice and Bob may either play the identity \emph{move} $I$ or perform
(unitary) \emph{actions} $S_{A}$ and $S_{B}$, respectively. The moves
$S_{A,B}$ (and $I$) consist of unique actions such as flipping a coin (or not)
and possibly reading it.

\item  The players agree upon \emph{payoff relations} $P_{A,B}(p_{A},p_{B})$
which determine their rewards as functions of their \emph{strategies}, that
is, the moves with probabilities $p_{A,B}$ assigned to them.

\item  The players fix their \emph{strategies} for repeated runs of the game.
In a \emph{mixed} strategy Alice plays the identity move $I$ with probability
$p_{A}$, say, while she plays $S_{A}$ with probability $\bar{p}_{A}=1-p_{A}$,
and similarly for Bob. In a \emph{pure} strategy, each player performs the
same action in each run.

\item  Whenever the players receive their part of the system, they perform a
move consistent with their strategy.

\item  The players inform an arbiter about their actions taken in each
individual run. After a large number of runs, they are rewarded according to
the agreed payoff relations $P_{A,B}$. The existence of the arbiter is for
clarity only: alternatively, the players can get together to decide on their payoffs.
\end{enumerate}

These conventions are sufficient to play a classical game. As an example,
consider the class of symmetric bimatrix games with payoff relations
\begin{align}
P_{A}(p_{A},p_{B})  &  =Kp_{A}p_{B}+Lp_{A}+Mp_{B}+N,\nonumber\\
P_{B}(p_{A},p_{B})  &  =Kp_{A}p_{B}+Mp_{A}+Lp_{B}+N, \label{payoffs}%
\end{align}
where $K,L,M,$ and $N$ are real numbers. Being functions of two real
variables, with $0\leq p_{A,B}\leq1$, the payoff relations $P_{A,B}$ reflect
the fact that each player may chose a strategy from a continuous one-parameter
set. The game is symmetric since
\begin{equation}
P_{A}(p_{A},p_{B})=P_{B}(p_{B},p_{A}). \label{symmetry}%
\end{equation}
Now consider pure strategies with $p_{A,B}=0$ or $1$ in Eq. (\ref{payoffs}):
we have%
\begin{align}
P_{A}(1,1)=P_{B}(1,1)  &  =r=K+L+M+N,\nonumber\\
P_{A}(1,0)=P_{B}(0,1)  &  =s=L+N,\nonumber\\
P_{A}(0,1)=P_{B}(1,0)  &  =t=M+N,\nonumber\\
P_{A}(0,0)=P_{B}(0,0)  &  =u=N, \label{constants}%
\end{align}
leading to the payoff \emph{matrix} for this game%

\begin{equation}%
\begin{array}
[c]{c}%
\text{Alice}%
\end{array}%
\begin{array}
[c]{c}%
I\\
S_{A}%
\end{array}
\overset{\overset{%
\begin{array}
[c]{c}%
\text{Bob}%
\end{array}
}{%
\begin{array}
[c]{cc}%
I & S_{B}%
\end{array}
}}{\left(
\begin{array}
[c]{cc}%
(r,r) & (s,t)\\
(t,s) & (u,u)
\end{array}
\right)  .} \label{matrixCorrel}%
\end{equation}
In words: If both Alice and Bob play the identity $I$, they are paid $r$
units; Alice playing the identity $I$ and Bob playing $S_{B}$ pays $s$ and $t$
units to them, respectively; etc. Knowledge of the payoff matrix
(\ref{matrixCorrel}) and the probabilities $p_{A,B}$ is, in fact, equivalent
to (\ref{payoffs}), since the expected payoffs $P_{A,B}$ are obtained by
averaging (\ref{matrixCorrel}) over many runs.

Let Alice and Bob act rationally: they will try to maximize their payoffs by
an appropriate strategy. If the entries of the matrix (\ref{matrixCorrel})
satisfy $s<u<r<t$, the Prisoners' Dilemma arises: the players opt for
strategies in which unilateral deviations are disadvantageous; nevertheless,
the resulting solution of the game, a Nash equilibrium, does \emph{not}
maximize their payoffs.

In view of the conditions (c1) and (c2), the form of the payoff relations
$P_{A,B}$ in (\ref{payoffs}) seems to leave no room to introduce quantum games
which would differ from classical ones. In the following we will introduce
payoff relations which \emph{are} sensitive to whether a game is played on
classical \emph{or} quantum objects. With a classical input they will
reproduce the classical game and the conditions (c1) and (c2) will be
respected throughout.

\section{EPR-type setting of matrix games}

\emph{Correlation games} will be defined in a setting which is inspired by
EPR-type experiments. Alice and Bob are spatially separated, and they share
information about a Cartesian coordinate system with axes $\mathbf{e}%
_{x},\mathbf{e}_{y},\mathbf{e}_{z}$. The physical input used in a correlation
game is a large number of identical systems with zero angular momentum,
$\mathbf{J}=0$. Each system decomposes into a pair of objects which carry
perfectly anti-correlated angular momenta $\mathbf{J}_{A,B}$, i.e.
$\mathbf{J}_{A}+\mathbf{J}_{B}=0$.

In each run, Alice and Bob will measure the dichotomic variable $\mathbf{e}%
\cdot\mathbf{J}/|\mathbf{e}\cdot\mathbf{J}|$ of their halves along the common
$z$-axis ($\mathbf{e}\rightarrow\mathbf{e}_{z}$) or along specific directions
$\mathbf{e}\rightarrow\mathbf{e}_{A}$ and $\mathbf{e}\rightarrow\mathbf{e}%
_{B}$ in two planes $\mathcal{P}_{A}$ and $\mathcal{P}_{B}$, respectively,
each containing the $z$-axis, as shown in Fig. (\ref{Fig1QCGs}). The vectors
$\mathbf{e}_{A}$ and $\mathbf{e}_{B}$ are characterized by the angles
$\theta_{A}$ and $\theta_{B}$ which they enclose with the $z$-axis:
\begin{equation}
\mathbf{e}_{z}\cdot\mathbf{e}_{A,B}=\cos\theta_{A,B}\text{ },\qquad0\leq
\theta_{A,B}\leq\pi. \label{directions}%
\end{equation}
In principle, Alice and Bob could be given the choice of both the directions
$\mathbf{e}_{A,B}$ \emph{and} the probabilities $p_{A,B}$. However, in
traditional matrix games each player has access to \emph{one} continuous
variable only, namely $p_{A,B}$. To remain within this framework, we impose a
relation between the probabilities $p_{A,B}\in\lbrack0,1]$, and the angles
$\theta_{A,B}\in\lbrack0,\pi]$:
\begin{equation}
p_{A,B}=g(\theta_{A,B})\,. \label{gfunction}%
\end{equation}
The function $g$ maps the interval $[0,\pi]$ to $[0,1]$, and it is specified
\emph{before} the game begins. This function is, in general, \emph{not}
required to be invertible or continuous. Relation (\ref{gfunction}) says that
Alice must play the identity with probability $p_{A}\equiv g(\theta_{A})$ if
she decides to select the direction $\mathbf{e}_{A}$ as her alternative to
$\mathbf{e}_{z}$; furthermore, she measures with probability $\overline{p}%
_{A}=1-g(\theta_{A})$ along $\mathbf{e}_{A}$. For an invertible function $g$,
Alice can choose either a probability $p_{A}$ or a direction $\theta_{A}$ and
find the other variable from Eq. (\ref{gfunction}). If the function $g$ is not
invertible, some values of probability are associated with more than one
angle, and it is more natural to have the players choose a direction first.
For simplicity we will assume the function $g$ to be invertible, if not
specified otherwise.

According to her chosen strategy, Alice will measure the quantity
$\mathbf{e}\cdot\mathbf{J}/|\mathbf{e}\cdot\mathbf{J}|$ with probability
$p_{A}$ along the $z$-axis, and with probability $\overline{p}_{A}=1-p_{A}$
along the direction $\mathbf{e}_{A}$. Similarly, Bob can play a mixed
strategy, measuring along the directions $\mathbf{e}_{z}$ or $\mathbf{e}_{B}$
with probabilities $p_{B}$ and $\overline{p}_{B}$, respectively. Hence,
Alice's moves consist of either $S_{A}$ (rotating a Stern-Gerlach type
apparatus from $\mathbf{e}_{z}$ to $\mathbf{e}_{A}$, followed by a
measurement) or of $I$ (a measurement along $\mathbf{e}_{z}$ with no previous
rotation). Bob's moves $I$ and $S_{B}$ are defined similarly. It is convenient
to denote the outcomes of measurements along the directions $\mathbf{e}%
_{A},\mathbf{e}_{B}$, and $\mathbf{e}_{z}$ by $a,b,$ and $c$, respectively. It
is emphasized that the move $I$ is not the same as the identity operation
$\hat{I}$ because measurements are always performed along $\mathbf{e}_{z}$ or
$\mathbf{e}_{A,B}$.

After each run, the players inform the arbiter about the chosen directions and
the result of their measurements. After $N\rightarrow\infty$ runs of the game,
the arbiter possesses a list $\mathcal{L}$ indicating the directions of the
measurements selected by the players and the measured values of the quantity
$\mathbf{e}\cdot\mathbf{J}/|\mathbf{e}\cdot\mathbf{J}|$. The arbiter uses the
list to determine the strategies played by Alice and Bob by simply counting
the number of times ($N_{A}$, say) that Alice measured along $\mathbf{e}_{A}$,
giving $p_{A}=\underset{N\rightarrow\infty}{\lim}(N-N_{A})/N$, etc. Finally,
the players are rewarded according to the payoff relations (\ref{payoffs}).%

%TCIMACRO{\FRAME{dtbpFU}{2.674in}{1.9943in}{0pt}{\Qcb{Figure $5$-$1$: The
%players' strategies consist of defining angles $\theta_{A,B}$ which the
%directions $\mathbf{e}_{A,B}$ make with the $z$-axis; for simplicity the
%planes $\mathcal{P}_{A,B}$ are chosen as the $x$-$z$ and $y$-$z$ planes,
%respectively.}}{\Qlb{Fig1QCGs}}{fig1.eps}%
%maintain-aspect-ratio TRUE;  display "PICT";  valid_file "F";  width 2.674in;
%height 1.9943in;  depth 0pt;  original-width 8.0678in;
%original-height 10.3259in;  cropleft "0.0826";  croptop "0.8654";
%cropright "0.7294";  cropbottom "0.5319";
%filename 'Fig1.eps';file-properties "XNPEU";}}}%
%BeginExpansion
\begin{center}
\includegraphics[
trim=0.666400in 5.492346in 2.183147in 1.389866in,
height=1.9943in,
width=2.674in
]%
{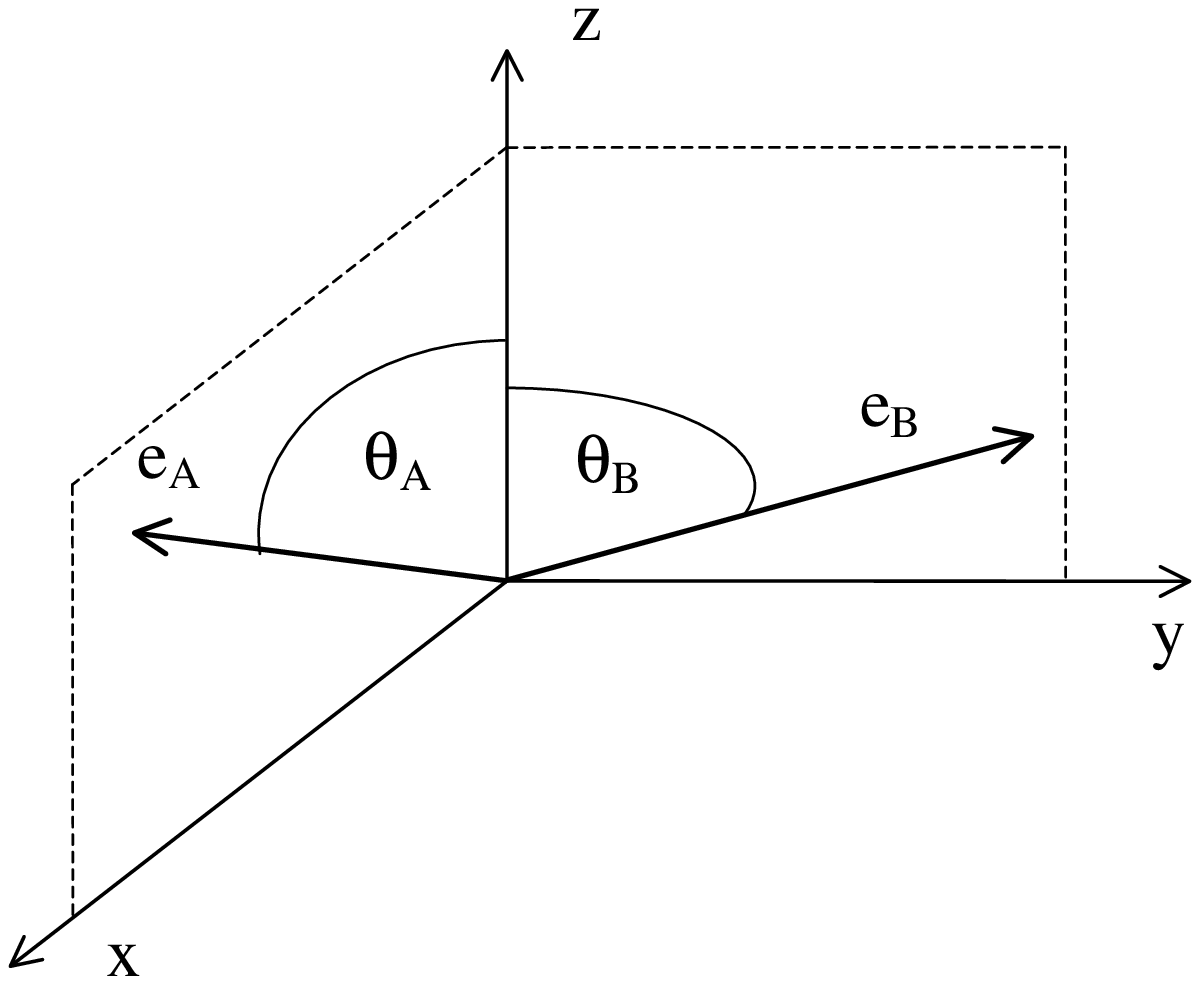}%
\\
Figure $5$-$1$: The players' strategies consist of defining angles
$\theta_{A,B}$ which the directions $\mathbf{e}_{A,B}$ make with the $z$-axis;
for simplicity the planes $\mathcal{P}_{A,B}$ are chosen as the $x$-$z$ and
$y$-$z$ planes, respectively.
\label{Fig1QCGs}%
\end{center}
%EndExpansion

\section{Correlation games}

We now develop a new perspective on matrix games in the EPR-type setting. The
basic idea is to define payoffs $P_{A,B}=P_{A,B}(\left\langle ac\right\rangle
,\left\langle cb\right\rangle )$ which depend explicitly on the
\emph{correlations} of the actual measurements performed by Alice and Bob. The
arbiter will extract the numerical values of the correlations $\left\langle
ac\right\rangle $ etc. from the list $\mathcal{L}$ in the usual way. Consider,
for example, all cases with Alice measuring along $\mathbf{e}_{A}$ and Bob
along $\mathbf{e}_{z}$. If there are $N_{ac}$ such runs, the correlation of
the measurements is defined by
\begin{equation}
\left\langle ac\right\rangle =\lim_{N_{ac}\rightarrow\infty}\left(  \sum
_{n=1}^{N_{ac}}\frac{a_{n}c_{n}}{N_{ac}}\right)  \,, \label{corraverage}%
\end{equation}
where $a_{n}$ and $c_{n}$ take the values $\pm1$. The correlations
$\left\langle ab\right\rangle $ and $\left\langle cb\right\rangle $ are
defined similarly.

A symmetric bimatrix correlation game is determined by a function $g$ in
(\ref{gfunction}) and by the relations
\begin{align}
P_{A}(\left\langle ac\right\rangle ,\left\langle cb\right\rangle )  &
=K\text{ }G(\left\langle ac\right\rangle )G(\left\langle cb\right\rangle
)+L\text{ }G(\left\langle ac\right\rangle )+M\text{ }G(\left\langle
cb\right\rangle )+N,\nonumber\\
P_{B}(\left\langle ac\right\rangle ,\left\langle cb\right\rangle )  &
=K\text{ }G(\left\langle ac\right\rangle )G(\left\langle cb\right\rangle
)+M\text{ }G(\left\langle ac\right\rangle )+L\text{ }G(\left\langle
cb\right\rangle )+N, \label{re-expressed2}%
\end{align}
where, in view of later developments, the function $G$ is taken to be
\begin{equation}
G(x)=g\left(  \frac{\pi}{2}(1+x)\right)  \,,\quad x\in\lbrack-1,1].
\label{Gfunction}%
\end{equation}
As they stand, the payoff relations (\ref{re-expressed2}) refer to neither a
classical nor a quantum mechanical input. Hence, condition (c2) from the
section ($5.1$) is satisfied: the payoff relations used in the classical and
the quantum version of the game are \emph{identical}, as given by Eqs.
(\ref{re-expressed2}). Furthermore, Alice and Bob choose from the same set of
moves in both versions of the game: they select directions $\mathbf{e}_{A}$
and $\mathbf{e}_{B}$ (with probabilities $p_{A,B}$ associated with
$\theta_{A,B}$ via (\ref{gfunction})) so that condition (c1) from the section
($5.1$) is satisfied. Nevertheless, the solutions of the correlation game
(\ref{re-expressed2}) will depend on the input being either a classical or a
quantum mechanical anti-correlated state.

\subsection{Classical correlation games}

Alice and Bob play a \emph{classical correlation game} if they receive
classically anti-correlated pairs and use the payoff relations
(\ref{re-expressed2}). In this case, the payoffs turn into
\begin{equation}
P_{A,B}^{cl}=P_{A,B}(\left\langle ac\right\rangle _{cl},\left\langle
cb\right\rangle _{cl}), \label{defineclcorrgame}%
\end{equation}
where the correlations, characteristic for classically anti-correlated
systems, are given by
\begin{align}
\left\langle ac\right\rangle _{cl}  &  =-1+2\theta_{A}/\pi,\nonumber\\
\left\langle cb\right\rangle _{cl}  &  =-1+2\theta_{B}/\pi. \label{correls}%
\end{align}
We use
%
%The angle $ \theta _{AB}=\arccos (\cos  \theta _{A}\cos  %\theta _{B})$ is defined by the directions %$\mathbf{e}_\alpha$ and $\mathbf{e}_\beta$.
%
%
%
%
%
%
%
%
%
%
%
%
%
%
%
%
%
%
%
%
%
%
%
%
%
%
%
%
%
%
%
%
%
%
%
%
%
%
%
%
%
%
%
%
%
%
%
%
%
%
%
%
%
%
%
%
%
%
%
%
%
%
%
%
%
%
%
%
%
%
%
%
%
%
%
%
%
%
%
%
%
%
%
%
%
%
%
%
%
%
%
%
%
%
%
%
%
%
%
%
%
%
%
%
%
%
%
%
%
%
%
%
%
%
%
%
%
%
%
%
%
%
%
%
%
%
%
%
%
%
%
%
%
%
%
%
%
%
%
%
%
%
%
%
%
%
%
%
%
%
%
%
%
%
%
%
%
%
%
%
%
%
%
%
%
%
%
%
%
%
%
%
%
%
%
%
%
%
%
%
%
%
%
%
%
%
%
%
%
%
%
%
%
%
%
%
%
%
%
%
%
%
%
%
%
%
%
%
%
%
%
%
%
%
%
%
%
%
%
%
%
%
%
%
%
%
%
%
%
%
%
%
%
%
%
%
%
%
%
%
%
%
%
%
%
%
%
%
%
%
%
%
%
%
%
the definition of the function $G$ in (\ref{Gfunction}) and the link
(\ref{gfunction}) between probabilities $p_{A,B}$ and angles $\theta_{A,B}$ to
obtain
\begin{align}
G(\langle ac\rangle)  &  =g(\theta_{A})=p_{A},\\
G(\langle cb\rangle)  &  =g(\theta_{B})=p_{B}. \label{G-p}%
\end{align}
Hence for classical input Eqs. (\ref{re-expressed2}) reproduce the payoffs of
a symmetric bimatrix game (\ref{payoffs}),
\begin{align}
P_{A}^{cl}(p_{A},p_{B})  &  =Kp_{A}p_{B}+Lp_{A}+Mp_{B}+N,\nonumber\\
P_{B}^{cl}(p_{A},p_{B})  &  =Kp_{A}p_{B}+Mp_{A}+Lp_{B}+N. \label{clpayoffs}%
\end{align}
The game-theoretic analysis of the classical correlation game is now
straightforward -- for example, appropriate values of the parameters
$(r,s,t,u)$ lead to the Prisoners' Dilemma, for \emph{any} invertible function
$g$.

\subsection{Quantum correlation games}

Imagine now that Alice and Bob receive quantum mechanical anti-correlated
singlet states
\begin{equation}
|\psi\rangle=\frac{1}{\sqrt{2}}\left(  |+,-\rangle-|-,+\rangle\right)  .
\label{singlet}%
\end{equation}
They are said to play a \emph{quantum correlation game} if again they use the
payoff relations (\ref{re-expressed2}), which in this case give
\begin{equation}
P_{A,B}^{q}=P_{A,B}(\left\langle ac\right\rangle _{q},\left\langle
cb\right\rangle _{q}). \label{defineqcorrgame}%
\end{equation}
As before, Alice and Bob transmit the results of their measurements (on their
quantum halves) to the arbiter who, after a large number of runs, determines
the correlations $\left\langle ac\right\rangle _{q}$ and $\left\langle
cb\right\rangle _{q}$ by the formula (\ref{corraverage})
\begin{align}
\left\langle ac\right\rangle _{q}  &  =-\cos\theta_{A},\nonumber\\
\left\langle cb\right\rangle _{q}  &  =-\cos\theta_{B}, \label{qcorrels}%
\end{align}
in contrast to (\ref{correls}).

The inverse of relation (\ref{gfunction}) links the probabilities and
correlations as follows:
\begin{align}
\left\langle ac\right\rangle _{q}  &  =-\cos\left(  g^{-1}(p_{A})\right)
,\nonumber\\
\left\langle cb\right\rangle _{q}  &  =-\cos\left(  g^{-1}(p_{B})\right)  .
\label{re-expressed correl}%
\end{align}
Inserting these expressions into the right-hand-side of (\ref{defineqcorrgame}%
), we obtain the \emph{quantum} payoffs:
\begin{align}
P_{A}^{q}(p_{A},p_{B})  &  =KQ_{g}(p_{A})Q_{g}(p_{B})+LQ_{g}(p_{A}%
)+MQ_{g}(p_{B})+N,\nonumber\\
P_{B}^{q}(p_{A},p_{B})  &  =KQ_{g}(p_{A})Q_{g}(p_{B})+MQ_{g}(p_{B}%
)+LQ_{g}(p_{A})+N. \label{Qpayoffs}%
\end{align}
where
\begin{equation}
Q_{g}(p_{A,B})=g\left(  \frac{\pi}{2}\left(  1-\cos\left(  g^{-1}%
(p_{A,B})\right)  \right)  \right)  \in\lbrack0,1]. \label{expquantum}%
\end{equation}
The payoffs $P_{A,B}^{q}$ turn out to be \emph{non-linear} functions of the
probabilities $p_{A,B}$, while the payoffs $P_{A,B}^{cl}$ of the classical
correlation game are \emph{bi-linear}. This modification has an impact on the
solutions of the game, as shown in the following section.

\section{Nash equilibria of quantum correlation games}

What are the properties of the quantum payoffs $P_{A,B}^{q}$ as compared to
the classical ones $P_{A,B}^{cl}$? The standard approach to `solving games'
consists of studying Nash equilibria. For a bimatrix game a pair of strategies
$(p_{A}^{\star},p_{B}^{\star})$ is a Nash equilibrium if each player's payoff
does not increase upon unilateral deviation from her strategy,
\begin{align}
P_{A}(p_{A},p_{B}^{\star})  &  \leq P_{A}(p_{A}^{\star},p_{B}^{\star}%
)\,,\quad\text{for all }p_{A},\nonumber\\
P_{B}(p_{A}^{\star},p_{B})  &  \leq P_{B}(p_{A}^{\star},p_{B}^{\star}%
)\,,\quad\text{for all }p_{B}. \label{NECorrel}%
\end{align}
In the following, we will study the differences between the classical and
quantum correlation games which are associated with two paradigmatic games:
the Prisoners' Dilemma (PD) and the Battle of Sexes (BoS).

The payoff matrix of the PD has been introduced in (\ref{matrixCorrel}). It
will be convenient to use the notation of game theory: $C\thicksim I$
corresponds to Cooperation, while $D\thicksim S_{A,B}$ is the strategy of
Defection. A characteristic feature of this game is that the condition
$s<u<r<t$ guarantees that the strategy $D$ dominates the strategy $C$ for both
players and that the unique equilibrium at $(D,D)$ is not Pareto optimal. An
outcome of a game is Pareto optimal if there is no other outcome that makes
one or more players better off and no player worse off. This can be seen in
the following way. The conditions (\ref{NECorrel}) give
\begin{align}
0  &  \leq\left(  Kp_{B}^{\star}+L\right)  (p_{A}^{\star}-p_{A})\text{ }%
,\quad\text{for all }p_{A},\nonumber\\
0  &  \leq\left(  Kp_{A}^{\star}+L\right)  (p_{B}^{\star}-p_{B})\text{ }%
,\quad\text{for all }p_{B}. \label{NE1Correl}%
\end{align}
with $K$ and $L$ from (\ref{constants}). The inequalities have only one
solution
\begin{equation}
p_{A}^{\star}=p_{B}^{\star}=0\text{ }, \label{clpureNE}%
\end{equation}
which corresponds to $(D,D)$, a pure strategy for both players. The PD is said
to have a pure Nash equilibrium.

The BoS is defined by the following payoff matrix:%

\begin{equation}%
\begin{array}
[c]{c}%
\text{Alice}%
\end{array}%
\begin{array}
[c]{c}%
I\\
S_{A}%
\end{array}
\overset{\overset{%
\begin{array}
[c]{c}%
\text{Bob}%
\end{array}
}{%
\begin{array}
[c]{cc}%
I & S_{B}%
\end{array}
}}{\left(
\begin{array}
[c]{cc}%
(\alpha,\beta) & (\gamma,\gamma)\\
(\gamma,\gamma) & (\beta,\alpha)
\end{array}
\right)  ,}%
\end{equation}
where $I$ and $S_{A,B}$ are pure strategies and $\alpha>\beta>\gamma$. Three
Nash equilibria arise in the classical BoS, two of which are pure: $(I,I)$ and
$(S_{A},S_{B})$. The third one is a mixed equilibrium where Alice and Bob play
$I$ with probabilities%

\begin{equation}
p_{A}^{\star}=\frac{\alpha-\gamma}{\alpha+\beta-2\gamma}\text{ },\qquad
p_{B}^{\star}=\frac{\beta-\gamma}{\alpha+\beta-2\gamma}\text{ }.
\label{BoSMixedNash}%
\end{equation}

For the \emph{quantum correlation game} associated with the generalized PD the
conditions (\ref{NECorrel}) turn into
\begin{align}
0  &  \leq\left(  KQ_{g}(p_{B}^{\star})+L\right)  \left(  Q_{g}(p_{A}^{\star
})-Q_{g}(p_{A})\right)  \,,\\
0  &  \leq\left(  KQ_{g}(p_{A}^{\star})+L\right)  \left(  Q_{g}(p_{B}^{\star
})-Q_{g}(p_{B})\right)  \,, \label{QNEqualities}%
\end{align}
where the range of $Q_{g}(p_{A,B})$ has been defined in (\ref{expquantum}).
Thus, the conditions for a Nash equilibrium of a quantum correlation game are
structurally similar to those of the classical game except for non-linear
dependence on the probabilities $p_{A,B}$. The only solutions of
(\ref{QNEqualities}) therefore read
\begin{equation}
Q_{g}({p_{A}^{\star}})=Q_{g}({p_{B}^{\star}})=0\,, \label{qpureNE}%
\end{equation}
generating upon inversion a Nash equilibrium at
\begin{equation}
(p_{A}^{\star})_{q}=(p_{B}^{\star})_{q}=Q_{g}^{-1}(0)=g\left(  \arccos\left(
1-\frac{2}{\pi}g^{-1}(0)\right)  \right)  , \label{QNEofPD}%
\end{equation}
where the transformed probabilities now come with a subscript $q$ indicating
the presence of quantum correlations. The location of this new equilibrium
depends on the actual choice of the function $g$, as is shown below.

Similar arguments apply to the pure Nash equilibria of the BoS game while the
mixed classical equilibrium (\ref{BoSMixedNash}) is transformed into%

\begin{align}
(p_{A}^{\star})_{q}  &  =Q_{g}^{-1}(p_{A}^{\star})=g\left(  \arccos\left(
1-\frac{2}{\pi}g^{-1}(\frac{\alpha-\gamma}{\alpha+\beta-2\gamma})\right)
\right)  ,\nonumber\\
(p_{B}^{\star})_{q}  &  =Q_{g}^{-1}(p_{B}^{\star})=g\left(  \arccos\left(
1-\frac{2}{\pi}g^{-1}(\frac{\beta-\gamma}{\alpha+\beta-2\gamma})\right)
\right)  .
\end{align}

When defining a quantum correlation game we need to specify a function $g$
which establishes the link between the probabilities $p_{A,B}$ and the angles
$\theta_{A,B}$. We will study the properties of quantum correlation games for
$g$-functions of increasing complexity. In the simplest case, the function $g$
is ($i$) continuous and invertible; next, we chose a function $g$ being ($ii$)
invertible and discontinuous or ($iii$) non-invertible and discontinuous. For
simplicity, all examples are worked out for piecewise linear $g$-functions.
The generalization to smooth $g$-functions turns out to be straightforward and
the results do not change qualitatively as long as the $g$-function preserves
its characteristic features.

\subsubsection{($i$) Continuous and invertible $g$-functions}

Consider the function $g_{1}(\theta)=\theta/\pi$ defined for $\theta\in
\lbrack0,\pi]$. We have $g_{1}(0,\pi)=0$ or $1$, and the classical and quantum
correlations coincide at $\theta=0,\pi/2,$ and $\pi$. In view of
(\ref{QNEofPD}) the function $g_{1}$ can have no effect on pure Nash
equilibria and the classical solution of PD is not modified in the quantum
game. Fig. (\ref{Fig2QCGs}) shows the function $g_{1}$.%

%TCIMACRO{\FRAME{dtbpFU}{2.7198in}{1.8654in}{0pt}{\Qcb{Figure $5$-$2$: The
%invertible and continuous $g$-function $g(\theta)=\theta/\pi$.}}%
%{\Qlb{Fig2QCGs}}{fig2.eps}{\special{ language "Scientific Word";
%type "GRAPHIC";  maintain-aspect-ratio TRUE;  display "PICT";
%valid_file "F";  width 2.7198in;  height 1.8654in;  depth 0pt;
%original-width 8.0652in;  original-height 10.3233in;  cropleft "0.2080";
%croptop "0.9067";  cropright "0.8257";  cropbottom "0.5772";
%filename 'Fig2.eps';file-properties "XNPEU";}}}%
%BeginExpansion
\begin{center}
\includegraphics[
trim=1.677562in 5.958609in 1.405764in 0.963164in,
height=1.8654in,
width=2.7198in
]%
{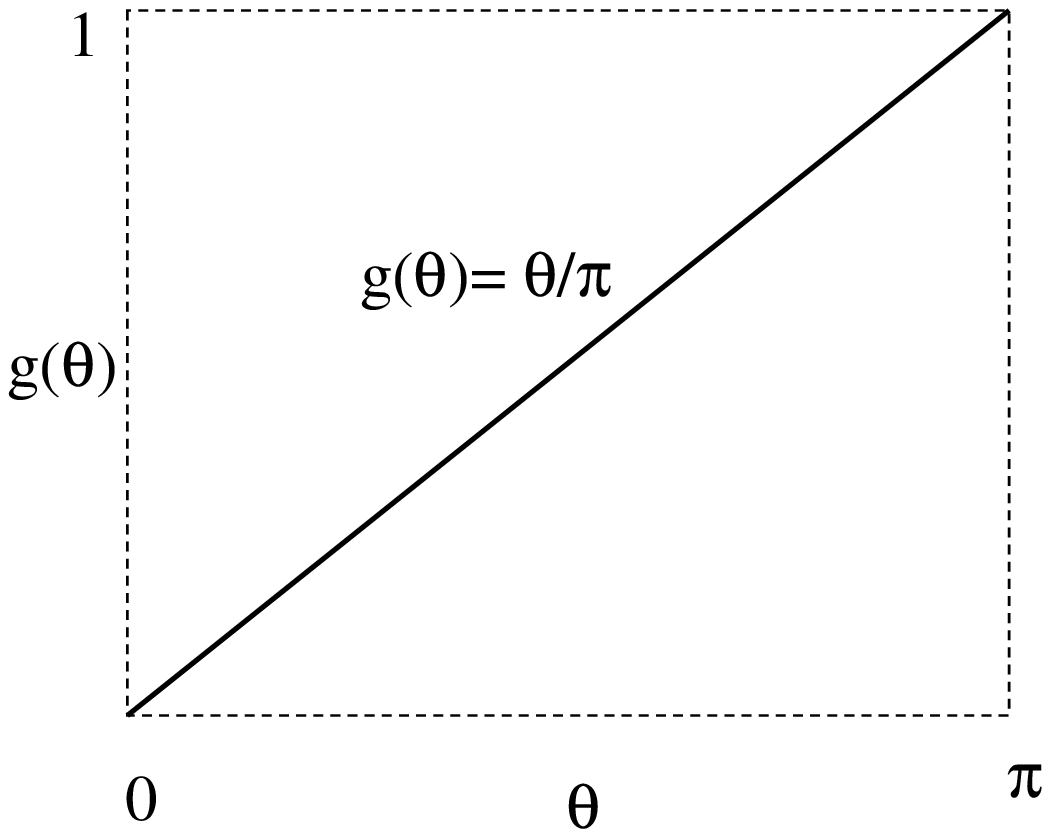}%
\\
Figure $5$-$2$: The invertible and continuous $g$-function $g(\theta
)=\theta/\pi$.
\label{Fig2QCGs}%
\end{center}
%EndExpansion

However, solutions $p_{A,B}^{\star}\in(0,1)$ correspond to a mixed classical
equilibrium, which will be modified if $g(\pi/2)\neq p_{A,B}^{\star}$ i.e.
when the angle associated with $p_{A,B}^{\star}$ is different from $\pi/2$.
For example with the function $g_{1}(\theta)$ the probabilities of the mixed
equilibrium of the quantum correlation BoS are $(p_{A}^{\star})_{q}%
=1-(1/\pi)\arccos\left\{  (\alpha-\gamma)/(\alpha+\beta-2\gamma)\right\}  $
and $(p_{B}^{\star})_{q}=1-(1/\pi)\arccos\left\{  (\beta-\gamma)/(\alpha
+\beta-2\gamma)\right\}  $. A similar result holds for the function
$g_{2}(\theta)=1-\theta/\pi$.

\subsubsection{($ii$) Invertible and discontinuous $g$-functions}

For simplicity we consider invertible functions that are discontinuous at one
point only. Piecewise linear functions are typical examples. One such
function, shown in Fig. (\ref{Fig3QCGs}), is
\begin{equation}
g_{3}(\theta)=\left\{
\begin{array}
[c]{rcl}%
\delta(1-\theta/\epsilon) & \text{if} & \theta\in\lbrack0,\epsilon]\text{ },\\
\delta+(1-\delta)(\theta-\epsilon)/(\pi-\epsilon) & \text{if} & \theta
\in(\epsilon,\pi)\text{ }.
\end{array}
\right.  \label{InvDis1}%
\end{equation}
The classical solution of PD $p_{A}^{\star}=p_{B}^{\star}=0$ disappears; the
new quantum solution is found at%

\begin{equation}
(p_{A}^{\star})_{q}=(p_{B}^{\star})_{q}=\left\{
\begin{array}
[c]{rcl}%
\delta+\frac{(1-\delta)}{(\pi-\epsilon)}\left\{  \arccos(1-2\epsilon
/\pi)-\epsilon\right\}  & \text{if} & \epsilon\in\lbrack0,\frac{\pi}{2}]\text{
},\\
\delta\left\{  1-\frac{1}{\epsilon}\arccos(1-2\epsilon/\pi)\right\}  &
\text{if} & \epsilon\in(\frac{\pi}{2},\pi]\text{ }.
\end{array}
\right.  \label{InvDis2}%
\end{equation}
If, for example, $\delta=1/2$ and $\epsilon=\pi/4,$ we obtain a mixed
equilibrium at $(p_{A}^{\star})_{q}=(p_{B}^{\star})_{q}=5/9.$ The appearance
of a mixed equilibrium in a quantum correlation PD game is an entirely
non-classical feature.%

%TCIMACRO{\FRAME{dtbpFU}{2.444in}{1.7331in}{0pt}{\Qcb{Figure $5$-$3$:
%Invertible and discontinuous function $g_{3}(\theta)$.}}{\Qlb{Fig3QCGs}%
%}{fig3.eps}{\special{ language "Scientific Word";  type "GRAPHIC";
%maintain-aspect-ratio TRUE;  display "PICT";  valid_file "F";  width 2.444in;
%height 1.7331in;  depth 0pt;  original-width 8.0652in;
%original-height 10.3233in;  cropleft "0.2354";  croptop "0.8805";
%cropright "0.9317";  cropbottom "0.4962";
%filename 'Fig3.eps';file-properties "XNPEU";}}}%
%BeginExpansion
\begin{center}
\includegraphics[
trim=1.898548in 5.122422in 0.550853in 1.233634in,
height=1.7331in,
width=2.444in
]%
{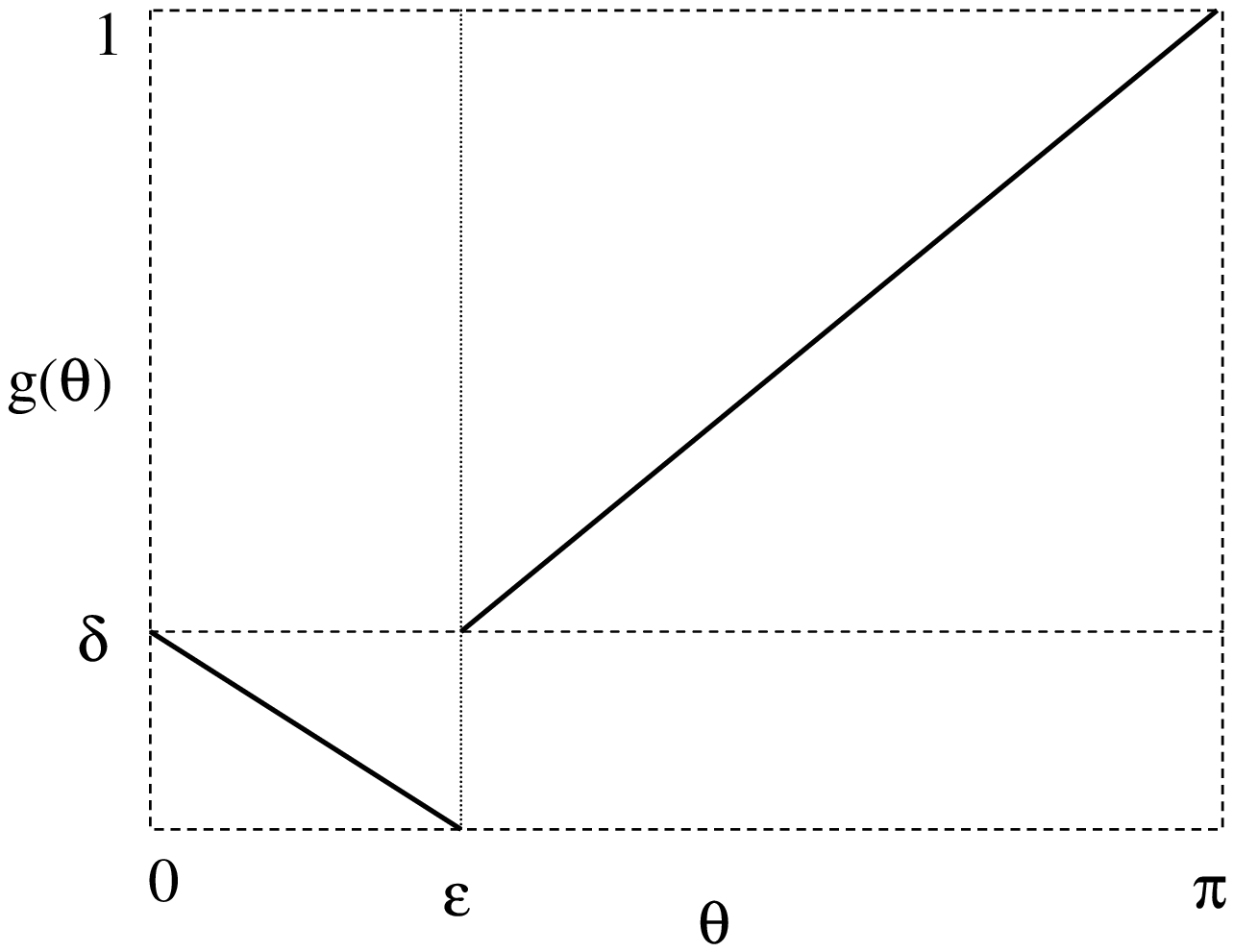}%
\\
Figure $5$-$3$: Invertible and discontinuous function $g_{3}(\theta)$.
\label{Fig3QCGs}%
\end{center}
%EndExpansion

The presence of a mixed equilibrium in the quantum correlation PD gives rise
to an interesting question: is there a Pareto-optimal solution of $(C,C)$ in a
quantum correlation PD with some invertible and discontinuous $g$-function? No
such solution exists for invertible and continuous $g$-functions. Also, the
$(C,C)$ equilibrium in PD cannot appear in a quantum correlation game played
with the function (\ref{InvDis1}): one has $g^{-1}(1)=\pi$ which cannot be
equal to $g^{-1}(0)$ when $g$ is invertible. As a matter of fact, the solution
$(C,C)$ for PD can be realized in a quantum correlation PD if one considers
$g$ from (\ref{InvDis2}) with $\epsilon=\pi/2$:%

\begin{equation}
g_{4}(\theta)=\left\{
\begin{array}
[c]{rcl}%
\delta(1-2\theta/\pi) & \text{if} & \theta\in\lbrack0,\frac{\pi}{2}]\text{
},\\
1-2(1-\delta)(\theta-\pi/2)/\pi & \text{if} & \theta\in(\frac{\pi}{2}%
,\pi]\text{ },
\end{array}
\right.  \label{InvDis3}%
\end{equation}
where $\delta\in(0,1)$, depicted in Fig. (\ref{Fig4QCGs}).%

%TCIMACRO{\FRAME{dtbpFU}{2.6524in}{1.7729in}{0pt}{\Qcb{Figure $5$-$4$:
%Invertible and discontinuous function $g_{4}(\theta)$.}}{\Qlb{Fig4QCGs}%
%}{fig4.eps}{\special{ language "Scientific Word";  type "GRAPHIC";
%maintain-aspect-ratio TRUE;  display "PICT";  valid_file "F";
%width 2.6524in;  height 1.7729in;  depth 0pt;  original-width 8.0652in;
%original-height 10.3233in;  cropleft "0.1967";  croptop "0.9057";
%cropright "0.9200";  cropbottom "0.5302";
%filename 'Fig4.eps';file-properties "XNPEU";}}}%
%BeginExpansion
\begin{center}
\includegraphics[
trim=1.586425in 5.473414in 0.645216in 0.973487in,
height=1.7729in,
width=2.6524in
]%
{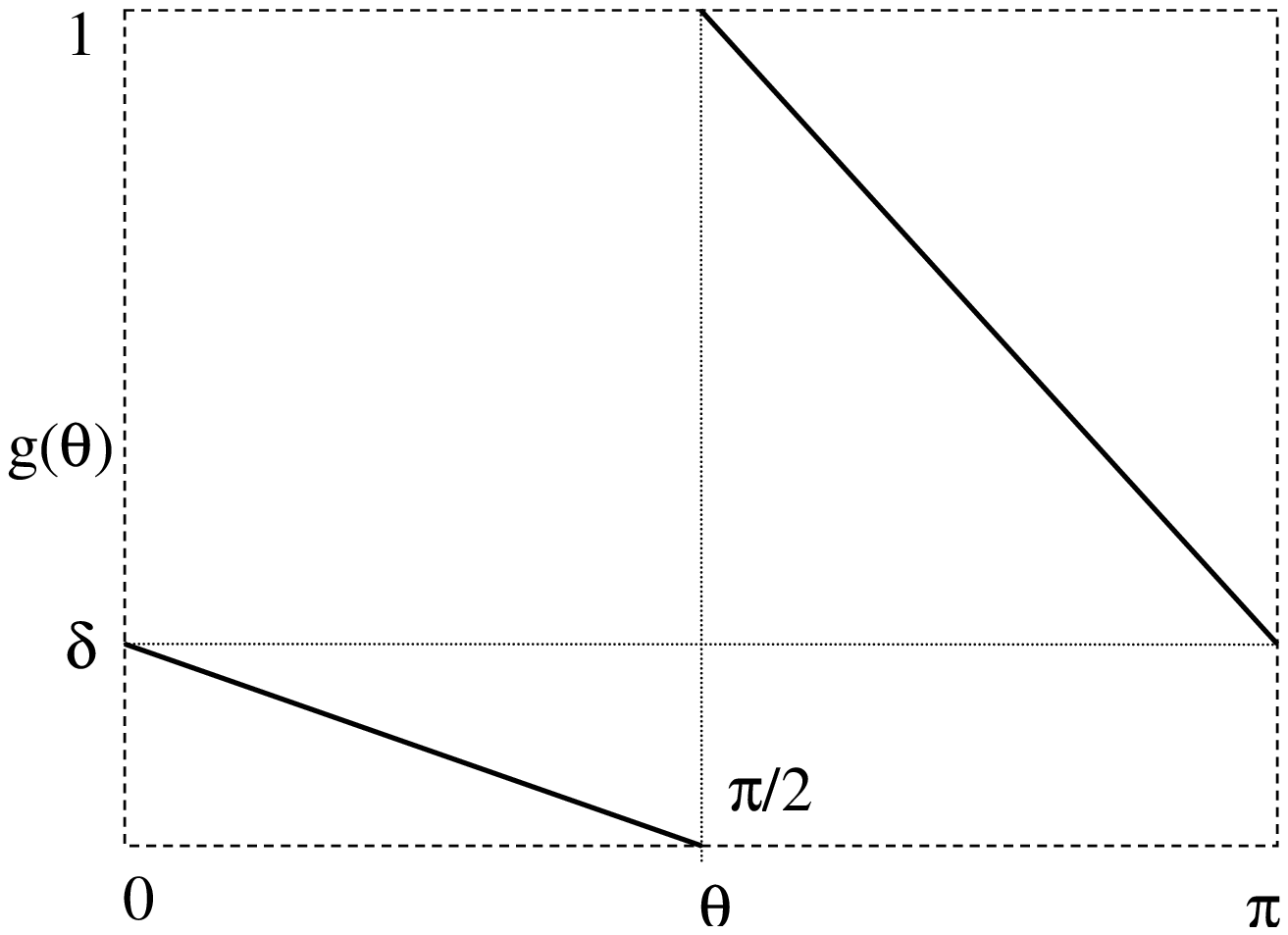}%
\\
Figure $5$-$4$: Invertible and discontinuous function $g_{4}(\theta)$.
\label{Fig4QCGs}%
\end{center}
%EndExpansion

This function satisfies $g^{-1}(0)=g^{-1}(1)=\pi/2$. Therefore, one has
$\cos\left\{  g^{-1}(1)\right\}  =1-2g^{-1}(0)/\pi$, which is the condition
for $(C,C)$ to be an equilibrium in PD. Cooperation $(C,C)$ will also be an
equilibrium in PD if the $g$-function is defined as%

\begin{equation}
g_{5}(\theta)=\left\{
\begin{array}
[c]{rcl}%
2(1-\delta)\theta/\pi+\delta & \text{if} & \theta\in\lbrack0,\frac{\pi}%
{2}]\text{ },\\
2\delta(\theta-\pi/2)/\pi & \text{if} & \theta\in(\frac{\pi}{2},\pi]\text{ },
\end{array}
\right.  \label{InvDis4}%
\end{equation}
where $\delta\in(0,1)$. Fig. (\ref{Fig5QCGs}) shows this function.%

%TCIMACRO{\FRAME{dtbpFU}{2.5218in}{1.849in}{0pt}{\Qcb{Figure $5$-$5$:
%Invertible and discontinuous function $g_{5}(\theta)$.}}{\Qlb{Fig5QCGs}%
%}{fig5.eps}{\special{ language "Scientific Word";  type "GRAPHIC";
%maintain-aspect-ratio TRUE;  display "PICT";  valid_file "F";
%width 2.5218in;  height 1.849in;  depth 0pt;  original-width 8.0652in;
%original-height 10.3233in;  cropleft "0.2436";  croptop "0.8816";
%cropright "0.9462";  cropbottom "0.4807";
%filename 'Fig5.eps';file-properties "XNPEU";}}}%
%BeginExpansion
\begin{center}
\includegraphics[
trim=1.964683in 4.962410in 0.433908in 1.222278in,
height=1.849in,
width=2.5218in
]%
{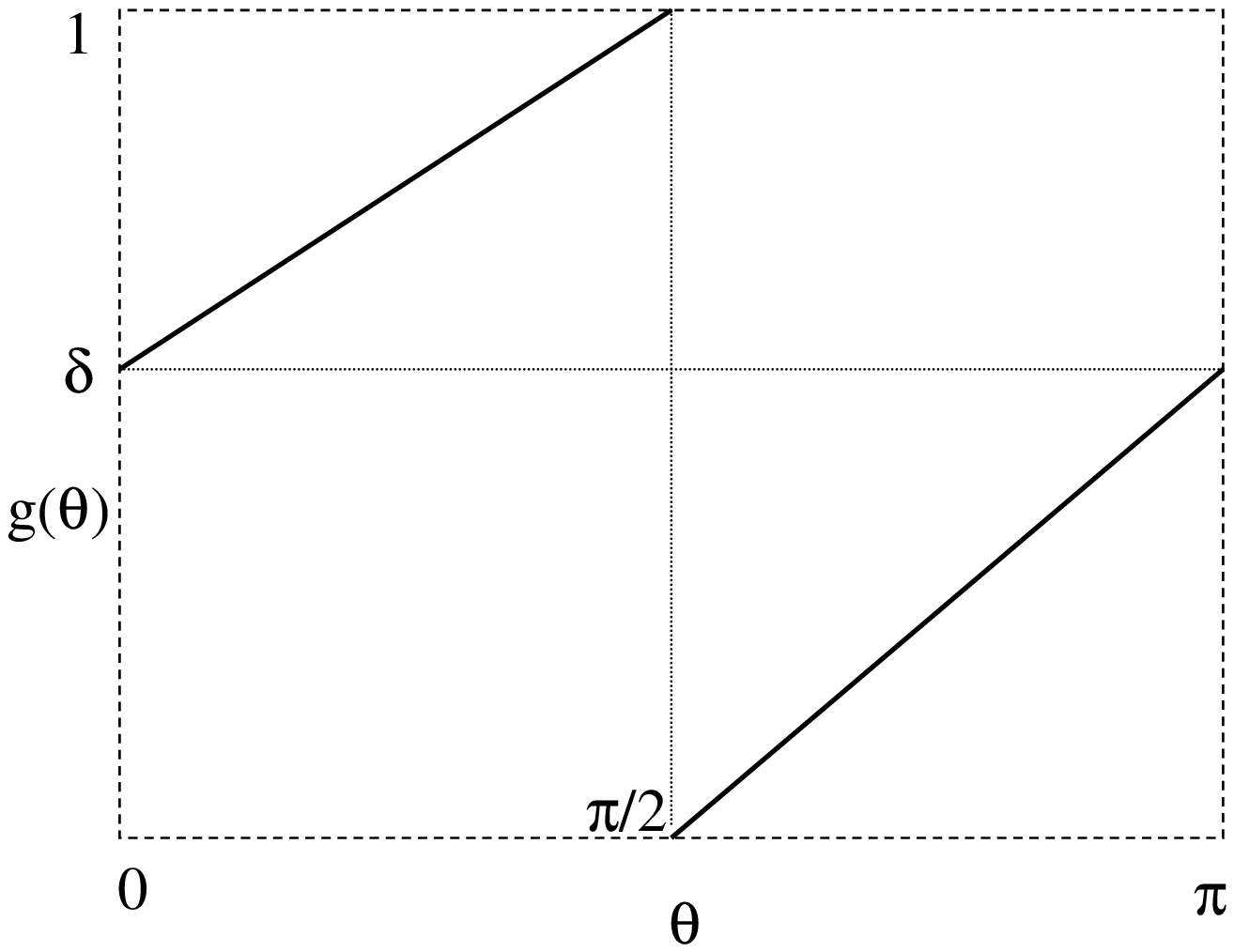}%
\\
Figure $5$-$5$: Invertible and discontinuous function $g_{5}(\theta)$.
\label{Fig5QCGs}%
\end{center}
%EndExpansion

In both cases (\ref{InvDis3}, \ref{InvDis4}), the $g$-function has a
discontinuity at $\theta=\pi/2$. With these functions both the pure and mixed
classical equilibria of BoS will also be susceptible to change. The shifts in
the pure equilibria in BoS will be similar to those of PD but the mixed
equilibrium of BoS will move, depending on the location of $\delta$.

Another example of an invertible and discontinuous function is given by%

\begin{equation}
g_{6}(\theta)=\left\{
\begin{array}
[c]{rcl}%
(1-\delta)\theta/\epsilon+\delta & \text{if} & \theta\in\lbrack0,\epsilon
]\text{ },\\
\delta(\pi-\theta)/(\pi-\epsilon) & \text{if} & \theta\in(\epsilon,\pi]\text{
},
\end{array}
\right.  \label{InvDis5}%
\end{equation}
where $\delta\in(0,1)$ and $\epsilon\in(0,\pi)$ and it is drawn in Fig.
(\ref{Fig6QCGs}).%

%TCIMACRO{\FRAME{dtbpFU}{2.7579in}{1.8896in}{0pt}{\Qcb{Figure $5$-$6$:
%Invertible and discontinuous function $g_{6}(\theta)$.}}{\Qlb{Fig6QCGs}%
%}{fig6.eps}{\special{ language "Scientific Word";  type "GRAPHIC";
%maintain-aspect-ratio TRUE;  display "PICT";  valid_file "F";
%width 2.7579in;  height 1.8896in;  depth 0pt;  original-width 8.0652in;
%original-height 10.3233in;  cropleft "0.1642";  croptop "0.8753";
%cropright "0.9002";  cropbottom "0.4832";
%filename 'Fig6.eps';file-properties "XNPEU";}}}%
%BeginExpansion
\begin{center}
\includegraphics[
trim=1.324306in 4.988219in 0.804907in 1.287316in,
height=1.8896in,
width=2.7579in
]%
{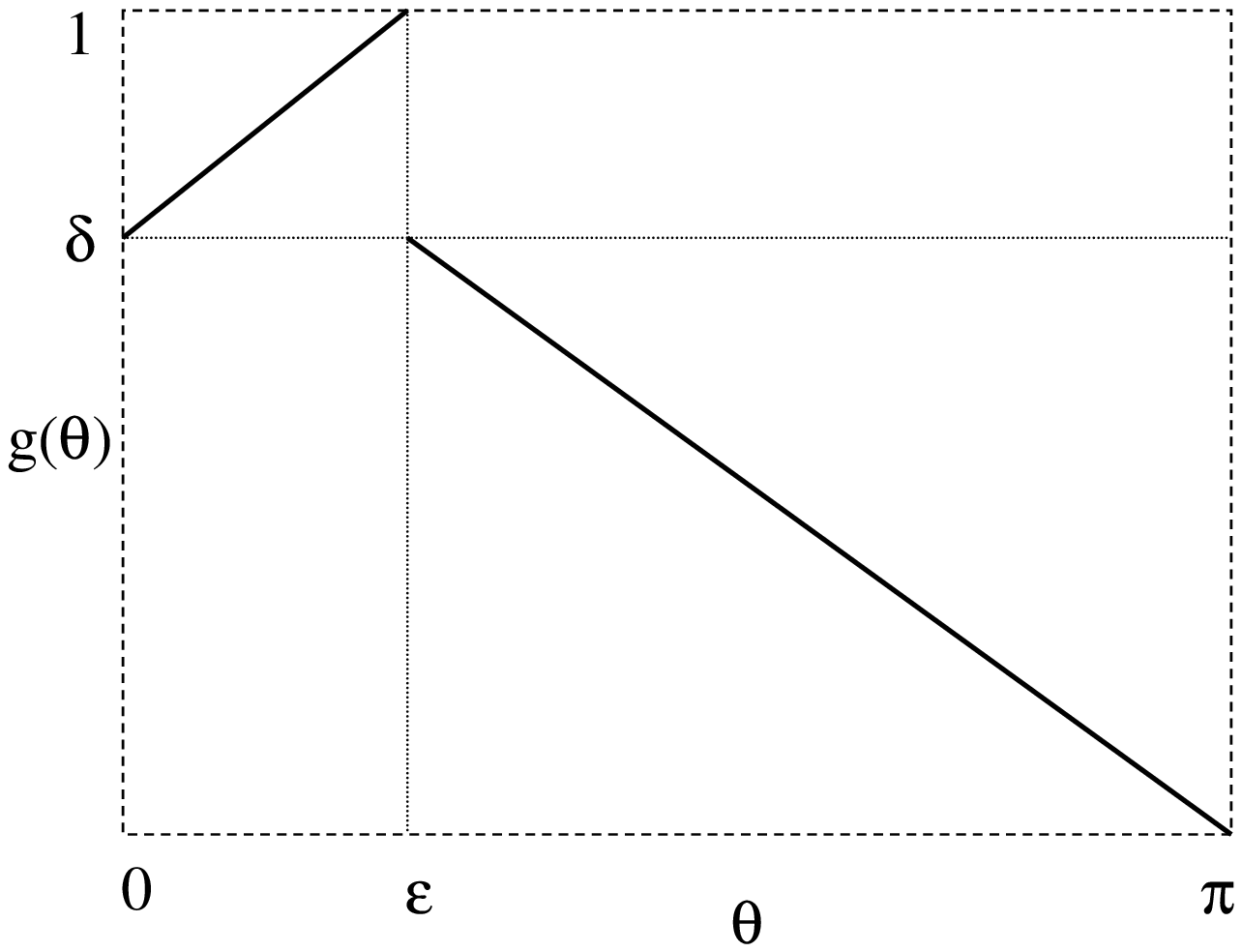}%
\\
Figure $5$-$6$: Invertible and discontinuous function $g_{6}(\theta)$.
\label{Fig6QCGs}%
\end{center}
%EndExpansion

With this function the pure classical equilibria $p_{A}^{\star}=p_{B}^{\star
}=0$ of PD as well as of BoS remain unaffected because these equilibria
require $\theta=\pi,$ and the function is not discontinuous at $\pi$. One
notices that if the angle corresponding to a classical equilibrium is
$0,\pi/2,$ or $\pi$, and there is no discontinuity at $\pi/2$, then the
quantum correlation game cannot change that equilibrium. With the function
(\ref{InvDis5}) in both PD or BoS the pure equilibrium with $p_{A}^{\star
}=p_{B}^{\star}=1$ corresponds to the angle $\theta=\epsilon$ where classical
and quantum correlations are different (for $\epsilon\neq\pi/2$).
Consequently, the equilibrium $p_{A}^{\star}=p_{B}^{\star}=1$ will be shifted
and the new equilibrium depends on the angle $\arccos(1-2\epsilon/\pi)$. The
mixed equilibrium of BoS will also be shifted by the function (\ref{InvDis5}).
Therefore, one of the pure equilibria and the mixed equilibrium may shift if
the $g$-function (\ref{InvDis5}) is chosen. The following function
\begin{equation}
g_{7}(\theta)=\left\{
\begin{array}
[c]{rcl}%
1-(1-\delta)\theta/\epsilon & \text{if} & \theta\in\lbrack0,\epsilon]\,,\\
\delta(\theta-\epsilon)/(\pi-\epsilon) & \text{if} & \theta\in(\epsilon
,\pi]\,,
\end{array}
\right.  \label{InvDis6}%
\end{equation}
where $\delta\in(0,1)$ and $\epsilon\in(0,\pi),$ cannot change the pure
equilibrium at $p_{A}^{\star}=p_{B}^{\star}=1$. However, it can affect the
equilibrium $p_{A}^{\star}=p_{B}^{\star}=1$, both in PD and BoS, and it can
shift the mixed equilibrium of BoS. Fig. (\ref{Fig7QCGs}) shows this function.%

%TCIMACRO{\FRAME{dtbpFU}{2.6308in}{1.9233in}{0pt}{\Qcb{Figure $5$-$7$:
%Invertible and discontinuous function $g_{7}(\theta)$.}}{\Qlb{Fig7QCGs}%
%}{fig7.eps}{\special{ language "Scientific Word";  type "GRAPHIC";
%maintain-aspect-ratio TRUE;  display "PICT";  valid_file "F";
%width 2.6308in;  height 1.9233in;  depth 0pt;  original-width 8.0652in;
%original-height 10.3233in;  cropleft "0.1888";  croptop "0.9057";
%cropright "0.8903";  cropbottom "0.5065";
%filename 'Fig7.eps';file-properties "XNPEU";}}}%
%BeginExpansion
\begin{center}
\includegraphics[
trim=1.522710in 5.228752in 0.884753in 0.973487in,
height=1.9233in,
width=2.6308in
]%
{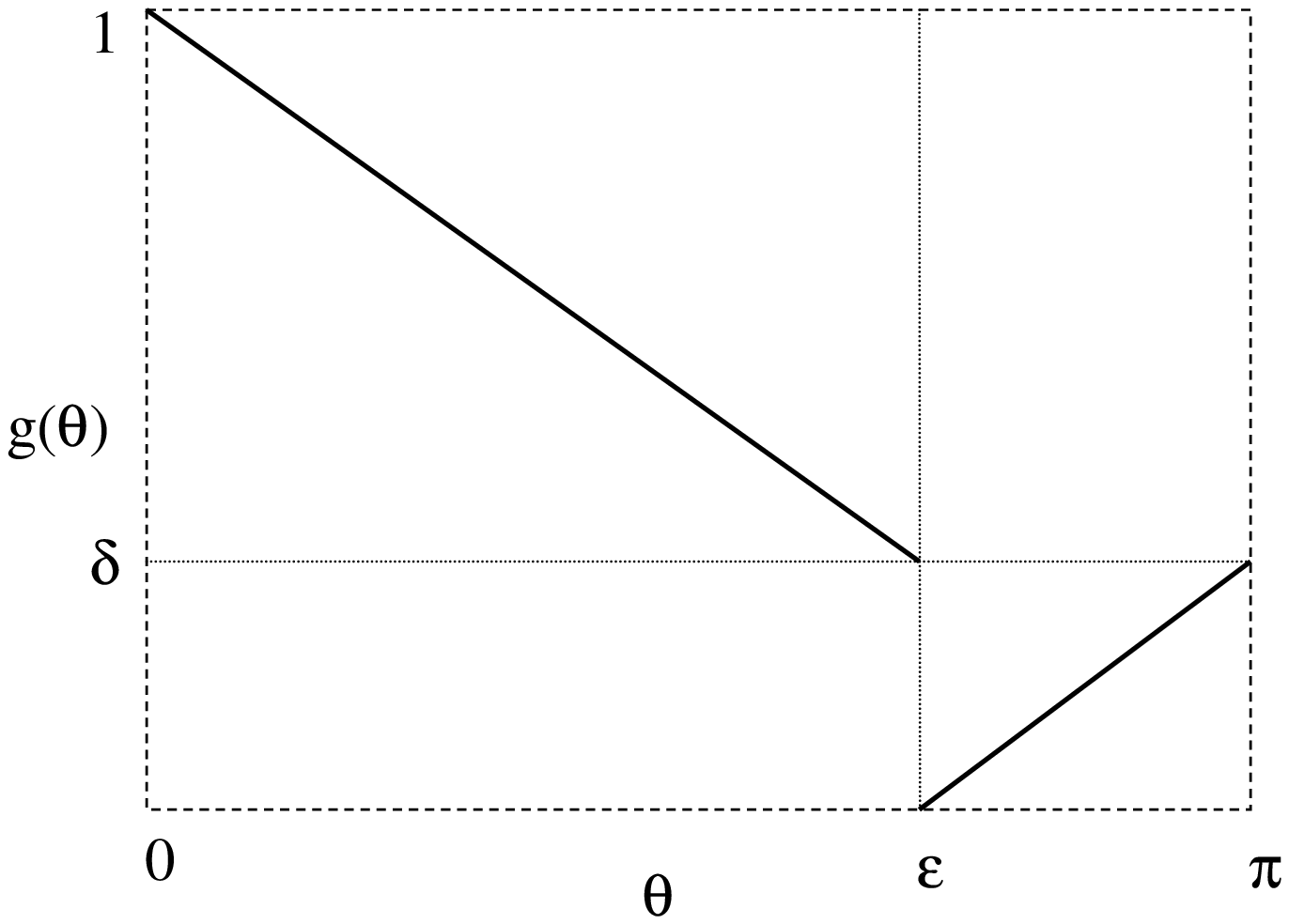}%
\\
Figure $5$-$7$: Invertible and discontinuous function $g_{7}(\theta)$.
\label{Fig7QCGs}%
\end{center}
%EndExpansion

\subsubsection{($iii$) Non-invertible and discontinuous $g$-functions}

A simple case of a continuous and non-invertible function (cf. Fig,
(\ref{Fig8QCGs})) is given by
\begin{equation}
g_{8}(\theta)=\left\{
\begin{array}
[c]{rcl}%
2\theta/\pi & \text{if} & \theta\in\lbrack0,\frac{\pi}{2}]\text{ ,}\\
1-2(\theta-\frac{\pi}{2})/\pi & \text{if} & \theta\in(\frac{\pi}{2},\pi]\text{
.}%
\end{array}
\right.  \label{NonInv1}%
\end{equation}
Consider a classical pure equilibrium with $p_{A}^{\star}=p_{B}^{\star}=0$.
Because $g^{-1}(0)=0$ or $\pi,$ two equilibria with $g\left\{  \arccos
(\pm1)\right\}  $ are generated in the quantum correlation game, but these
coincide and turn out to be same as the classical ones. Similarly, the
function (\ref{NonInv1}) does not shift the pure classical equilibrium at
$p_{A}^{\star}=p_{B}^{\star}=1$. However, if $p_{A,B}^{\star}\in(0,1)$
corresponds to a mixed equilibrium such that $g^{-1}(p^{\star})=\theta
1^{\star},\theta2^{\star}\neq\pi/2$ then in the quantum correlation game,
$p_{A,B}^{\star}$ will not only shift but will also bifurcate. The resulting
values will differ from $p_{A,B}^{\star}$.%

%TCIMACRO{\FRAME{dtbpFU}{2.7406in}{1.9199in}{0pt}{\Qcb{Figure $5$-$8$:
%Non-invertible and continuous function $g_{8}(\theta)$.}}{\Qlb{Fig8QCGs}%
%}{fig8.eps}{\special{ language "Scientific Word";  type "GRAPHIC";
%maintain-aspect-ratio TRUE;  display "PICT";  valid_file "F";
%width 2.7406in;  height 1.9199in;  depth 0pt;  original-width 8.0652in;
%original-height 10.3233in;  cropleft "0.1745";  croptop "0.9058";
%cropright "0.8899";  cropbottom "0.5158";
%filename 'Fig8.eps';file-properties "XNPEU";}}}%
%BeginExpansion
\begin{center}
\includegraphics[
trim=1.407377in 5.324759in 0.887978in 0.972455in,
height=1.9199in,
width=2.7406in
]%
{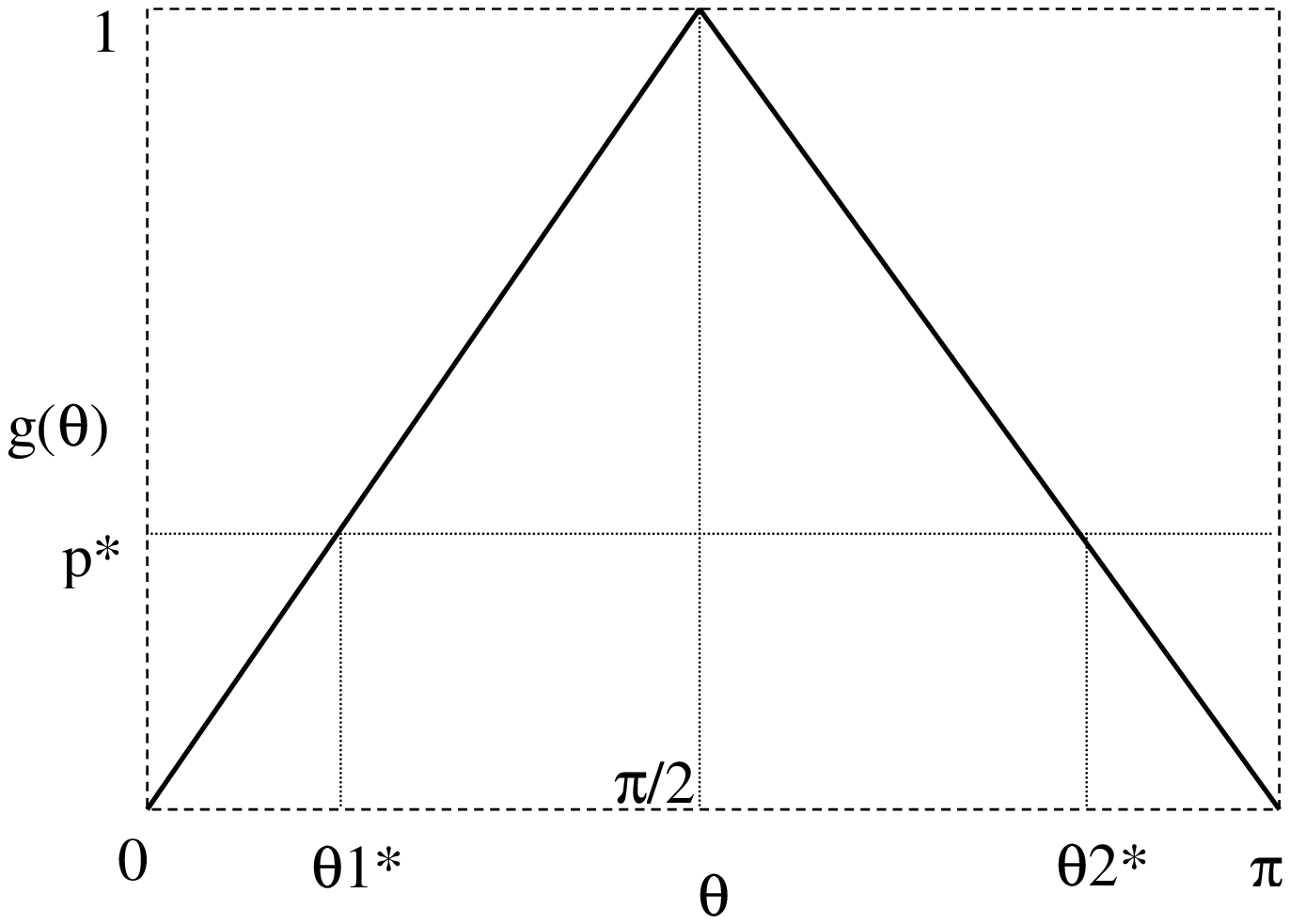}%
\\
Figure $5$-$8$: Non-invertible and continuous function $g_{8}(\theta)$.
\label{Fig8QCGs}%
\end{center}
%EndExpansion

We now ask whether the equilibria in a classical correlation game already
(without quantum effects) show a similar sensitivity in the presence of a
non-invertible and continuous $g$-function like (\ref{NonInv1})? When the
players receive the classical pairs of objects the angles $\theta1^{\star
},\theta2^{\star}$ are mapped to themselves, resulting in the same probability
$p^{\star}$, obtained now using the non-invertible and continuous $g$-function
(\ref{NonInv1}). Therefore, in a classical correlation game played with the
function (\ref{NonInv1}) the bifurcation observed in the quantum correlation
game does not show up, in spite of the fact that there are \emph{two} angles
associated with one probability.

\section{Discussion}

\label{sec:Discussion}In this chapter, we propose a new approach to the
introduction of a quantum mechanical version of two-player games. One of our
main objectives has been to find a way to respect two constraints when
`quantizing': on the one hand, no new moves should emerge in the quantum game
(c1) and, on the other hand, the payoff relations should remain unchanged
(c2). The constraints c1 and c2 have been introduced in the section $5.1$. In
this way, we hope to circumvent objections \cite{EnkPike}\ which have been
raised against the existing procedures \cite{Eisert} used to quantize games.
It has been pointed out in the literature that new quantum moves or modified
payoff relations do not necessarily indicate a true quantum character of a
game since their emergence can be understood in terms of a modified
\emph{classical} game. Our intention is to avoid this type of criticism.

\emph{Correlation games} are based on payoff relations which are sensitive to
whether the input is anti-correlated classically or quantum mechanically. The
players' allowed moves are fixed once and for all, and a setting inspired by
EPR-type experiments is used. Alice and Bob are both free to select a
direction in prescribed planes $\mathcal{P}_{A,B}$; subsequently they
individually measure, on their respective halves of the supplied system, the
value of a dichotomic variable either along the selected axis or along the
$z$-axis. When playing mixed strategies they must use probabilities which are
related to the angles by a function $g$ which is made public in the beginning.
After many runs the arbiter establishes the correlations between the
measurement outcomes and rewards the players according to fixed payoff
relations $P_{A,B}$. The rewards depend only on the numerical values of the
\emph{correlations} -- by definition, they do not make reference to classical
or quantum mechanics.

If the incoming states are classical, correlation games reproduce classical
bimatrix games. The payoffs $P_{A,B}^{cl}$ and $P_{A,B}^{q}$ correspond to one
single game, since both expressions emerge from the same payoff relation
$P_{A,B}$. If the input consists of quantum mechanical singlet states,
however, the correlations turn quantum and the solutions of the correlation
game change. For example, in a generalized Prisoners' Dilemma a \emph{mixed}
Nash equilibrium can be found. This is due to an effective \emph{non-linear}
dependence of the payoff relations on the probabilities, since the comparison
of Eqs. (\ref{clpayoffs}) and (\ref{Qpayoffs}) shows that `quantization' leads
to the substitution%

\begin{equation}
p_{A,B}\rightarrow Q_{g}(p_{A,B})\,. \label{quantize!}%
\end{equation}
As the payoffs of traditional bimatrix games are bi-linear in the
probabilities, it is difficult, if not impossible, to argue that the quantum
features of the quantum correlation game would arise from a disguised
classical game: there is no obvious method to let the payoffs of a classical
matrix game depend non-linearly on the strategies of the players.

Our analysis of the Prisoners' Dilemma and the Battle of Sexes as quantum
correlation games shows that, typically, both the structure and the location
of classical Nash equilibria are modified. The location of the quantum
equilibria depends sensitively on the properties of the function $g$ but,
apart from exceptional cases, the modifications are structurally stable. It is
\emph{not} possible to create any desired type of solution for a bimatrix game
by a smart choice of the function $g$.

Finally, we would like to comment on the link between correlation games and
Bell's inequality. In spite of their similarity to an EPR-type experiment, it
is not obvious how correlation games can directly exploit Bell's inequality.
Actually, violation of the inequality is \emph{not} crucial for the emergence
of the modifications in the quantum correlation game, as one can see from the
following argument. Consider a correlation game played on a mixture of quantum
mechanical anti-correlated \emph{product} states,
\begin{equation}
\hat{\rho}=\frac{1}{4\pi}\int_{\Omega}\,d\Omega\,|\mathbf{{e}_{\Omega}^{+}%
,{e}_{\Omega}^{-}\rangle\langle{e}_{\Omega}^{+},{e}_{\Omega}^{-}|\,,}
\label{mixture}%
\end{equation}
where the integration is over the unit sphere. The vectors $\mathbf{e}%
_{\Omega}^{\pm}$ are of unit length, and $|\mathbf{e}_{\Omega}^{\pm}\rangle$
denote the eigenstates of the spin component $\mathbf{e}_{\Omega}\cdot
\hat{\mathbf{S}}$ with eigenvalues $\pm1$, respectively. The correlations in
this entangled mixture are weaker than for the singlet state $|\psi\rangle$,
\begin{equation}
\left\langle ac\right\rangle _{\rho}=-\frac{1}{3}\cos\theta_{A}\,,\qquad
\text{\mbox{etc.}} \label{rhocorrels}%
\end{equation}
The factor $1/3$ makes a violation of Bell's inequality impossible.
Nevertheless, a classical bimatrix game is modified as before if $\hat{\rho}$
is chosen as the input state of the correlation game. To put this observation
differently: the payoffs introduced in Eq. (\ref{re-expressed2}) depend on the
two correlations $\langle ac\rangle$ and $\langle cb\rangle$ only, not on the
third one $\left\langle ab\right\rangle $ which is present in Bell's inequality.

\chapter{\label{HV NPs in QGs}Hidden variables and negative probabilities in
quantum games}

\section{Introduction}

Quantum mechanical predictions are non-deterministic i.e. the theory generally
does not predict outcome of any measurement with certainty. Quantum theory
merely tells what the probabilities of the outcomes are. This sometimes leads
to the strange situation in which measurements of a certain property done on
two identical systems can give different answers. The question naturally
arises whether there might be some `deeper objective reality' which is hidden
beneath quantum mechanics and which can be described by a more fundamental
theory that can always predict the outcome of each measurement with certainty.

The search for more fundamental theories led physicists to the \emph{Hidden
Variable Theories }(HVTs) \cite{Belinfante} which assume that the
probabilistic features of quantum mechanical predictions are due to the
complicated random behaviour of a classical substructure. That is, the outcome
of measurements are statistical averages over hidden variables. The variables
are hidden in the sense that we don't know anything about them, although no
fundamental principle forbids us from knowing them.

In the early 1980s, while pointing to a deep link existing between hidden
variables and probabilistic relations, Fine \cite{Fine}\ observed that the
existence of a deterministic hidden variables model is \emph{strictly
equivalent} to the existence of a joint probability distribution function for
the four observables involved in a correlation experiment, like the EPR-type
experiment. The joint probability is such that it returns the probabilities of
the experiments as marginals.

Apart from Fine's indication\footnote{In the later part of his paper Fine
\cite{Fine} commented ``\emph{hidden variables and Bell inequalities are all
about imposing requirements to make well defined precisely those probability
distributions for non-commuting observables whose rejection is the very
essence of quantum mechanics}''.} of the close relationship between hidden
variables models and the existence of joint probability distributions, there
is also another viewpoint on the EPR paradox. During recent years, some
authors \cite{Han et al,Cereceda,Rothman} have provided explicit proofs
showing how the assumption of hidden variables forces certain probability
measures (involved in the EPR paradox) to assume negative values. Negative
probability is indeed a stark departure from Kolmogorov's axioms of
probability, invoked in an attempt to explain the EPR paradox.

Interestingly, even negative probability is not the only possible way in which
quantum mechanics has been shown to depart from the classical world. Atkinson
\cite{Atkinson} has identified two other ways in which quantum mechanics can
depart from Kolmogorov's axioms:

\begin{enumerate}
\item  By allowing that $\Pr(A\cup B)\neq\Pr(A)+\Pr(B)$ although $A\cap
B=\phi$.

\item  By allowing $\Pr(A\cap B)\neq\Pr(A)\Pr(B)$ although $A$ and $B$ are
physically independent.
\end{enumerate}

The first way was proclaimed by Dirac and Feynman as the most striking feature
of quantum mechanics \cite{Atkinson}. In \cite{Atkinson98} Atkinson discussed
the second way.

Historically, M\"{u}ckenheim \cite{Muckenheim} used negative probability in
1982 in an attempt to resolve the EPR paradox. It is, of course, impossible to
give physical meaning to a negative probability within relative frequency
interpretation of probability. But negative probabilities always come together
with positive probabilities, to which they are added in such a way that final
predictions are always positive in the relative-frequency interpretation of
probability. Quoting M\"{u}ckenheim \cite{Muckenheim1}: ``\textit{Kolmogorov's
axiom may hold or not; the probability for the existence of negative
probabilities is not negative.}'' M\"{u}ckenheim is not alone in this approach
to the EPR paradox. For example, Feynman \cite{FenymanNProb,FeynmanNProb1}
once cleverly anticipated \cite{Cereceda,Rothman} that `` \textit{The
\emph{only} difference between the classical and quantum cases is that in the
former we assume that the probabilities are positive-definite.}'' For obvious
reasons, this particular approach to resolve the EPR paradox is taken
\cite{Muckenheim2} to be ``\textit{as unattractive as all others.''}.

In the present chapter we adapt the positive attitude towards negative
probabilities by observing that, in spite of their unattractiveness, they seem
to have undisputed value in terms of providing an indication as to what can
perhaps be taken as the ``true quantum character'' in certain quantum
mechanical experiments. Secondly, the negative probabilities, though labelled
`unattractive', seem to have the potential to provide alternative (and perhaps
much shorter) routes to the construction of simple examples of quantum games.
These constructions, designed for the physical implementation and realization
of quantum games, will, of course, again have the EPR experiments as their
underlying physical structure.

The approach towards quantum games which is developed in this chapter can be
divided into two steps. In the first step elementary probability theory is
used to analyze a hypothetical physical implementation of a two-player game.
The implementation is motivated by, and has close similarities with, the
experimental set-up of the EPR experiments. From an alternative viewpoint this
approach can also be taken as a procedure that re-defines the classical game
in a way that justifies and facilitates the next step in the present approach.
In the second step the peculiar probabilities arising out of the EPR-type
experiments are introduced in order to see their resulting effects on players'
payoff functions and on the solutions of the game.

Apart from being a short-cut towards the demonstration and construction of
simple quantum games, this approach seems to us to be presently needed both in
game theory and in economics \cite{Piotrowski2}. Recent years have witnessed
serious efforts to introduce the methods and notation of quantum mechanics
into these domains. In our view, in spite of these developments, it remains a
fact that, in these domains, the concepts of a wavefunction, of the ket bra
notation, and of Hermitian operators still have an ``alien'' appearance, with
quantum mechanics being believed to be their \emph{only} rightful place. The
present chapter tries to fill this gap by looking at how quantum games can
also be understood by using the peculiar probabilities that appear in certain
well-known quantum mechanical experiments, without recourse to the
mathematical tools of quantum mechanics. In other words, we try to show how
the unusual probabilities arising in the EPR paradox have a potential to leave
their mark on game theory.

We compare the playing of a two-player game with the setting of EPR-type
experiments in order to motivate a four-coin physical implementation of the
game. We then develop such a hypothetical physical implementation. Building on
this construction we look at how the game is affected when, instead of four
coins, two correlated particles are used to play the game.

\section{\label{Physical implementation of a bi-matrix game}Physical
implementation of playing a two-player game}

We consider a two-player two-strategy non-cooperative game that is given as a
bimatrix. Two players Alice and Bob are not allowed to communicate and each
player has to go for one of the two available strategies. A usual physical
implementation of this game consists of giving two coins to Alice and Bob,
each receiving one. Both receive the coin in a ``head'' state. Each player
plays his/her strategy, which consists of flipping/not flopping the coin. The
players then return their coins to a referee. The referee observes the coins
and rewards the players according to the strategies that they have played and
the game under consideration.

Consider now the well-known setting of EPR experiments. Once again, Alice and
Bob are spatially separated and are unable to communicate with each other.
Both receive one half of a pair of particles that originate from a common
source. In an individual run both choose one from the two options (strategies)
available to him/her. The strategies are usually two directions in space along
which measurements can be made. Each measurement generates $+1$ or $-1$ as the
outcome, which can be associated with the head and tail states of a coin,
respectively. Experimental results are recorded for a large number of
individual runs of the experiment.

Apparent similarities between the two-coin physical implementation of a
two-player game and EPR experiments can immediately be noticed. The
similarities hint at the possibility of using the EPR experiments to play
two-player games. However, before moving further along that direction, we can
make the following observations:

\begin{enumerate}
\item [ a)]In a two-coin implementation of a bimatrix game a player knows the
head or tail state of his/her coin after he/she has played his/her strategy.

\item[ b)] In an EPR experiment when a player decides his/her strategy as one
of two available directions along which a measurement is to be made, he/she
does not know whether the measurement is going to result in $+1$ or $-1$ until
the measurement has actually been made.
\end{enumerate}

This shows that a two-coin physical implementation of a two-player game is not
strictly analogous to an EPR experiment. In an individual run of the EPR
experiment each player has to chose from one of two available directions.
After the player makes a choice between the two directions the measurement
generates $+1$ or $-1$ as outcome. This suggests a four-coin implementation of
the game.

\section{\label{Games with four coins}Two-player games with four coins}

A two-player game can also be played using four coins instead of the two used
in the game described above. A procedure that physically implements the game
with four coins is suggested as follows. Its motivation comes from the EPR
experiments and it serves to make possible, in the next step, a smoother
transition towards a situation in which the \emph{same} game is played using
those experiments.

The players Alice and Bob are given two coins each. It is not known, and it
does not matter, whether the given coins are in head or tail states. Before
the game starts the referee announces that a player's strategy is to choose
one out of the two coins in his/her possession. After playing their
strategies, the players give the two chosen coins, representing their
strategies, to the referee. The referee tosses both the coins together and
records the outcomes. The tossing of the coins is repeated a large number of
times, while each player plays his/her move each time he/she is asked to
choose one of the two coins in his/her possession. After a large number of
individual runs the referee rewards the players according to the strategies
they have played and the probability distributions followed by the four coins
during their tossing.

After stating the general idea, we consider in the following an example of a
symmetric game.%

\begin{equation}%
\begin{array}
[c]{c}%
\text{Alice}%
\end{array}%
\begin{array}
[c]{c}%
X_{1}\\
X_{2}%
\end{array}
\overset{\overset{%
\begin{array}
[c]{c}%
\text{Bob}%
\end{array}
}{%
\begin{array}
[c]{cc}%
\acute{X}_{1} & \acute{X}_{2}%
\end{array}
}}{\left(
\begin{array}
[c]{cc}%
(K,K) & (L,M)\\
(M,L) & (N,N)
\end{array}
\right)  } \label{matrix}%
\end{equation}
where $X_{1}$, $X_{2}$, $\acute{X}_{1}$ and $\acute{X}_{2}$ are the players'
strategies. Entries in parentheses are Alice's and Bob's payoffs, respectively.

We want to physically implement the game (\ref{matrix}) using repeated tossing
of four coins which follows the following probability distribution%

\begin{equation}%
\begin{array}
[c]{c}%
\text{Alice}%
\end{array}
\overset{%
\begin{array}
[c]{c}%
\text{Bob}%
\end{array}
}{%
\begin{array}
[c]{cc}%
\begin{array}
[c]{c}%
S_{1}%
\end{array}%
\begin{array}
[c]{c}%
H\\
T
\end{array}
\overset{\overset{%
\begin{array}
[c]{c}%
\acute{S}_{1}%
\end{array}
}{%
\begin{array}
[c]{cc}%
H & T
\end{array}
}}{\left(
\begin{array}
[c]{cc}%
p_{1} & p_{2}\\
p_{3} & p_{3}%
\end{array}
\right)  } & \overset{\overset{%
\begin{array}
[c]{c}%
\acute{S}_{2}%
\end{array}
}{%
\begin{array}
[c]{cc}%
H & T
\end{array}
}}{\left(
\begin{array}
[c]{cc}%
p_{5} & p_{6}\\
p_{7} & p_{8}%
\end{array}
\right)  }\\%
\begin{array}
[c]{c}%
S_{2}%
\end{array}%
\begin{array}
[c]{c}%
H\\
T
\end{array}
\left(
\begin{array}
[c]{cc}%
p_{9} & p_{10}\\
p_{11} & p_{12}%
\end{array}
\right)  & \left(
\begin{array}
[c]{cc}%
p_{13} & p_{14}\\
p_{15} & p_{16}%
\end{array}
\right)
\end{array}
} \label{4 Coin Stats}%
\end{equation}
where, for example, $S_{1}$ is Alice's pure strategy to ``always select the
coin $1$'' etc. The pure strategies $S_{2},\acute{S}_{1}$ and $\acute{S}_{2}$
can be interpreted similarly. Also, $H\thicksim$Head and $T\thicksim$Tail and,
for obvious reasons, we have%

\begin{equation}
\overset{4}{\underset{1}{\sum}}p_{i}=\overset{8}{\underset{5}{\sum}}%
p_{i}=\overset{12}{\underset{9}{\sum}}p_{i}=\overset{16}{\underset{13}{\sum}%
}p_{i}=1 \label{Constraints}%
\end{equation}
In the construction of the four-coin statistics (\ref{4 Coin Stats}) the
following points should be taken into consideration.

\begin{enumerate}
\item  The statistics (\ref{4 Coin Stats}) may convey the impression that in
an individual run both players forward both of their coins to the referee, who
then tosses the four coins together. In fact, in an individual run the referee
tosses only two coins. The statistics (\ref{4 Coin Stats}) are generated under
the \emph{assumption} that there is randomness involved in the players'
strategies to go for one or the other coin.

\item  Associated with the above impression is the fact that in every
individual run the statistics (\ref{4 Coin Stats}) assign head or tail states
to the two coins that have \emph{not} been tossed. Thus, in each individual
run two tosses are \emph{counterfactual }\cite{AsherPeres}.
\end{enumerate}

With reference to our coin game, being counterfactual means that two coins out
of four are not tossed in each individual turn, but these untossed coins are
still assigned head or tail states in the mathematical steps used in the
derivation of Bell-CHSH inequality. This assignment is often justified by an
argument which goes under the label of \emph{realism}.

To play the game (\ref{matrix}) with four-coin statistics (\ref{4 Coin
Stats}), we assume that the referee uses the following recipe\footnote{The
recipe (\ref{Payoffs}) is not unique and others may be suggested. Any recipe
is justified if it is able to reproduce the game under consideration within
the description of the statistical experiment involved.} to reward the players:%

\begin{equation}
\left.
\begin{array}
[c]{c}%
P_{A}(S_{1},\acute{S}_{1})=Kp_{1}+Lp_{2}+Mp_{3}+Np_{4}\\
P_{A}(S_{1},\acute{S}_{2})=Kp_{5}+Lp_{6}+Mp_{7}+Np_{8}\\
P_{A}(S_{2},\acute{S}_{1})=Kp_{9}+Lp_{10}+Mp_{11}+Np_{12}\\
P_{A}(S_{2},\acute{S}_{2})=Kp_{13}+Lp_{14}+Mp_{15}+Np_{16}%
\end{array}
\right\}  \label{Payoffs}%
\end{equation}
Here $P_{A}(S_{1},\acute{S}_{2})$, for example, is Alice's payoff when she
plays $S_{1}$ and Bob plays $\acute{S}_{2}$. The corresponding payoff
expressions for Bob can be found by the transformation $L\leftrightarrows M$
in Eqs. (\ref{Payoffs}). The recipe, of course, makes sense if repeated tosses
are made with four coins. $S_{1},S_{2},\acute{S}_{1}$and $\acute{S}_{2}$ are
taken as players' pure strategies; a mixed strategy for Alice, for example, is
a convex linear combination of $S_{1}$ and $S_{2}$.

We now find constraints on the four-coin statistics (\ref{4 Coin Stats}) such
that each equation in (\ref{Payoffs}) represents a mixed strategy payoff
function for the bimatrix game (\ref{matrix}) and that can be written in a
bi-linear form. To allow this interpretation for the payoff functions
(\ref{Payoffs}) four probabilities $r,$ $s,$ $\acute{r}$ and $\acute{s}$ are
required that can give a bi-linear representation to the payoff functions
(\ref{Payoffs}) i.e.%

\begin{equation}
\left.
\begin{array}
[c]{c}%
P_{A}(S_{1},\acute{S}_{1})=Kr\acute{r}+Lr(1-\acute{r})+M\acute{r}%
(1-r)+N(1-r)(1-\acute{r})\\
P_{A}(S_{1},\acute{S}_{2})=Kr\acute{s}+Lr(1-\acute{s})+M\acute{s}%
(1-r)+N(1-r)(1-\acute{s})\\
P_{A}(S_{2},\acute{S}_{1})=Ks\acute{r}+Ls(1-\acute{r})+M\acute{r}%
(1-s)+N(1-s)(1-\acute{r})\\
P_{A}(S_{2},\acute{S}_{2})=Ks\acute{s}+Ls(1-\acute{s})+M\acute{s}%
(1-s)+N(1-s)(1-\acute{s})
\end{array}
\right\}  \label{bilinear payoffs}%
\end{equation}
This then allows us to make the association%

\begin{equation}
S_{1}\thicksim r\text{, \ \ }S_{2}\thicksim s\text{, \ \ }\acute{S}%
_{1}\thicksim\acute{r}\text{, \ \ }\acute{S}_{2}\thicksim\acute{s}
\label{correspondences}%
\end{equation}
where $r,s,\acute{r}$ and $\acute{s}$ are the probabilities of heads for coins
$S_{1},S_{2},\acute{S}_{1}$ and $\acute{S}_{2}$, respectively. If a consistent
set of these four probabilities can be found, as is always the case, then each
equation in (\ref{Payoffs}) can be interpreted in terms of a mixed strategy
game between the two players. Thus the four pairs%

\begin{equation}
(r,\acute{r}),\text{ \ }(r,\acute{s}),\text{ \ }(s,\acute{r}),\text{
\ }(s,\acute{s}) \label{probability pairs}%
\end{equation}
represent the four possible situations that may result when each player has
two strategies to choose from. For example, the strategy pair $(S_{1}%
,\acute{S}_{2})$ is associated with the pair $(r,\acute{s})$ and it
corresponds to the mixed strategy game given as%

\begin{equation}%
\begin{array}
[c]{c}%
\text{Alice}%
\end{array}%
\begin{array}
[c]{c}%
r\\
(1-r)
\end{array}
\overset{\overset{%
\begin{array}
[c]{c}%
\text{Bob}%
\end{array}
}{%
\begin{array}
[c]{ccc}%
\acute{s} &  & (1-\acute{s})
\end{array}
}}{\left(
\begin{array}
[c]{cc}%
(K,K) & (L,M)\\
(M,L) & (N,N)
\end{array}
\right)  }%
\end{equation}
In this game Alice plays $S_{1}$ with probability of heads $r$, and Bob plays
$\acute{S}_{2}$ with probability of heads $\acute{s}$. The other equations in
(\ref{bilinear payoffs}) can be given similar interpretation.

We now find constraints on the four-coin statistics (\ref{4 Coin Stats}) that
make the payoff functions (\ref{Payoffs}) identical to the bi-linear payoff
functions (\ref{bilinear payoffs}), for any real numbers $K,L,M$ and $N$. A
comparison of these equations shows that this happens when $r,s,\acute{r}$ and
$\acute{s}$ depend on $p_{i}$, for $16\geqslant i\geqslant0$, as follows:%

\begin{equation}
\left.
\begin{array}
[c]{c}%
r\acute{r}=p_{1},\text{ \ \ }r(1-\acute{r})=p_{2},\text{ \ \ }\acute
{r}(1-r)=p_{3},\text{ \ \ }(1-r)(1-\acute{r})=p_{4}\\
r\acute{s}=p_{5},\text{ \ \ }r(1-\acute{s})=p_{6},\text{ \ \ }\acute
{s}(1-r)=p_{7},\text{ \ \ }(1-r)(1-\acute{s})=p_{8}\\
s\acute{r}=p_{9},\text{ \ \ }s(1-\acute{r})=p_{10},\text{ \ \ }\acute
{r}(1-s)=p_{11},\text{ \ \ }(1-s)(1-\acute{r})=p_{12}\\
s\acute{s}=p_{13},\text{ \ \ }s(1-\acute{s})=p_{14},\text{ \ \ }\acute
{s}(1-s)=p_{15},\text{ \ \ }(1-s)(1-\acute{s})=p_{16}%
\end{array}
\right\}  \label{probability definitions}%
\end{equation}
The probabilities $r,s,\acute{r}$ and $\acute{s}$ can be extracted from
(\ref{probability definitions}) as follows:%

\begin{equation}
r=p_{1}+p_{2}\text{, \ \ }s=p_{9}+p_{10}\text{, \ \ }\acute{r}=p_{1}%
+p_{3}\text{, \ \ }\acute{s}=p_{5}+p_{7} \label{discrete prob}%
\end{equation}
provided that the $p_{i}$ satisfy the relations%

\begin{equation}
\left.
\begin{array}
[c]{c}%
p_{1}+p_{2}=p_{5}+p_{6},\text{ \ \ }p_{1}+p_{3}=p_{9}+p_{11}\\
p_{9}+p_{10}=p_{13}+p_{14},\text{ \ \ }p_{5}+p_{7}=p_{13}+p_{15}%
\end{array}
\right\}  \label{constraints on probs}%
\end{equation}
With the defining relations (\ref{discrete prob}), and the constraints
(\ref{constraints on probs}) on the four-coin statistics, each pair in
$(S_{1},\acute{S}_{1}),(S_{1},\acute{S}_{2}),(S_{2},\acute{S}_{1})$ and
$(S_{2},\acute{S}_{2})$ has an interpretation as a mixed strategy game. The
correspondence (\ref{correspondences}) means that, for example, Alice's payoff
functions become%

\begin{equation}
\left.
\begin{array}
[c]{c}%
P_{A}(S_{1},\acute{S}_{1})=P_{A}(r,\acute{r})\text{, \ \ }P_{A}(S_{1}%
,\acute{S}_{2})=P_{A}(r,\acute{s})\\
P_{A}(S_{2},\acute{S}_{1})=P_{A}(s,\acute{r})\text{, \ \ }P_{A}(S_{2}%
,\acute{S}_{2})=P_{A}(s,\acute{s})
\end{array}
\right\}
\end{equation}
Now, suppose that $(s,\acute{s})$ is a Nash equilibrium (NE) i.e.%

\begin{equation}
\left\{  P_{A}(s,\acute{s})-P_{A}(r,\acute{s})\right\}  \geqslant0\text{,
\ \ }\left\{  P_{B}(s,\acute{s})-P_{B}(s,\acute{r})\right\}  \geqslant0
\label{NE}%
\end{equation}
Using Eqs. (\ref{bilinear payoffs}) one gets%

\begin{equation}
\left.
\begin{array}
[c]{c}%
P_{A}(s,\acute{s})-P_{A}(r,\acute{s})=(s-r)\left\{  (K-M-L+N)\acute
{s}+(L-N)\right\}  \geqslant0\\
P_{B}(s,\acute{s})-P_{B}(s,\acute{r})=(\acute{s}-\acute{r})\left\{
(K-M-L+N)s+(L-N)\right\}  \geqslant0
\end{array}
\right\}  \label{NE1}%
\end{equation}
Consider the game of Prisoners' Dilemma which is produced when $M>K>N>L$ in
the matrix (\ref{matrix}). We select our first representation of PD by taking%

\begin{equation}
K=3,\text{ }L=0,\text{ }M=5\text{ \ and \ }N=1 \label{PDConstants1}%
\end{equation}
and the inequalities (\ref{NE1}) are then reduced to%

\begin{equation}
0\geqslant(s-r)(1+\acute{s}),\text{ \ \ }0\geqslant(\acute{s}-\acute{r})(1+s)
\label{NEPD}%
\end{equation}
Now from (\ref{NE1}) the pair $(s,\acute{s})$ is a NE when both inequalities
in (\ref{NEPD}) are true for all $r,\acute{r}\in\lbrack0,1]$. Because
$(1+s)\geqslant1$ and $(1+\acute{s})\geqslant1$, $s=\acute{s}=0$ is the
equilibrium. In the present set-up, to play PD this equilibrium appears if,
apart from the constraints of Eqs. (\ref{discrete prob}, \ref{constraints on
probs}), we also have%

\begin{equation}
s=p_{9}+p_{10}=0,\text{ \ \ }\acute{s}=p_{5}+p_{7}=0
\label{four-coin constraints2}%
\end{equation}
which are other constraints on the four-coins statistics (\ref{4 Coin Stats})
to hold true if the PD produces the NE $(s,\acute{s})=(0,0)$.

The above analysis can be reproduced for other probability pairs in
(\ref{probability pairs}). For example, when $(r,\acute{s})$ is NE i.e.%

\begin{equation}
\left\{  P_{A}(r,\acute{s})-P_{A}(s,\acute{s})\right\}  \geqslant0\text{,
\ \ }\left\{  P_{B}(r,\acute{s})-P_{B}(r,\acute{r})\right\}  \geqslant0
\label{2ndNE}%
\end{equation}
we again get $(r,\acute{s})=(0,0)$. However, the relations (\ref{discrete
prob}) now say that it will exist as a NE when%

\begin{equation}
p_{1}+p_{2}=0\text{, \ \ }p_{5}+p_{7}=0
\end{equation}
which should, of course, be true along with the relations (\ref{constraints on
probs}). That is, in order to reproduce a particular NE in the two-player game
the probabilities of heads of the four coins representing the game need to be
fixed. Also, from the bi-linear payoff functions (\ref{bilinear payoffs}) it
is clear that at the equilibria $(s,\acute{s})=(0,0)$ and $(r,\acute
{s})=(0,0)$ the reward for both the players is $N$.

Summarizing, we have shown that when the four-coin statistics (\ref{4 Coin
Stats}) satisfy the constraints of Eqs. (\ref{Constraints}, \ref{constraints
on probs}), the payoff functions (\ref{Payoffs}) can be interpreted in terms
of a mixed strategy version of a bimatrix game. In this setting four
strategies $S_{1},S_{2},\acute{S}_{1}$ and $\acute{S}_{2}$ available to the
players are associated with the probabilities $r,s,\acute{r}$ and $\acute{s}$,
respectively. This association allows us to interpret the payoff recipe of Eq.
(\ref{Payoffs}) in terms of a mixed strategy game. We showed that when
$r,s,\acute{r}$ and $\acute{s}$ are expressed in terms of the probabilities
$p_{i}$ for $16\geqslant i\geqslant1$ (as is the case in the Eqs.
(\ref{discrete prob})) the bi-linear payoff functions (\ref{bilinear payoffs})
become identical to the payoff functions (\ref{Payoffs}). This procedure is
designed to re-express the playing of a two-player game with four coins in
such a way that choosing which coin to toss is part of a player's strategy.

\section{\label{Games with perfectly correlated}Games with perfectly
correlated particles}

The re-expression of the playing of a two-player game in terms of a four-coin
tossing experiment opens the way to seeing what happens when the four-coin
statistics become correlated, particularly what if the correlations go beyond
what is achievable with the so-called classical `coins'.

At present there exists general agreement in the quantum physics community
that the EPR-type experiments, when performed on correlated pairs of
particles, violate the predictions of LHV models. Negative probabilities are
found to emerge when certain LHV models are forced to predict the experimental
outcomes of EPR-type experiments. For example, Han, Hwang and Koh \cite{Han et
al} showed the need for negative probabilities when explicit solutions for
probability measures can reproduce quantum mechanical predictions for some
spin-measurement directions and for all entangled states. In their analysis a
special basis is used to show the appearance of negative probabilities for a
class of LHV models.

Rothman and Sudarshan \cite{Rothman} demonstrated that quantum mechanics does
predict a set of probabilities that violate the Bell-CHSH inequality. The
predicted probabilities, however, are not positive-definite but are physically
meaningful in that they give the usual quantum-mechanical predictions in
physical situations. Rothman and Sudarshan observed that all derivations of
Bell's inequalities assume that LHV theories produce a set of
positive-definite probabilities for detecting a particle with a given spin orientation.

Using a similar approach, Cereceda \cite{Cereceda} proved independently the
necessity of negative probabilities in \textit{all} instances where the
predictions of the LHV model violate the Bell-CHSH inequality. Interestingly,
Cereceda's proof does not rely on any particular basis states or measurement
directions. In the concluding section of his paper Cereceda analyzes the case
of pairs of particles that have perfect correlation between them and proceeds
to show the necessity of negative probabilities for those pairs.

The necessity of involving negative probability measures in order to explain
the experimental outcomes in EPR-type experiments motivates questions about
the consequences that these may have for the solutions of a game which is
physically implemented using such experiments. This question can be expressed
as follows. What happens to the players' payoff functions and the solutions of
a game that is physically implemented by using pairs of perfectly correlated
particles? It seems quite reasonable to demand that when the predictions of an
LHV model agree with the Bell-CHSH inequality then the game attains a
classical interpretation.

One clear advantage of the above approach towards quantum games appears to be
that it is possible to see, without using the machinery of quantum mechanics,
how game-theoretical solutions are affected when a game is physically
implemented using quantum-mechanically correlated pairs of particles.

We follow Cereceda's notation \cite{Cereceda} for probabilities in EPR-type
experiments which are designed to test Bell-CHSH inequality. Two correlated
particles $1$ and $2$ are emitted in opposite directions from a common source.
Afterwards, each of the particles enters its own measuring apparatus which can
measure either one of two different physical variables at a time. We denote
these variables $S_{1}$ or $S_{2}$ for particle $1$ and $\acute{S}_{1}$ or
$\acute{S}_{2}$ for particle $2$. These variables can take possible values of
$+1$ and $-1$. The source emits a very large number of particle pairs.

To describe this experiment Cereceda \cite{Cereceda} considers a deterministic
hidden variable model as follows. The model assumes that there exists a hidden
variable $\lambda$ for every pair of particles emitted by the source.
$\lambda$ has a domain of variation $\Lambda$ and it determines locally (for
example, at the common source) the response of the particle to each of the
measurements to which it can be subjected. It is possible to partition the set
of all $\lambda$ into $16$ disjoint subsets $\Lambda_{i}$ (with respect to a
probability measure $m_{i}$) according to the outcomes of the four possible
measurements, $S_{1}$ or $S_{2}$ for particle $1$ and $\acute{S}_{1}$ or
$\acute{S}_{2}$ for particle $2$. Table $1$ is reproduced from \cite{Cereceda}%
. It shows the $16$ rows characterizing the subsets $\Lambda_{i}$. The $i$-th
row gives the outcome of different measurements when the particle pair is
described by a hidden variable pertaining to the subset $\Lambda_{i}$.

Table $1$: The set $\Lambda$ is partitioned into $16$ possible subsets. The
hidden variables in each subset $\Lambda_{i}$ uniquely determine the outcomes
for each of the four possible single measurements $S_{1},\acute{S}_{1},S_{2},$
and $\acute{S}_{2}$. The table is reproduced from Ref. \cite{Cereceda}.%

\[%
\begin{tabular}
[c]{lll}%
Subset of $\Lambda$ & $%
\begin{array}
[c]{cccc}%
S_{1} & \acute{S}_{1} & S_{2} & \acute{S}_{2}%
\end{array}
$ & {\small Probability measure}\\
$%
\begin{array}
[c]{c}%
\Lambda_{1}\\
\Lambda_{2}\\
\Lambda_{3}\\
\Lambda_{4}\\
\Lambda_{5}\\
\Lambda_{6}\\
\Lambda_{7}\\
\Lambda_{8}\\
\Lambda_{9}\\
\Lambda_{10}\\
\Lambda_{11}\\
\Lambda_{12}\\
\Lambda_{13}\\
\Lambda_{14}\\
\Lambda_{15}\\
\Lambda_{16}%
\end{array}
$ & \multicolumn{1}{c}{$%
\begin{array}
[c]{cccc}%
+ & + & + & +\\
+ & + & + & -\\
+ & + & - & +\\
+ & + & - & -\\
+ & - & + & +\\
+ & - & + & -\\
+ & - & - & +\\
+ & - & - & -\\
- & + & + & +\\
- & + & + & -\\
- & + & - & +\\
- & + & - & -\\
- & - & + & +\\
- & - & + & -\\
- & - & - & +\\
- & - & - & -
\end{array}
$} & $%
\begin{array}
[c]{c}%
m_{1}\\
m_{2}\\
m_{3}\\
m_{4}\\
m_{5}\\
m_{6}\\
m_{7}\\
m_{8}\\
m_{9}\\
m_{10}\\
m_{11}\\
m_{12}\\
m_{13}\\
m_{14}\\
m_{15}\\
m_{16}%
\end{array}
$%
\end{tabular}
\]
The probabilities $p_{i}$ are given below in obvious notation \cite{Cereceda}.%

\begin{align}
p_{1}  &  \equiv p(S_{1}+;\acute{S}_{1}+)=m_{1}+m_{2}+m_{3}+m_{4}%
,\label{Eq1}\\
p_{2}  &  \equiv p(S_{1}+;\acute{S}_{1}-)=m_{5}+m_{6}+m_{7}+m_{8}%
,\label{Eq2}\\
p_{3}  &  \equiv p(S_{1}-;\acute{S}_{1}+)=m_{9}+m_{10}+m_{11}+m_{12}%
,\label{Eq3}\\
p_{4}  &  \equiv p(S_{1}-;\acute{S}_{1}-)=m_{13}+m_{14}+m_{15}+m_{16},\\
p_{5}  &  \equiv p(S_{1}+;\acute{S}_{2}+)=m_{1}+m_{3}+m_{5}+m_{7}%
,\label{Eq5}\\
p_{6}  &  \equiv p(S_{1}+;\acute{S}_{2}-)=m_{2}+m_{4}+m_{6}+m_{8}%
,\label{Eq6}\\
p_{7}  &  \equiv p(S_{1}-;\acute{S}_{2}+)=m_{9}+m_{11}+m_{13}+m_{15},\\
p_{8}  &  \equiv p(S_{1}-;\acute{S}_{2}-)=m_{10}+m_{12}+m_{14}+m_{16},\\
p_{9}  &  \equiv p(S_{2}+;\acute{S}_{1}+)=m_{1}+m_{2}+m_{9}+m_{10},\\
p_{10}  &  \equiv p(S_{2}+;\acute{S}_{1}-)=m_{5}+m_{6}+m_{13}+m_{14},\\
p_{11}  &  \equiv p(S_{2}-;\acute{S}_{1}+)=m_{3}+m_{4}+m_{11}+m_{12},\\
p_{12}  &  \equiv p(S_{2}-;\acute{S}_{1}-)=m_{7}+m_{8}+m_{15}+m_{16},\\
p_{13}  &  \equiv p(S_{2}+;\acute{S}_{2}+)=m_{1}+m_{5}+m_{9}+m_{13},\\
p_{14}  &  \equiv p(S_{2}+;\acute{S}_{2}-)=m_{2}+m_{6}+m_{10}+m_{14},\\
p_{15}  &  \equiv p(S_{2}-;\acute{S}_{2}+)=m_{3}+m_{7}+m_{11}+m_{15},\\
p_{16}  &  \equiv p(S_{2}-;\acute{S}_{2}-)=m_{4}+m_{8}+m_{12}+m_{16}
\label{Eq16}%
\end{align}
Combining Eqs. (\ref{Constraints}) with Table $1$ gives%

\begin{equation}
\underset{i=1}{\overset{16}{\sum}}m_{i}=1 \label{Sum m's}%
\end{equation}
Continuing with Cereceda's description \cite{Cereceda}, an example is now
considered. Suppose that the particle pair is described by a given $\lambda
\in\Lambda_{2}$; then the particles must behave as follows. If $S_{1}$ is
measured on particle $1$ the result will be $+1$, if $S_{2}$ is measured on
particle $1$ the result will be $+1$, if $\acute{S}_{1}$ is measured on
particle $2$ the result will be $+1$, if $\acute{S}_{2}$ is measured on
particle $2$ the result will be $-1$. Also for each of the plans the results
of measurements made on particle $1$ are independent of the results of
measurements made on particle $2$.

For perfectly correlated particles two of the probabilities $p_{2}$ and
$p_{3}$ can be set equal to zero. Physically this means that the results for
the joint measurement of two observables, one for each particle, must both be
either $+1$ or $-1$. From a physical point of view, it is reasonable to
suppose that, for the case in which $p_{2}=0$ and $p_{3}=0$, the probability
measures $m_{5},m_{6},m_{7},m_{8},m_{9},m_{10},m_{11},$ and $m_{12}$ also
vanish. This can be verified from Eqs. (\ref{Eq2}, \ref{Eq3}). If this is not
the case then joint detection events will be generated by the LHV model which
do not happen, according to the assumptions made. Cereceda showed that in this
case, when the predictions of the LHV model violate the Bell-CHSH inequality,
the negativity of either $m_{4}$ or $m_{13}$ can be proved.

We assume now that the probabilities $p_{i}$ appearing in Eqs. (\ref{Payoffs})
correspond to the LHV model of the EPR-type experiments performed on perfectly
correlated particles to test the Bell-CHSH inequality. As is the case with the
four-coin tossing experiment, the payoff functions (\ref{Payoffs}) can be
interpreted as bi-linear payoff functions. To do this we use the Eqs.
(\ref{discrete prob}, \ref{constraints on probs}) to get%

\begin{equation}
r=p_{1}=p_{5}+p_{6},\text{ \ \ }\acute{r}=p_{1}=p_{9}+p_{11},\text{
\ \ }s=p_{9}+p_{10},\text{ \ \ }\acute{s}=p_{5}+p_{7}
\label{Probabilities Referee}%
\end{equation}
But $p_{2}=p_{3}=0$, so that from the first Eq. of (\ref{probability
definitions}) we have $r(1-\acute{r})=\acute{r}(1-r)$ which gives $p_{1}=0$ or
$1$ because $r=\acute{r}=p_{1}$ from (\ref{Probabilities Referee}). We
select\footnote{Selecting $p_{1}=0$ makes $p_{4}=1$ because $\underset
{i=1}{\overset{4}{\sum}}p_{i}=1$. It will result in different but analogous
expressions for $r,\acute{r},s$ and $\acute{s}$ given in terms of $m_{i}$
without affecting the present argument.} $p_{1}=1$. Now, from Eqs. (\ref{Eq5},
\ref{Eq6}) we have $p_{5}+p_{6}=\underset{i=1}{\overset{8}{\sum}}m_{i}$ and
because $\underset{i=1}{\overset{16}{\sum}}m_{i}=1$ from Eq. (\ref{Sum m's})
and $m_{i}=0$ for $12\geqslant i\geqslant5$ we get%

\begin{equation}
\left.
\begin{array}
[c]{c}%
r=\acute{r}=1=m_{1}+m_{2}+m_{3}+m_{4}\\
s=m_{1}+m_{2}+m_{13}+m_{14}\\
\acute{s}=m_{1}+m_{3}+m_{13}+m_{15}%
\end{array}
\right\}  \label{ProbsCorrelParticles}%
\end{equation}
With these relations the bi-linear payoff functions (\ref{bilinear payoffs})
are written as%

\begin{align}
P_{A}(S_{1},\acute{S}_{1})  &  =K\label{CorrelPayoff1}\\
P_{A}(S_{1},\acute{S}_{2})  &  =L+(K-L)(m_{1}+m_{3}+m_{13}+m_{15}%
)\label{CorrelPayoff2}\\
P_{A}(S_{2},\acute{S}_{1})  &  =M+(K-M)(m_{1}+m_{2}+m_{13}+m_{14})
\label{CorrelPayoff3}%
\end{align}%

\begin{equation}
\left.
\begin{array}
[c]{c}%
P_{A}(S_{2},\acute{S}_{2})=(K-L-M+N)\times\\
(m_{1}+m_{2}+m_{13}+m_{14})(m_{1}+m_{3}+m_{13}+m_{15})+\\
(L-N)(m_{1}+m_{2}+m_{13}+m_{14})+\\
(M-N)(m_{1}+m_{3}+m_{13}+m_{15})+N
\end{array}
\right\}  \label{CorrelPayoff4}%
\end{equation}
Each of the correlated payoff functions (\ref{CorrelPayoff2},
\ref{CorrelPayoff3}, \ref{CorrelPayoff4}) can be split into two parts i.e.%

\begin{equation}
\left.
\begin{array}
[c]{c}%
P_{A}(S_{1},\acute{S}_{2})=P_{A_{a}}(S_{1},\acute{S}_{2})+P_{A_{b}}%
(S_{1},\acute{S}_{2})\\
P_{A}(S_{2},\acute{S}_{1})=P_{A_{a}}(S_{2},\acute{S}_{1})+P_{A_{b}}%
(S_{2},\acute{S}_{1})\\
P_{A}(S_{2},\acute{S}_{2})=P_{A_{a}}(S_{2},\acute{S}_{2})+P_{A_{b}}%
(S_{2},\acute{S}_{2})
\end{array}
\right\}  \label{payoff splitting}%
\end{equation}
where%

\begin{align}
P_{A_{a}}(S_{1},\acute{S}_{2})  &  =L+(K-L)(m_{1}+m_{3}%
)\label{Alice's split payoffs1}\\
P_{A_{b}}(S_{1},\acute{S}_{2})  &  =(K-L)(m_{13}+m_{15}%
)\label{Alice's split payoffs2}\\
P_{A_{a}}(S_{2},\acute{S}_{1})  &  =M+(K-M)(m_{1}+m_{2}%
)\label{Alice's split payoffs3}\\
P_{A_{b}}(S_{2},\acute{S}_{1})  &  =(K-M)(m_{13}+m_{14})
\label{Alice's split payoffs41}%
\end{align}%

\begin{equation}
\left.
\begin{array}
[c]{c}%
P_{A_{a}}(S_{2},\acute{S}_{2})=(K-L-M+N)(m_{1}+m_{2})(m_{1}+m_{3})+\\
(L-N)(m_{1}+m_{2})+(M-N)(m_{1}+m_{3})+N
\end{array}
\right\}  \label{Alice's split payoffs5}%
\end{equation}%

\begin{equation}
\left.
\begin{array}
[c]{c}%
P_{A_{b}}(S_{2},\acute{S}_{2})=(K-L-M+N)\{(m_{1}+m_{2})(m_{13}+m_{15})+\\
(m_{1}+m_{3})(m_{13}+m_{14})+(m_{13}+m_{14})(m_{13}+m_{15})\}+\\
(L-N)(m_{13}+m_{14})+(M-N)(m_{13}+m_{15})
\end{array}
\right\}  \label{Alice's split payoffs6}%
\end{equation}
The significance of this splitting is that components $P_{A_{b}}(S_{1}%
,\acute{S}_{2}),$ $P_{A_{b}}(S_{2},\acute{S}_{1})$ and $P_{A_{b}}(S_{2}%
,\acute{S}_{2})$ of Alice's payoffs in Eqs. (\ref{payoff splitting}) become
zero when the predictions of the LHV model agree with the Bell-CHSH
inequality. Cheon and Tsutsui \cite{Cheon} have also shown a similar
splitting, using correlated payoff operators whose expectation values are the
players' payoffs.

Consider again the PD with the selection of the constants given in
(\ref{PDConstants1}). Let the game be played using perfectly correlated
particles, for which substitutions can be made from
(\ref{ProbsCorrelParticles}) into the Nash equilibrium condition (\ref{NEPD}).
This gives%

\begin{equation}
0\geqslant(s-1)(1+\acute{s}),\text{ \ \ }0\geqslant(\acute{s}-1)(1+s)
\label{NEPD1}%
\end{equation}
where%

\begin{align}
s  &  =m_{1}+m_{2}+m_{13}+m_{14}\label{def of s}\\
\acute{s}  &  =m_{1}+m_{3}+m_{13}+m_{15} \label{def of s'}%
\end{align}
It can be noticed that:

\begin{itemize}
\item  The predictions of the LHV model agree with the Bell-CHSH inequality.
It makes $m_{i}\geqslant0$ for all $16\geqslant i\geqslant1$. Combining it
with (\ref{ProbsCorrelParticles}), i.e.$\underset{i=1}{\overset{4}{\text{
}\sum m_{i}}}=1,$ gives%

\begin{equation}
m_{13}=m_{14}=m_{15}=m_{16}=0 \label{m13 to m16=0}%
\end{equation}
\end{itemize}

so that $s$ and $\acute{s}$ in (\ref{def of s}, \ref{def of s'}) are reduced to%

\begin{equation}
s_{1}=m_{1}+m_{2},\text{ \ \ }\acute{s}_{1}=m_{1}+m_{3} \label{New Def s s'}%
\end{equation}

\begin{itemize}
\item  The requirement $(s_{1},\acute{s}_{1})=(0,0)$ says that when the
predictions of the LHV model agree with the Bell-CHSH inequality then the pair
$(0,0)$ is a NE. Eq. (\ref{New Def s s'}) gives
\end{itemize}%

\begin{equation}
m_{1}=m_{2}=m_{3}=0 \label{00 is NE when LHV agrees chsh}%
\end{equation}
Before proceeding with a further question we make following observations:

\begin{enumerate}
\item  Eqs. (\ref{Eq1}) to (\ref{Eq16}) give the probabilities $p_{i}$ in
terms of $m_{i}$, for $16\geqslant i\geqslant1$, corresponding to the EPR-type
experiments. These probabilities satisfy the constraints (\ref{constraints on
probs}) which emerge when Eqs. (\ref{Payoffs}) are interpreted in terms of
bi-linear payoffs of Eqs. (\ref{bilinear payoffs}).

\item  The expressions (\ref{ProbsCorrelParticles}) for $r,s,\acute{r}$ and
$\acute{s}$ are obtained from the corresponding expressions for the coin
tossing case (\ref{discrete prob}), while taking into consideration the
constraints on the probabilities for perfectly correlated particles.

\item  The Eqs. (\ref{m13 to m16=0}, \ref{00 is NE when LHV agrees chsh})
together make $s=\acute{s}=0$ in the definitions (\ref{def of s}, \ref{def of
s'}). These definitions correspond to the representation (\ref{PDConstants1})
of PD, for which the constraints (\ref{four-coin constraints2}) should be true
in the case that $s=\acute{s}=0$ is a NE when the game is played with coins.
It is observed that both $(p_{9}+p_{10})$ and $(p_{5}+p_{7})$ become zero from
the definitions of $p_{i}$ in terms of $m_{i}$ given in Table $1$, when Eqs.
(\ref{m13 to m16=0}, \ref{00 is NE when LHV agrees chsh}) are both true. It
means that when PD gives $s=\acute{s}=0$ as an equilibrium the constraints on
the probabilities become identical in the following two cases:

a) The game is played using repeated tosses with four coins.

b) The game is played with perfectly correlated particles such that the
predictions of the LHV model agree with the Bell-CHSH inequality.
\end{enumerate}

Two-player games other than PD, presumably with different Nash equilibria,
would give rise to different but analogous constraints on the probabilities
$s,\acute{s},r$ and $\acute{r}$. In the light of these observations the
following question immediately arises. What happens to the Nash conditions
(\ref{NEPD1}) when the predictions of the LHV model disagree with the
Bell-CHSH inequality? To answer this question we consider the Nash conditions
(\ref{NEPD1}) with a substitution from (\ref{00 is NE when LHV agrees chsh}),
thus obtaining%

\begin{equation}
\left.
\begin{array}
[c]{c}%
0\geqslant(s_{2}-1)(1+\acute{s}_{2})\\
0\geqslant(\acute{s}_{2}-1)(1+s_{2})
\end{array}
\right\}  \label{NEPD2}%
\end{equation}
where%

\begin{equation}
s_{2}=m_{13}+m_{14},\text{ \ \ }\acute{s}_{2}=m_{13}+m_{15}
\label{NE when LHV violates chsh}%
\end{equation}
We recall that in Cereceda's analysis $m_{13}$ can take a negative value when
the predictions of the LHV model disagree with the Bell-CHSH inequality. Both
$(s_{2}-1)$ and $(\acute{s}_{2}-1)$ in (\ref{NEPD2}) remain negative whether
$m_{13}$ is positive or negative. Similarly, both $(1+\acute{s}_{2})$ and
$(1+s_{2})$ remain positive whether $m_{13}$ is positive or negative.
Therefore, the Nash conditions (\ref{NEPD2}), which correspond to the
representation (\ref{PDConstants1}) of PD, are \emph{not} violated whether the
predictions of the LHV model agree or disagree with the Bell-CHSH inequality.

The players' payoffs at the equilibrium $(s_{2},\acute{s}_{2})$ can be found
from Eq. (\ref{Alice's split payoffs6}) as%

\begin{equation}
\left.
\begin{array}
[c]{c}%
P_{A_{a}}(S_{2},\acute{S}_{2})=P_{B_{a}}(S_{2},\acute{S}_{2})=N\\
P_{A_{b}}(S_{2},\acute{S}_{2})=(K-L-M+N)s_{2}\acute{s}_{2}+(L-N)s_{2}%
+(M-N)\acute{s}_{2}\\
P_{B_{b}}(S_{2},\acute{S}_{2})=(K-L-M+N)s_{2}\acute{s}_{2}+(M-N)s_{2}%
+(L-N)\acute{s}_{2}%
\end{array}
\right\}
\end{equation}
where $s_{2}$ and $\acute{s}_{2}$ are found from (\ref{NE when LHV violates
chsh}). For example, with the PD representation (\ref{PDConstants1}), the
players' payoffs are obtained from Eqs. (\ref{Alice's split payoffs5},
\ref{Alice's split payoffs6}) as%

\begin{equation}
\left.
\begin{array}
[c]{c}%
P_{A}(S_{2},\acute{S}_{2})=\acute{s}_{2}(4-s_{2})-s_{2}+1\\
P_{B}(S_{2},\acute{S}_{2})=s_{2}(4-\acute{s}_{2})-\acute{s}_{2}+1
\end{array}
\right\}  \label{PD(1) payoffs (Q)}%
\end{equation}

The NE $(s_{2},\acute{s}_{2})$ in Eqs. (\ref{NE when LHV violates chsh})
becomes a solution when the predictions of LHV model disagree with the
Bell-CHSH inequality. Its defining inequalities (\ref{NEPD2}) show that the NE
exists even when either $s_{2}$ or $\acute{s}_{2}$ take negative values, which
can be realized when $m_{13}$ is negative. Accordingly, the NE in the PD
representation (\ref{PDConstants1}) can be `displaced' when the predictions of
LHV model disagree with the Bell-CHSH inequality. Here `displacement' means
that either $s_{2}$ or $\acute{s}_{2}$ can take negative values. However, this
extra freedom of assuming negative values does \emph{not} disqualify
$(s_{2},\acute{s}_{2})$ from existing as a NE. From Eq. (\ref{PD(1) payoffs
(Q)}) it can be seen that $P_{A}(S_{2},\acute{S}_{2})$ and $P_{B}(S_{2}%
,\acute{S}_{2})$ cannot be greater than $1$ when both $s_{2}$ and $\acute
{s}_{2}$ take negative values.

We now show that this may not be the case with another representation of PD.
That is, the extra freedom for $s_{2}$ and $\acute{s}_{2}$ to take negative
values, which is granted when the predictions of the LHV model disagree with
Bell-CHSH inequality, leads to the disqualification of $(s_{2},\acute{s}_{2})$
as a possible NE in that representation of PD.

Consider the PD again, but with a slightly different value assigned to the
constant $N$ of the game \cite{Cheon}:%

\begin{equation}
K=3,\text{ \ }L=0,\text{ \ }M=5\text{ \ and \ }N=0.2 \label{PDConstants2}%
\end{equation}
In this representation the inequalities (\ref{NE1}) are reduced to%

\begin{equation}
0\geqslant(s-r)(1.8\acute{s}+0.2),\text{ \ \ }0\geqslant(\acute{s}-\acute
{r})(1.8s+0.2)
\end{equation}
The substitution of $r=\acute{r}=1$ from (\ref{ProbsCorrelParticles}) and then
the addition of both the inequalities gives%

\begin{equation}
\frac{1}{9}\left\{  4(s+\acute{s})+1\right\}  \geqslant s\acute{s}
\label{CheonGameNE}%
\end{equation}
Suppose that the predictions of the LHV model disagree the Bell-CHSH
inequality i.e. both $s$ and $\acute{s}$ are to be replaced by $s_{2}$ and
$\acute{s}_{2}$ in (\ref{CheonGameNE}):%

\begin{equation}
\frac{1}{9}\left\{  4(s_{2}+\acute{s}_{2})+1\right\}  \geqslant s_{2}\acute
{s}_{2} \label{CheonGameNE1}%
\end{equation}
where $s_{2}$ and $\acute{s}_{2}$ are given by (\ref{NE when LHV violates
chsh}). Interestingly, it is observed that the inequality (\ref{CheonGameNE1})
is violated if $-0.25>(s_{2}+\acute{s}_{2})$ and the NE of Eq. (\ref{NE when
LHV violates chsh}) then ceases to exist. Of course, this applies to the
representation of PD given by (\ref{PDConstants2}). The players' payoffs are
given as%

\begin{equation}
\left.
\begin{array}
[c]{c}%
P_{A}(S_{2},\acute{S}_{2})=\acute{s}_{2}(4.8-1.8s_{2})+0.2(1-s_{2})\\
P_{B}(S_{2},\acute{S}_{2})=s_{2}(4.8-1.8\acute{s}_{2})+0.2(1-\acute{s}_{2})
\end{array}
\right\}  \label{PD(2) payoffs (Q)}%
\end{equation}
As it is the case with the first representation of PD, the payoffs
$P_{A}(S_{2},\acute{S}_{2}),P_{B}(S_{2},\acute{S}_{2})$ cannot be greater than
$0.2$ when both $s_{2}$ and $\acute{s}_{2}$ take negative values.

This shows that in the physical implementation of PD, using perfectly
correlated particles, the two representations (\ref{PDConstants1},
\ref{PDConstants2}) behave differently from each other. In representation
(\ref{PDConstants1}) the disagreement of the predictions of the LHV model with
the Bell-CHSH inequality leads to displacement of the NE $(s,\acute{s})$ such
that $s$ and $\acute{s}$ can assume negative values. Displacement occurs but
$(s,\acute{s})$ continues to exist as a NE.

On the other hand, in the representation (\ref{PDConstants2}) the disagreement
of the predictions of the LHV model with the Bell-CHSH inequality leads to the
disappearance of the NE $(s,\acute{s})$ when both $s$ and $\acute{s}$ assume
negative values and their sum becomes less than $-0.25$.

A `minimalist' interpretation of the present approach can also be given as
follows. Constraints (\ref{constraints on probs}, \ref{four-coin
constraints2}) are required on the four-coin statistics (\ref{4 Coin Stats})
to make $(s,\acute{s})=(0,0)$ a NE when the PD is played in representation
(\ref{PDConstants1}) with repeated tosses of four coins. When the \emph{same}
game is played with pairs of perfectly correlated particles and the
predictions of the LHV model disagree with the Bell-CHSH inequality, we can
have $(p_{9}+p_{10})$ or $(p_{5}+p_{7})$ becoming negative, which contradicts
(\ref{four-coin constraints2}). If it affects the solution of the game then
one can say that this is because of the change in the underlying probabilities
of our physical system. The present approach can be defended by observing that
it is \emph{not }the question of changing the underlying probabilities which
has been addressed here. On the other hand, we have introduced a new procedure
which addresses the problem of finding the true `quantum content' of a quantum
game in the following two steps:

\begin{enumerate}
\item [ a)]For perfectly correlated particles, developing an association such
that the results are guaranteed to be those of the classical game when the
predictions of the LHV model agree with the Bell-CHSH inequality.

\item[ b)] With the above association retained, finding how solutions of a
game are affected when the predictions of the LHV model disagree with the
Bell-CHSH inequality.
\end{enumerate}

When these steps are taken into consideration, the possibility of the
construction of a classical game which is able to reproduce the overall effect
of a quantum game cannot be taken to support the argument that quantum games
have no quantum content \cite{Enk}. In our opinion, the question which quantum
game theory asks is: How the quantum mechanical aspects of a physical system
can leave their mark on game-theoretic solutions? The possibility of a
classical construction of a quantum game does not make this question disappear.

\chapter{Summary and Future Perspectives}

Using entanglement to obtain new solutions of games has come out to be the
general theme of the new field of quantum game theory. The
Einstein-Podolsky-Rosen (EPR) paradox occupies the central position in the
phenomenon of quantum entanglement. The paradox motivated the EPR experiments
which confirmed the departure of quantum mechanical predictions from the
hidden variable models. These experiments are widely believed in the quantum
physics community to provide the strongest evidence of the `true quantum
character', even though what lessons such experiments teach about local
realism continues to attract active debate.

Both of the two suggestions developed in chapters \ref{Quantum Correlation
Games} and \ref{HV NPs in QGs} use EPR-type experiments to play quantum games.
These suggestions, however, are distinct from each another in the following
sense. Firstly, the suggestions use two different definitions of players'
strategies. Secondly, these suggestions are motivated by two different
expressions of the true quantum character, even though both suggestions use
EPR experiments for their physical implementation. The suggestion of chapter
\ref{Quantum Correlation Games} exploits the explicitly different classical
and quantum correlations for anti-correlated pairs of objects. The suggestion
of the chapter \ref{HV NPs in QGs} exploits the recently reported results that
some LHV models from the outcome of EPR experiments predict the assignment of
negative values to certain probability measures that are involved in such experiments.

In chapter \ref{Quantum Correlation Games} a quantization scheme for
two-player games is proposed, using different mathematical forms of the
correlations corresponding to entangled and product states, respectively, for
anti-correlated pairs of objects. The set-up uses EPR-type experiments such
that players' strategies are defined by unit vectors in the $x$-$z$ and
$y$-$z$ planes. Experimental measurements are performed in EPR-type setting
along the players' two chosen directions and the $z$-axis. Players'
strategies, represented by unit vectors, appear in the experimentally measured
correlations, allowing players' strategies to be obtained from the
experimental outcomes. In effect, this opens the way to re-express the playing
of a two-player game in terms of the experimental outcomes of EPR-type
experiments. In such a re-expression, the explicit form of the payoff
relations remain exactly the same, whether the game is played using classical
or quantum correlated pairs of objects. Also, in both the classical and
quantum version of a game the players' strategies remain the same, i.e.
rotations given to unit vectors. The set-up of a correlation game makes it
very difficult to argue that the quantum version of a game consists of
`another' classical game.

The following steps can be identified in the set-up for defining a correlation game:

\begin{enumerate}
\item [ S1]Making available a large number of anti-correlated objects, such
that each player has access to one half of the objects.

\item[ S2] Devise the players' strategies consisting of the rotations which
they may give to their respective unit vectors $a$ and $b$ residing in the
$x$-$z$ and $y$-$z$ planes, respectively. The $z$-axis is shared between the
players as the common direction.

\item[ S3] Referee uses an invertible function $g:[0,\pi]\rightarrow
\lbrack0,1]$ that translates players' rotations to probabilities $p_{A,B}%
\in\lbrack0,1]$.

\item[ S4] After the players have played their strategies, the referee
measures the correlation $\left\langle ac\right\rangle ,\left\langle
cb\right\rangle ,$ and $\left\langle ab\right\rangle $.

\item[ S5] To give a meaning to the experimental results the referee uses a
function $\theta:[-1,1]\rightarrow\lbrack0,\pi]$ that translates the
experimentally measured correlations into angles. This function is set to be
$\theta=\frac{\pi}{2}(1+\left\langle ij\right\rangle )$, independently of the
nature of any correlations which may exist between the anti-correlated pairs.

\item[ S6] The referee uses the function $g:[0,\pi]\rightarrow\lbrack0,1]$ to
get new probabilities $p_{A,B}^{\prime}$, which now depend on the correlations.

\item[ S7] Referee finds payoffs from the new probabilities $p_{A,B}^{\prime}$
by using the classical payoff relations for a two-player game.
\end{enumerate}

Clearly when the players receive classically correlated objects the step S5
becomes an identity operation and $p_{A,B}^{\prime}=p_{A,B}$, which means that
the correlation game will produce payoffs identical to those in a classical
matrix game. For quantum correlated objects we have $p_{A,B}^{\prime}\neq
p_{A,B}$ and the players' payoffs do not remain classical. The question of
interest now becomes whether the new payoffs can lead to new game-theoretical
equilibria. We showed that this indeed happens for the game of Prisoners'
Dilemma. We found that the new equilibria depend sensitively not only on the
correlations but also on the form of the invertible $g$-function, which is
used in playing the game. A careful selection of the $g$-function can result
in some equilibria in a game that has no classical analogue.

We now look at the quantization procedure suggested for a correlation game as
developed in this thesis, from the viewpoint of what quantization means for a
given classical system. For such a system in some specific state $S$, a
physical quantity can be associated with a (Borel) function\footnote{A Borel
function is measurable function, with respect to the Borel $\sigma$-algebras,
from one topological space to another.} $\Im:S\rightarrow R$, where $R$ is the
set of real numbers. `Quantization' can then be viewed \cite{Isham} as a map
$\Im\longmapsto\hat{\Im}$ associating to each such function $\Im$ a
self-adjoint operator $\hat{\Im}$ on the quantum state space $\mathcal{H}$
(Hilbert space). A quantum game can then be viewed as a triple $(\left|
\varphi\right\rangle ,\hat{\Im},\mho)$ where $\left|  \varphi\right\rangle $
is a quantum state in $\mathcal{H}$, $\hat{\Im}$ is a self-adjoint operator
and $\mho$\ is a function from the spectrum of $\hat{\Im}$ to $R$. That is,
the function $\mho$\ determines the payoffs in the quantum game.

Within this view of quantization a triple $(\left|  \varphi\right\rangle
,\hat{h},\mho)$ can be associated to a quantum correlation game proposed in
chapter \ref{Quantum Correlation Games} as follows. A classical correlation
game redefines a classical game using EPR-type experiments. The quantum
correlation game retains the essential structure of the corresponding
classical correlation game, except that:

\begin{enumerate}
\item [ a)]the state of an anti-correlated pair is in Hilbert space
$\mathcal{H}$

\item[ b)] the correlations $\left\langle ij\right\rangle $ are obtained as
eigenvalues of self-adjoint operators
\end{enumerate}

Thus in a quantum correlation game the step S5 finds the correlations
$\left\langle ij\right\rangle $ as eigenvalues of self-adjoint operators. All
the other steps remain the same both for classical and quantum correlation
games. It leads to the players obtaining payoffs, which are different from
those of the classical payoffs, when the pairs of objects they receive are
quantum correlated. Essentially this is because in both the classical and
quantum games the function $\theta=\frac{\pi}{2}(1+\left\langle
ij\right\rangle )$ is used to find angles from their experimentally determined
correlations $\left\langle ij\right\rangle $. The function guarantees that
players are rewarded classically whenever the correlations $\left\langle
ij\right\rangle $ become classical, because then $p_{A,B}^{\prime}$ become
identical to $p_{A,B}$. It is to be noticed that the function $\mho$, mapping
correlations into payoffs,\ remains the same for both the classical and
quantum correlation games.

A possible extension of the proposal of chapter \ref{Quantum Correlation
Games} is to the case in which a player's strategy involves choosing
\emph{any} direction in space, instead of one which is confined to the $x$-$z$
or $y$-$z$ planes. That is, with players having freedom on their choice of
directions, the payoffs of a correlation game are defined in such a way that
they become sensitive to a violation of Bell's inequality. In this case, the
construction would ensure that the game involves non-classical probabilities,
which are impossible to obtain by any classical game.

It appears that three-player games can be more appropriate for the proposed
extension of correlation games. For example, consider three players $A,B$ and
$C$, whose strategies are given by the unit vectors $\vec{a},\vec{b}$ and
$\vec{c}$, respectively. Let a Greenberger-Horne-Zeilinger (GHZ) state:%

\begin{equation}
\left|  \psi\right\rangle =(\left|  0\right\rangle _{1}\left|  0\right\rangle
_{2}\left|  0\right\rangle _{3}+\left|  1\right\rangle _{1}\left|
1\right\rangle _{2}\left|  1\right\rangle _{3})/\sqrt{2} \label{GHZ state}%
\end{equation}
be shared among the players, where $\left|  i\right\rangle _{j}$ is the $i$-th
state of the $j$-th qubit. Each player measures the dichotomic observable
$\vec{n}.\vec{\sigma}$ where $\vec{n}=\vec{a},\vec{b},\vec{c}$ and
$\vec{\sigma}$ is a vector, the components of which are the standard Pauli
matrices. The family of observables $\vec{n}.\vec{\sigma}$ covers all possible
dichotomic observables for a qubit system. Kaszlikowski and \.{Z}ukowski
\cite{Kaszlikowski} have reported that the probability of obtaining the result
$m=\pm1$ for the player $A$ when he plays the strategy $\vec{a}$, the result
$l=\pm1$ for the player $B$, when she plays the strategy $\vec{b}$ and the
result $k=\pm1$ for the player $C$ when she plays the strategy $\vec{c}$ is
equal to%

\begin{equation}
\Pr_{QM}(m,l,k;\vec{a},\vec{b},\vec{c})=\frac{1}{8}(1+mla_{3}b_{3}%
+mka_{3}c_{3}+lkb_{3}c_{3}+mlk\sum_{r,p,s=1}^{3}M_{rps}a_{r}b_{p}c_{s})
\label{Kaszlikowski equation}%
\end{equation}
where $a_{r}$, $b_{p}$, $c_{s}$ are components of vectors $\vec{a},\vec
{b},\vec{c}$ and where the nonzero elements of the tensor $M_{rps}$ are
$M_{111}=1,$ $M_{122}=-1,$ $M_{212}=-1,$ $M_{221}=-1$. Now, from Eq.
(\ref{Kaszlikowski equation}) it can be observed that in the absence of
three-qubit correlations the third components of the players' strategies can
be expressed in terms of the experimentally measured probabilities on the left
of (\ref{Kaszlikowski equation}) corresponding to a selection of values for
the triples $(m,l,k)$. Assume that a classical game is re-expressed so that
the players' strategies are the third components of the unit vectors which
they chose and their payoffs are functions of their strategies and of the
experimental probabilities on the left of (\ref{Kaszlikowski equation}). Such
a re-expression will then allow the three qubit correlations to reveal
themselves in the solutions of the game when the game is played with a GHZ
state (\ref{GHZ state}) of three qubits. Analysis along these lines seems to
be a natural extension of the approach developed in chapter \ref{Quantum
Correlation Games}. It appears that this extension will be free from the
weakness of the proposal made in the chapter \ref{Quantum Correlation Games},
namely that solutions different from classical ones can emerge even when input
states do not violate the Bell inequality. This feature can be regarded as a
weakness, because the emergence of a completely classical game when input
states do not violate the Bell inequality is a natural requirement.

In contrast to defining strategies in terms of directions, a different
definition of strategy is adopted in chapter \ref{HV NPs in QGs}. The argument
derives from the reported results that, when perfect correlations exist
between the two particles that are forwarded to the two players, the violation
of the Bell-CHSH inequality by the predictions of a class of LHV models forces
certain probability measures to take negative values. We investigated how can
this affects the solution of a two-player game which is physically implemented
using an EPR-type set-up. To establish better comparison with EPR-type
experiments, a hypothetical physical implementation of a two-player game is
developed that uses repeated tosses with four coins. This opens the way to
directly incorporate the peculiar probabilities which are involved in the
EPR-type experiments designed to test the Bell-CHSH inequality.

In the proposed set-up, two choices are made available to each player; each
player then decides on a strategy. The players' payoffs depend on the outcomes
of repeated measurements and on the constants defining the game. It is
required that a classical game results whenever a LHV model does not predict
negative probability.

We find that a consistent set of probabilities can be obtained, given in terms
of the statistics involved in the four-coin tossing experiment, such that the
game between the players is interpretable as a classical two-player
noncooperative game allowing mixed strategies. The proposal is designed in a
way that allows, in the next step, the introduction of the peculiar
probabilities emerging in the EPR-type experiments.

We find that when the game is played with perfectly correlated pairs of
particles the players' payoffs are observed to split into two parts,
corresponding to the two situations arising when the predictions of the class
of LHV models do and do not violate the Bell-CHSH inequality, respectively.

Apart from the splitting of the payoffs, we showed that the implementation
using perfectly correlated particles distinguishes between two representations
of the game that are completely equivalent in the classical context. In
general there is a linkage between the game's solutions and the determination
of whether the predictions of the LHV model do or do not violate the Bell-CHSH
inequality. We found that the degree of this linkage is sensitive to the
particular representation used for the game.

A possible extension of the approach presented in chapter \ref{HV NPs in QGs}
is to analyze three-player two-strategy games in the same set-up. For such
games, interestingly, instead of negative probabilities, HVTs predict
algebraic contradictions such as those appearing in the CHSH version of the
Bell theorem without inequalities \cite{CHSH,CHSH1,CHSH2}. It appears that
such games will present a stronger departure of their quantum solutions from
the classical ones.

Another possible line of investigation to extend the approach developed in
chapter \ref{HV NPs in QGs} is to use Fine's results which link the absence of
a joint probability distribution with the violation of the Bell inequalities.
That involves looking at the solutions of a game when it is implemented by
players sharing a physical system for which there is no joint probability
distribution able to produce the measurement outcomes as marginals. It appears
that, once again, three-player games will offer more convincing constructions.

Here it seems appropriate to mention that GHZ \cite{GHZ} used a so-called
deterministic hidden variable model in their approach to the GHZ paradox.
Khrennikov \cite{Khrennikov1} has presented a probabilistic analysis in the
contextualist framework, namely under the assumption that the distributions of
hidden variables depend on the settings of measurement devices. Khrennikov has
found that in the contextualist framework, there exist classes of probability
distributions of hidden variables such that the GHZ scheme does not imply a
contradiction between local realism and the quantum formalism. It seems that
Khrennikov's results may have a direct relevance for those proposals of three
player games in which the mentioned contradiction shows itself in the
game-theoretical solutions.

In this thesis we have referred to the continuing debate about the lessons
regarding local realism that arise from a violation of the Bell inequalities,
because it might have a bearing on the understanding of what constitutes the
true quantum character and content of quantum games. For example, one view
\cite{Enk,Pseudo Telepathy} interprets their `true' quantum character as being
their non-local aspect, which in the majority view \cite{AsherPeres} stands as
the unavoidable conclusion of the violation of the Bell inequalities. However,
another view \cite{Fine,Malley,Gustafson,DeBaere} says that non-locality is
not the sole or principal feature in deciding quantum character. Even the
presence of entanglement has been claimed as not describing everything that
may be called a true quantum character. Interestingly, entanglement and
non-locality, widely believed to be closely inter-linked, have now been
claimed \cite{GisinNew}\ to be different resources.

\begin{acknowledgement}
During work on this thesis I was supported by a research scholarship from the
University of Hull (UoH). Also, the Department of Mathematics (DoM) at the UoH
provided its support. I gratefully acknowledge supports from the UoH and the
DoM. I am thankful to the staff members at the DoM for providing both help and
a friendly environment.
\end{acknowledgement}

\end{document}